\documentclass[twocolumn,tighten]{aastex61}
\usepackage{graphicx}
\usepackage{hyperref}
\usepackage{color}
\usepackage[toc,page]{appendix}
\newcommand{\msun}{M$_{\odot}$\,}

\shorttitle{Precessing Disk in the Nucleus of M31}
\shortauthors{Lockhart et al.}

\begin{document}

\title{A Slowly Precessing Disk in the Nucleus of M31 as the Feeding
Mechanism for a Central Starburst}

\author{K. E. Lockhart}
\affiliation{Institute for Astronomy, University of Hawaii, Manoa,
Honolulu, HI 96822}

\author{J. R. Lu}
\affiliation{Dept. of Astronomy, University of California, Berkeley, CA 94720}

\author{H. V. Peiris}
\affiliation{Dept. of Physics and Astronomy, University College London, London WC1E 6BT, UK}
\affiliation{The Oskar Klein Centre for Cosmoparticle Physics,
  Stockholm University, AlbaNova, Stockholm, SE-106 91, Sweden}

\author{R. M. Rich}
\affiliation{Dept. of Physics and Astronomy, University of California, Los Angeles, CA 90095}

\author{A. Bouchez}
\affiliation{Giant Magellan Telescope, Pasadena, CA 91107}

\author{A. M. Ghez}
\affiliation{Dept. of Physics and Astronomy, University of California, Los Angeles, CA 90095}

\correspondingauthor{Jessica R. Lu}
\email{jlu.astro@berkeley.edu}

\begin{abstract}
We present a kinematic study of the nuclear stellar disk in M31 at
infrared wavelengths using high spatial resolution integral field
spectroscopy.
The spatial resolution achieved, FWHM =
0\farcs12 (0.45 pc at the distance of M31), has only previously
been equaled in spectroscopic studies by space-based
long-slit observations. 
Using adaptive optics-corrected integral field
spectroscopy from the OSIRIS instrument at the W.~M.~Keck Observatory, we map the line-of-sight
kinematics over the entire old stellar eccentric disk orbiting the
supermassive black hole (SMBH) at a distance of $r <$ 4 pc. The peak
velocity dispersion is 
381$\pm$55 km s$^{-1}$, offset by 0\farcs13 $\pm$ 0\farcs03 from the SMBH, 
consistent with previous high-resolution long-slit observations. There
is a lack of near-infrared (NIR) emission at the 
position of the SMBH and young nuclear cluster, suggesting a spatial separation between the young and 
old stellar populations within the nucleus. We compare
the observed kinematics with dynamical models from \citet{peiris2003eccentric-disk}. The best-fit disk
orientation to the NIR flux is [$\theta_l$, $\theta_i$, $\theta_a$] =
[$-$33$^{\circ}\pm$4$^{\circ}$, 44$^{\circ}\pm$2$^{\circ}$, $-$15$^{\circ}\pm$15$^{\circ}$],
which is tilted with respect to both the larger-scale galactic disk and the best-fit orientation derived from optical observations.
The precession rate of the old
disk is $\Omega_P = 0.0 \pm 3.9$ km s$^{-1}$ pc$^{-1}$, lower than the majority of
previous observations. This slow precession rate suggests that stellar
winds from the disk will collide and shock, driving rapid gas inflows
and fueling an episodic central starburst as suggested in
\citet{chang2007the-origin}. 
\end{abstract}

\keywords{black hole physics -- Galaxy:center --
infrared: stars -- techniques: high angular resolution -- galaxies:
star clusters -- galaxies: kinematics and dynamics -- galaxies: star formation}

\section{Introduction}
\label{sec:intro}

Many galaxies harbor not only supermassive black holes (SMBH), 
but also young stellar populations in
the central few parsecs. The origin of these young stars is unusual 
given that extreme tidal forces near the SMBH will
shear typical molecular clouds apart before they can collapse to form
stars \citep[e.g.,][]{Sanders:1992,Morris:1993}. 
A young or intermediate age stellar population has been
detected in nearly all nuclear star clusters found in nearby galaxies  
\citep{Allen:1990, rossa2006hubble, Seth:2006, Paumard:2006}, including in the nuclear star cluster of 
M31.
The SMBH in M31
\citep[distance $d$=785 pc, black hole mass M$_\bullet \sim 10^8$
\msun][]{mcconnachie2005distances,bender2005hst-stis}
is surrounded by a young ($<200$ Myr) nuclear cluster
designated P3, which is visible in the ultraviolet (UV), has a total mass 
of $\sim 10^4$ \msun, is confined to the central  
0\farcs1 (0.4 pc), and is orbiting in a disk coplanar with the rest of the nuclear star cluster
\citep[e.g.,][]{lauer1998m32-/--1,bender2005hst-stis,lauer2012the-cluster}. 
The abundance of young stars around nearby SMBHs suggests
that whatever physical mechanism deposits or forms stars in this
region may be important in many galactic nuclei. 

One explanation for the existence of young stars near a SMBH is {\em in
situ} star formation in a massive self-gravitating gas disk around the
SMBH, such as has been postulated for the Milky Way from kinematic
studies of its young stellar population 
\citep{Levin:2003,Lu:2009,yelda2014properties}. Such a disk would be sufficiently
dense to overcome the strong tidal forces and fragment to form stars,
as has been suggested in the context of active galactic nuclei (AGN) accretion disks
\citep[e.g.,][]{Kolykhalov:1980,Goodman:2003}. 
Unlike the Milky Way, which hosts $\geq$10$^4$ M$_{\odot}$ of cold gas and dust in the inner 5 pc \citep{genzel1985the-neutral-gas}, M31 is
relatively gas-poor in the central few kpc 
\citep{Sofue:1993,Nieten:2006} and the origin of the gas that formed the young
nuclear star cluster is not yet known.

\begin{figure}[t]
\begin{center}
\includegraphics[width=\columnwidth]{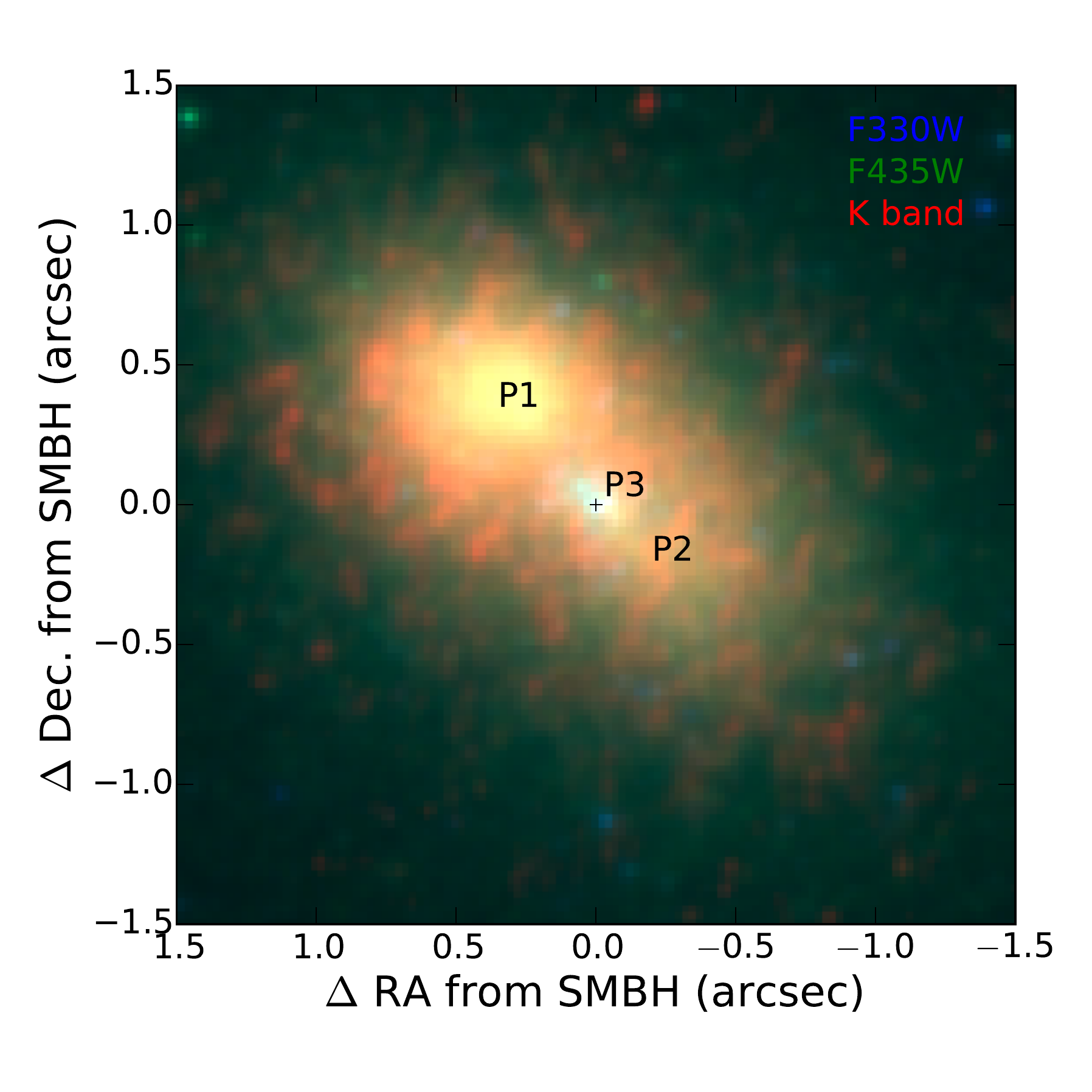}
\caption{Three-color image of the nucleus of M31 with components labeled. The eccentric disk is demarcated by 
P1 at apoapse and P2 at periapse and contains old stars. The young nuclear cluster, P3, is centered
on the SMBH and is prominent at blue wavelengths.  1$''$ is $\sim$4 pc
at M31's distance of 785 pc.} 
\label{fig:3colorm31}
\end{center}
\end{figure}

In M31, there is an eccentric disk
of old stars extending a few
parsecs from the central SMBH \citep{tremaine1995an-eccentric-disk}
that has been postulated
to both be the source of the molecular gas and the means for ushering that gas into the central parsec 
where the young stars are observed
\citep{chang2007the-origin}. The mass loss from the red giants and asymptotic
giant branch stars in this disk is high enough to drive a new \emph{in situ}
star formation event every 500 Myr, if there are crossing orbits in
the eccentric disk of old stars. The crossing orbits would lead to gas
collisions, shocks, and inflows on timescales shorter than typical
viscous times. The presence of such crossing orbits depends critically
on a low precession speed ($\lesssim$10 km s$^{-1}$ pc$^{-1}$) for the eccentric
disk of old stars.

The structure and kinematics of the eccentric nuclear disk have been
studied at multiple wavelengths. Optical high-spatial-resolution
observations with the \emph{Hubble Space Telescope} (HST)
\citep{lauer1993planetary, lauer1998m32-/--1} reveal the structure of
the eccentric disk: a brighter stellar concentration, P1, at apoapse, and
a fainter concentration, P2, at 
periapse. The old stellar disk surrounds the young nuclear cluster,
P3, and extends roughly 1$''$ (4 pc) from the
SMBH (Figure \ref{fig:3colorm31}). The red color of the disk suggests an
older stellar population, assumed to be roughly the age of the
surrounding bulge, or $>$4 Gyr \citep{olsen2006the-star, saglia2010the-old-and-heavy}. 
Current dynamical models \citep{peiris2003eccentric-disk} suggest 
the orientation of the 
eccentric disk is inclined $\sim$20$^{\circ}$ to the
larger-scale galactic disk. CO observations with the HEterodyne
Receiver Array (HERA) on the IRAM 30m telescope show that a dusty 0.7
kpc ring is also misaligned with the 
larger-scale galactic disk, with a position angle (PA) of $-$66$^{\circ}$ versus 37$^{\circ}$ for the larger disk 
\citep{melchior2013a-cold-gas}, which is not aligned with the
eccentric nuclear disk. 
Recent dynamical models of the eccentric nuclear
disk also show that it is thick \citep[$h/r$ $\sim$ 0.4,][]{peiris2003eccentric-disk}, and that the razor-thin
models \citep{sambhus2000the-pattern, 
jacobs2001long-lived, 
salow2001eccentric, sambhus2002dynamical,salow2004self-gravitating} cannot properly fit the 
observations. 

A wide range of conflicting precession values, $\Omega_P$, have been
measured for the eccentric
disk, from 3 km s$^{-1}$ pc$^{-1}$ to over 30 km s$^{-1}$ pc$^{-1}$
\citep{sambhus2000the-pattern, 
bacon2001the-m-31-double,jacobs2001long-lived, 
salow2001eccentric, sambhus2002dynamical,salow2004self-gravitating}. Thus the origin of the gas
that formed P3 remains unresolved.
The existence of the eccentric disk also poses another
puzzle: realistic models that explain its existence are difficult to compute.
Simulations suggest that gas inflows can drive large-scale gas instabilities which can set up long-lived nuclear eccentric disks \citep{hopkins2010the-nuclear}, or that a pair of counter-rotating massive star clusters can decompose into an eccentric disk \citep{kazandjian2013the-doubling}. Either method predicts that the resulting structure will be slowly precessing ($\Omega_P$ = 1--5 km s$^{-1}$ pc$^{-1}$). However, the issue is by no means settled observationally or theoretically.

Complete modeling of the eccentric disk, 
including a precise measurement of the precession
rate, has been limited by the fact that 
many spectroscopic studies of the eccentric disk have used long-slit
spectroscopy, often in combination with imaging \citep{kormendy1999the-double, statler1999stellar, 
bender2005hst-stis}. While several of these long-slit studies have been at high spatial resolution ($\lesssim$ 0\farcs1), they have been limited by lack of coverage of the field of view. 
Several integral field
spectroscopy studies have been conducted \citep{bacon2001the-m-31-double,menezes2013discovery},
but the limited spatial resolution ($\sim$0\farcs4) likely contributes to
some of the scatter in the precession rate measurements. In addition, all spectroscopic studies of the eccentric disk to date have been conducted in the optical, though the eccentric disk is brighter in the near-infrared (NIR).
With the advent of laser guide star adaptive optics (LGS AO)
feeding integral field unit (IFU) spectrographs,
high spatial resolution 2D kinematics can now be obtained and used 
to test precessing eccentric disk models and hypotheses for the formation of the
young nuclear star cluster in M31.

We present new observations of the nucleus of M31 obtained with the
OSIRIS IFU spectrograph and the LGS AO system at the Keck Observatory.
With a spatial resolution of 0\farcs13 (0.45 pc), we map the
line-of-sight kinematics over the entire old stellar eccentric disk.
We model the observed 2D flux, velocity, and dispersion maps with
dynamical models from \citet{peiris2003eccentric-disk}, 
adding rigid-body precession to the model. The measured slow precession rate
supports the \citet{chang2007the-origin} theory for the formation of
the young nuclear cluster from the winds of the old eccentric disk.
We discuss the observations in \S\ref{sec:obs}, including new ground-based NIR data in \S
\ref{ssec:ao_obs}, the data reduction and analysis in \S\ref{sec:an}, and results from the data in \S
\ref{sec:res}. The results from the analysis are compared to models in \S\ref{sec:mod} and discussed 
in \S\ref{sec:disc}. We summarize our conclusions in \S\ref{sec:con}.

\section{Observations}
\label{sec:obs}
New high-spatial resolution observations of the nucleus of M31 ($\alpha=$ 00 42 44.3, 
$\delta=$ +41 16 09, J2000) 
were taken in the NIR with the adaptive optics system 
at the W.~M.~Keck Observatory. Archival optical HST imaging was obtained to place the new NIR observations in context. 
While HST provides high-resolution imaging in the optical, the only means of obtaining high spatial 
resolution integral field spectroscopy is using ground-based adaptive optics (AO)
in the IR. Details of the AO-assisted infrared integral field spectroscopy and imaging
are described in \S\ref{ssec:ao_obs}. Optical imaging with HST is described in 
\S\ref{sec:hst_obs}.

\subsection{Ground-based observations}
\label{ssec:ao_obs}
NIR observations of the nucleus of M31 were taken with the LGS AO
 system on the W. M. Keck II 10 m telescope
\citep{van-dam2006the-w.-m.-keck, wizinowich2006the-w.-m.-keck}. 
The laser was positioned on the nucleus to correct high-order atmospheric
aberrations. Low-order aberrations were corrected using two different
tip-tilt stars during different observing runs. 
During good seeing, a close and faint globular cluster was used for
tip-tilt correction; it is located 35$''$ 
to the southwest from the nucleus with $R$=16.2 at 00 42 42.203 +41 15 46.01 (J2000).
A second globular cluster, NB95, with $R$=14.9 and located 53$''$ southeast from the nucleus 
at 00 42 47.973 +41 15 37.07, was used as the tip-tilt reference when seeing was
poor or variable or if clouds were present
\citep{battistini1993new-globular,galleti2007an-updated}.
These tip-tilt objects are located in a region of M31 with high surface
brightness and a strong gradient in the unresolved galaxy light. Thus,
in order to properly background-subtract the tip-tilt wave front
sensor, 
the sky background was measured at a manually-selected sky position such 
that the light from M31's bulge had comparable intensity to that of
the tip-tilt object positions.

\subsubsection{Keck/OSIRIS}

\begin{deluxetable}{lcc}
\tablecaption{M31 Keck/OSIRIS Data}
\tablehead{
\colhead{Night} & \colhead{No. of frames} & \colhead{Position} 
}
\startdata
2008 Oct 21 & 13 & center \\
2010 Aug 15 & 18 & center \\
2010 Aug 28$^a$ & 22 & NW, SE \\
2010 Aug 29 & 23 & NW, SE\\
\enddata
\tablenotetext{a}{Night removed from analysis due to poor spatial resolution}
\label{tab:data}
\end{deluxetable}

Spectroscopic measurements of the M31 nucleus were made
using OSIRIS \citep{larkin2006osiris:}, an integral field unit spectrograph,
on 2008 October 21 (PI: M. Rich), 2010 August 15 (PI: J. Lu), and 2010 August 28--29 
(PI: A. Ghez, Table \ref{tab:data}).
The individual OSIRIS data cubes cover 0\farcs8 $\times$ 3\farcs2 and were
oriented at a PA of 56$^\circ$ along the major axis of the 
eccentric stellar disk. The field was sampled with a pixel scale of 0\farcs05 pix$^{-1}$
and each spatial pixel (spaxel) provides an independent spectrum across the $K$-broadband filter 
($Kbb$: $\lambda = 2.18~\micron, \Delta\lambda = 0.4~\micron$) with an 
average spectral resolution of $R\sim$3600. A total of 76 individual exposures
were taken with t$_{\text{exp}}$ = 900 s in a 3$\times$1 mosaic
pattern. Observations taken the night of 2010 August 28 were subject
to poor weather; the spatial resolution of these frames was generally
poor (see \S\ref{ssec:err_res}) and thus these frames were
removed from the analysis, leaving 54 total exposures. 
Observations on two of the remaining three nights were centered on the
field, 
while the northwest and southeast pointings of the mosaic were
observed on the other night.
On each night, observations of telluric standard stars and blank sky
were also obtained for use in calibration. 

\begin{figure}[h]
\begin{center}
\includegraphics[width=\columnwidth]{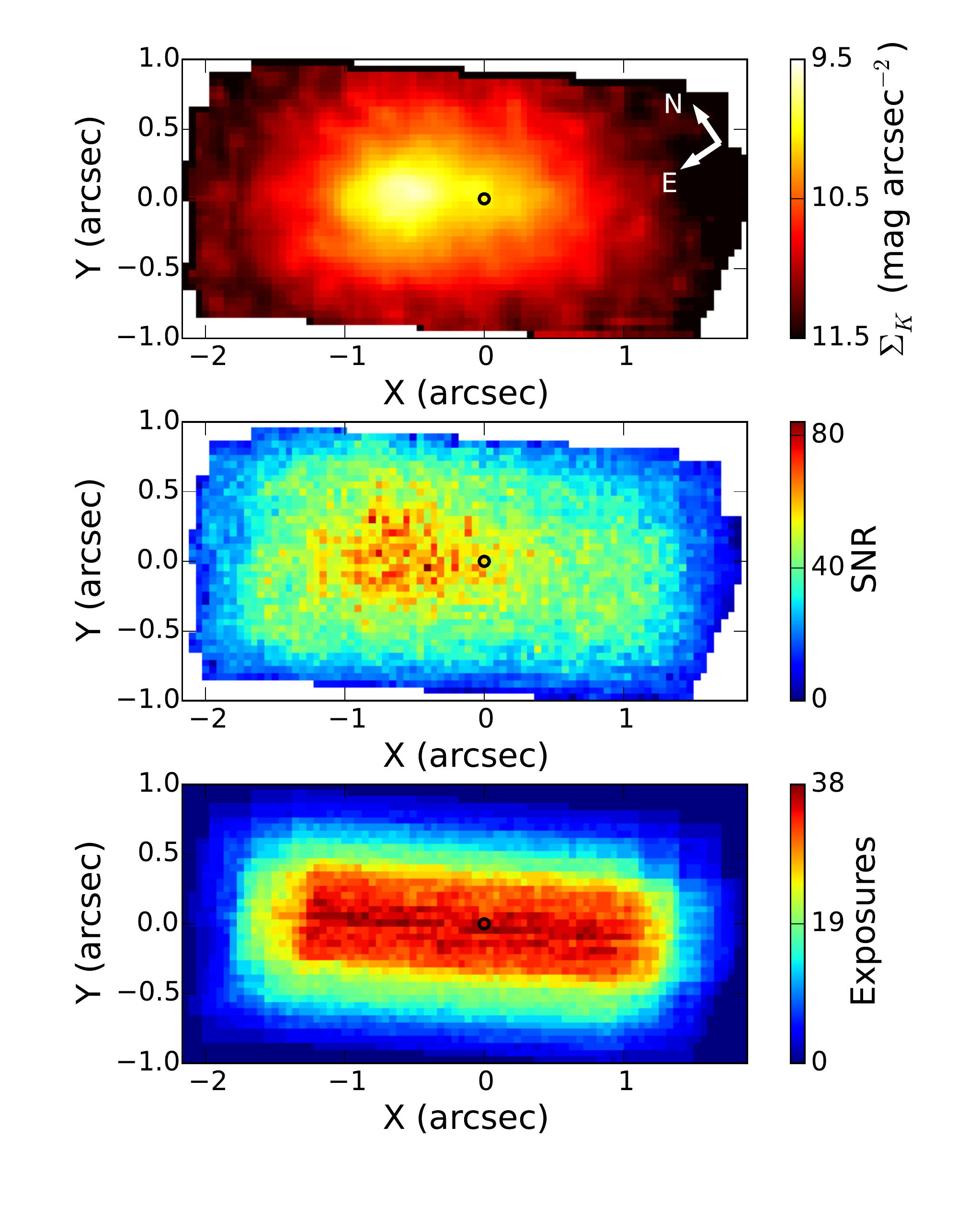}
\caption{Data quality maps of the OSIRIS mosaic, using data from 2008
  Oct 21, 2010 Aug 15, and 2010 Aug 29. \emph{Top:} Mosaic of all
  OSIRIS frames from these three nights, collapsed along the
  wavelength direction. \emph{Center:} Signal-to-noise ratio map
  calculated using a line-free portion of the
  continuum. \emph{Bottom:} Number of exposures combined at each
  spaxel in the mosaic. }
\label{fig:dataquality}
\end{center}
\end{figure}

Total integration
time on the center of the field consists of $\sim$35 exposures (9
hours) while the edges of the field of view (FOV) were observed in 5--20 exposures
(1.25--5 hours; see Figure \ref{fig:dataquality}, bottom panel).

\subsubsection{Keck/NIRC2}
New supporting NIR imaging of the M31 nucleus was taken with 
NIRC2, the facility NIR camera on Keck. On 2005 July 29, a total of 8
images were taken in the $K$-prime ($K'$)-band
filter ($\lambda = 2.12~\micron, \Delta\lambda = 0.35~\micron$) with
30 s exposure times (PI: K. 
Matthews). 

On 2009 September 10, the nucleus was imaged with NIRC2 using the $J$
($\lambda = 1.25~\micron, \Delta\lambda = 0.16~\micron$) and $H$
($\lambda = 1.63~\micron, \Delta\lambda = 0.30~\micron$) band
filters. Exposure times were 120 s and 60 s, respectively. A total of
15 frames were combined for the final $J$ image, and 14 frames were used
for the $H$ image (PI: A. Ghez). 

The $J$, $H$, and $K'$ images described above were all taken using the
narrow camera, which provides a pixel scale of 0\farcs01 pix$^{-1}$. Wide
camera images, with a pixel scale of 0\farcs04 pix$^{-1}$, were obtained
from the Keck Observatory Archive from observations taken on 2007
October 19 during engineering time with NIRC2 using the $K'$ filter. One
exposure was used, with 5 coadds of 1 s integration time each, for a
total exposure time of 5 s.

\subsection{HST}
\label{sec:hst_obs}
Supporting archival imaging of the nucleus from HST was also used. Optical images of the M31 nucleus were obtained with HST and
downloaded from the Hubble Legacy Archive. The nucleus was imaged on
2006 June 15--16 using the Advanced Camera for Surveys (ACS) High
Resolution Channel (HRC) with the F330W and F435W filters \citep[PI:
T. Lauer,][]{lauer2012the-cluster}. Total exposure time with the F330W filter was 8120 s over 12
exposures, while the total exposure time with the F435W filter was
2384 s over 8 exposures.

The nucleus was imaged on 1994 Oct 02 using the Wide Field and
Planetary Camera 2 (WFPC2) with the F555W, F814W, and F1042M filters
\citep[PI: M. Rich,][]{rich1996local}. Total exposure time for the F555W filter was 1680 s over 6
exposures, for the F814W filter was 1280 s over 5 exposures, and for
the F1042M filter was 5000 s over 10 exposures.

\section{Analysis}
\label{sec:an}
 
\subsection{OSIRIS data reduction}
The OSIRIS spectroscopic data were reduced using the OSIRIS reduction
pipeline\footnote{\url{github.com/Keck-DataReductionPipelines/OsirisDRP}} \citep{krabbe2004data}
to subtract a dark frame, correct bad pixels and cosmic rays, perform the
data cube assembly and wavelength calibration, and remove the
background sky and telluric OH emission. The pipeline version utilized
for the reduction was v3.2. The pipeline includes internal logic to account for the
hardware changes that have occurred since the data were taken and
uses the modules appropriate for the date of observations.

For each data cube, the telluric absorption spectrum was modeled and
removed using a combination of empirical and theoretical telluric
absorption spectra. The empirical telluric spectrum was created using
two standard stars with different spectral types as described in
\citet{hanson1996a-spectral}. A hot, early-type A0V star was observed
(HD 209932) each night. This star has a featureless black body spectrum in the
$K-$band, except for a wide Br$\gamma$ absorption feature at 2.166
\micron. A solar-type star was also observed (HD 218633) each night and divided
by a solar spectrum constructed from NSO data by ESO and convolved to
match the OSIRIS spectral resolution
\citep{livingston1991an-atlas,maiolino1996correction} to obtain a
telluric spectrum. The solar-type telluric standard was used to
replace the region around Br$\gamma$, from 2.151 \micron~to 2.181
\micron, in the A0V spectrum. The corrected A0V spectrum was then
divided by a 9500 K blackbody to obtain the final empirical telluric
absorption spectrum.

However, changing atmospheric conditions throughout each night and a
variable spectral resolution across the OSIRIS FOV led to
large residuals when correcting for telluric absorption using the
empirical telluric spectrum, which was observed at only a single
detector location. This was specifically an issue where blue-shifted
stellar CO bandheads in the M31 spectra overlap with the large
telluric residuals. To better model the changing telluric line depths,
we created model telluric spectra with the \texttt{molecfit} package
\citep{smette2015molecfit:, kausch2015molecfit:}, which generates
synthetic telluric absorption spectra. Given an atmospheric profile
for a given night and observatory location, \texttt{molecfit} uses a radiative
transfer code to obtain absorption line depths and fits the output
directly to the telluric features in a science spectrum. Before using
\texttt{molecfit}, the OSIRIS M31 science frames were first mosaicked by night
and by pointing, roughly corresponding with first and second
half-nights, to better capture temporal telluric variations. A subset
of 3$\times$3 spaxels was extracted from a region of each mosaic where
the stellar absorption lines did not overlap with the telluric
lines. The median of these spectra was taken to create a single typical
spectrum per half-night, and \texttt{molecfit} was run on each of these
representative spectra to obtain the telluric absorption line
depths. However, these model spectra could not be used directly in the
telluric correction, as OSIRIS introduces a shape to the continuum
that is not fit well by the \texttt{molecfit} modeling. The resulting \texttt{molecfit}
line depths were combined with the continuum from the empirical
telluric spectrum to produce a final telluric absorption spectrum,
which was used to telluric-correct the M31 data.


The final mosaic was assembled by calculating shifts between frames
using cross-correlations between each OSIRIS frame and the NIRC2
$K-$band image, rebinned to the OSIRIS spatial scale. Frames within a
single dithered set (typically 4--9 frames) were first mosaicked
together, as the offsets between these frames are given by the input
dithers. Each of these mosaics was then cross-correlated with the NIRC2
image to obtain its relative shift. The frames were then combined
using a mean clipping method. The final OSIRIS data set consists of
one fully combined data cube and three subset cubes, each containing
1/3 of the data, used for determining uncertainties.

\subsection{Flux errors and data quality}
\label{ssec:err_res}
The formal errors output by the pipeline include real spatial and
spectral variations caused by removed OH sky lines, telluric
absorption, interpixel sensitivity, and cosmic rays. However, the
overall error is too large by nearly an order of magnitude relative to
empirical error estimates, determined via
the standard error on the mean of three subset cubes at each spaxel
and spectral channel. As the magnitude of the flux errors impacts the errors on the kinematic measurements (\S\ref{ssec:ppxf}), the errors need to be properly scaled. To combine the error variation captured by
the formal pipeline errors with the more accurate magnitude of the
empirical errors, we scale the pipeline errors by the ratio of the
median of the two error arrays at every spaxel. We adopt these scaled
flux errors for the remainder of the analysis.

\begin{figure}[h]
\begin{center}
\includegraphics[width=\columnwidth]{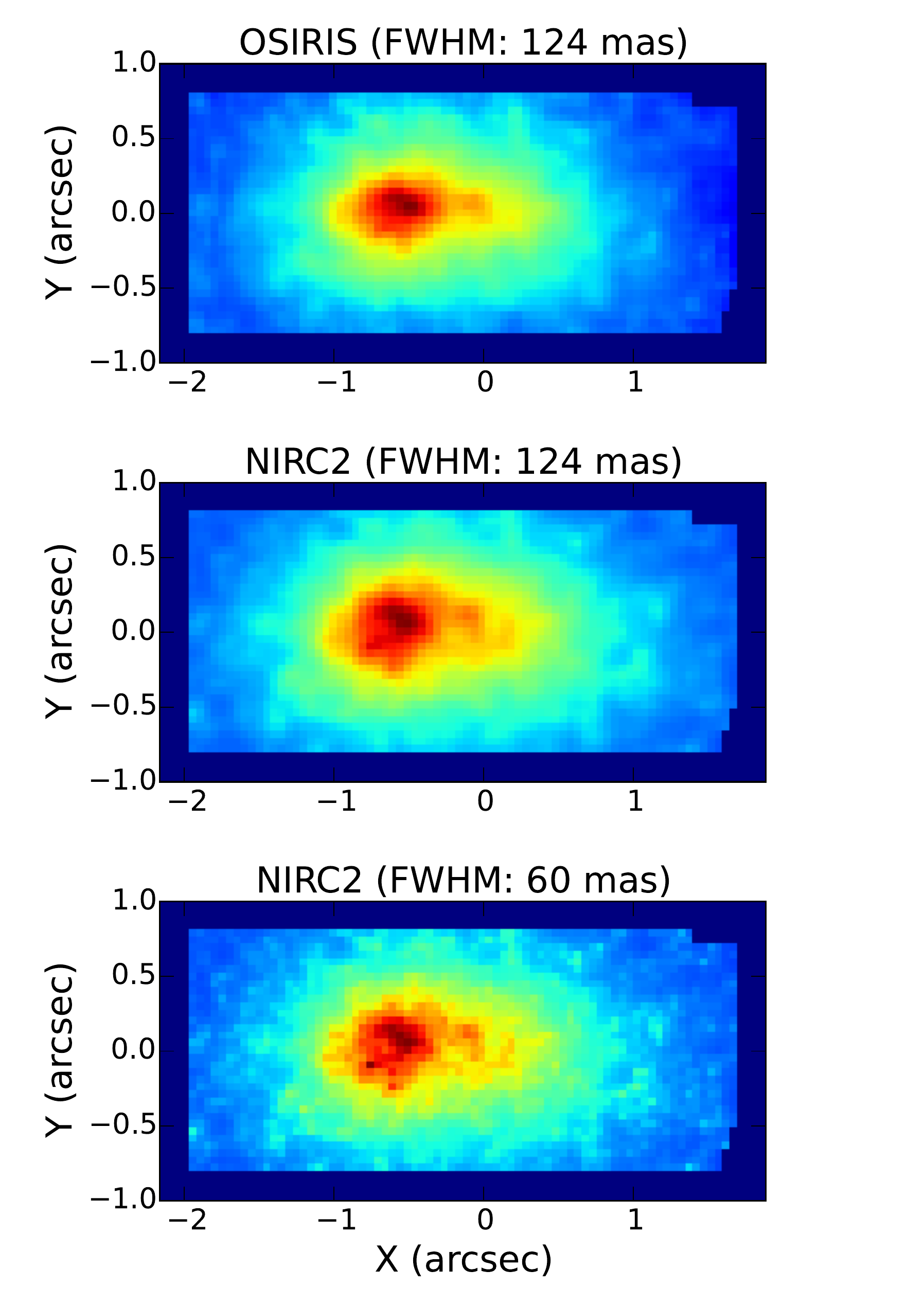}
\caption{The spatial resolution of the combined OSIRIS mosaic was
  determined by comparing it with a high resolution $K-$band NIRC2
  image convolved with a Gaussian. \emph{Top:} OSIRIS data cube,
  collapsed in the wavelength direction. \emph{Middle:} NIRC2 $K-$band
  image, convolved to the same spatial resolution as the OSIRIS
  cube. \emph{Bottom:} Original NIRC2 image, binned to the same
  pixel scale as the OSIRIS data. } 
\label{fig:spatres}
\end{center}
\end{figure}

The OSIRIS image quality was estimated by
first collapsing the cube along the spectral direction. Then the 
OSIRIS image was compared to the NIRC2 image by convolving the 
NIRC2 image with an iteratively determined kernel that reduced the 
difference between the NIRC2 and OSIRIS images. A Gaussian kernel was used to represent the 
diffraction-limited core. The
amplitude was set to 1 and not allowed to vary, while
the width of the core Gaussian was allowed to vary. This gives an estimate of the
spatial
resolution of the OSIRIS images compared to the NIRC2 images.
The resulting spatial resolution ranges between 50 and 350 mas FWHM for the
individual exposures and the median spatial resolution across all frames is 260 mas. 
The majority of the lower quality frames are in the southeast pointing
of the mosaic, due to worse observing conditions during the nights
this pointing was observed. We experimented with removing various
combinations of the lowest quality frames, striving to maintain even
coverage across the FOV. Ultimately, all frames from 2010 August 28, the
night with the worst weather, were removed, resulting in a median
spatial resolution of 205 mas amongst the remaining frames and a
resolution of 124$\pm$6 mas for the full combined mosaic
(Figure \ref{fig:spatres}). We proceeded with the analysis using this
clipped set.

\subsection{Bulge subtraction}
\label{ssec:bulge}

\begin{figure}
\begin{center}
\includegraphics[width=\columnwidth]{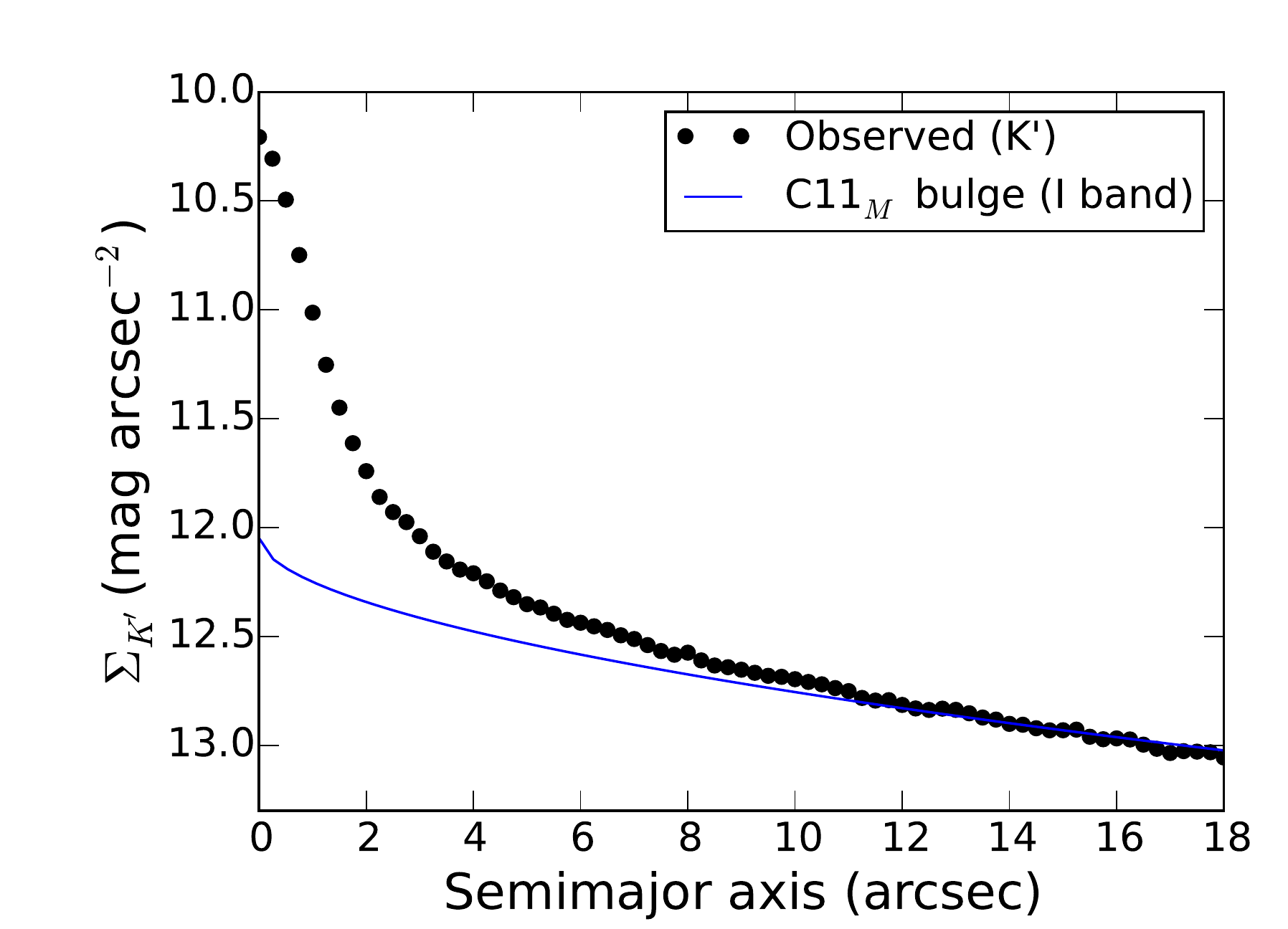}
\caption{Profiles for the bulge surface brightness \citep{courteau2011the-luminosity} match
  measured $K-$band surface brightness profiles well beyond 11$''$. The flux-calibrated $K-$band
  surface brightness profile is measured from the NIRC2 wide camera
  image and is calculated in elliptical apertures using the
  ellipticity given by \citet{courteau2011the-luminosity}. The bulge
  profile is given in \citet[their model M]{courteau2011the-luminosity}; an
  additive offset has been applied to the bulge profile so it
  matches the flux level in the $K-$band at 15$''$, necessary because of
  the flux difference between the two bandpasses. }
\label{fig:NIRC2bulgeprofile}
\end{center}
\end{figure}

\begin{figure}
\begin{center}
\includegraphics[width=\columnwidth]{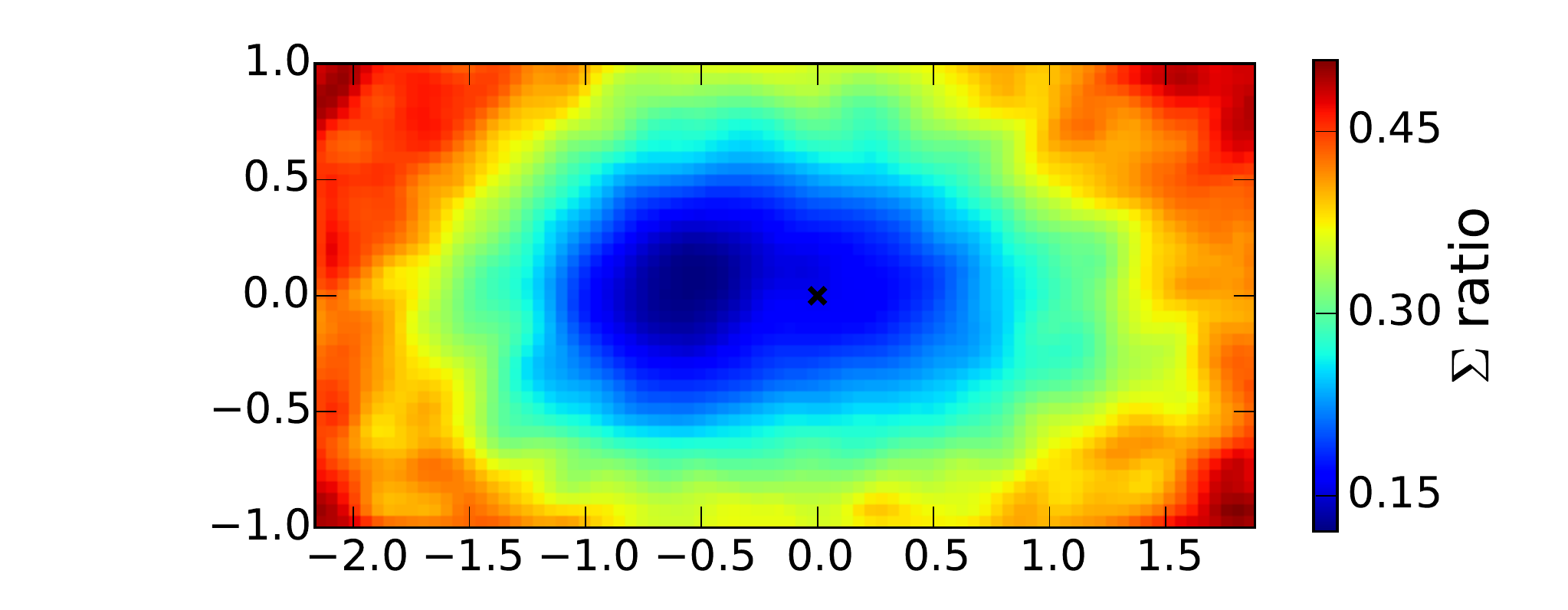}
\caption{Ratio of the bulge luminosity to the total $K-$band
  luminosity within the nuclear region, derived using the $I-$band
  Sersic profile (model M) from \citet{courteau2011the-luminosity} in
  comparison with the NIRC2 $K-$band image. The SMBH position is
  marked with the black cross, and the orientation is as in
  Figure \ref{fig:dataquality}.}
\label{fig:sbratio}
\end{center}
\end{figure}

The old stellar population of the bulge is the dominant source of
light at 5$\farcs$5 $<$ $r$ $<$ 300$''$ \citep[hereafter KB99]{kormendy1999the-double} and
is a source of foreground contamination in observations of the nuclear
disk. In addition to being a source of excess surface brightness, the
bulge's slow rotation and internal dispersion can distort the
kinematic signature of the nuclear disk. Both the surface brightness
and the kinematic contribution of the bulge must be removed before
extracting the nuclear disk kinematics. 

First, we determined the fraction of bulge light in each
spaxel. Previous studies have found that both the optical and the NIR
surface brightness profile of the bulge is well-fit by a Sersic
profile \citep[KB99,][]{courteau2011the-luminosity,
  dorman2013a-new-approach} of index $\sim$2, half-light radius
$\sim$1 kpc, and half-light surface brightness 17.55--17.85 mag
arcsec$^{-2}$ (optical $V-$band, $\lambda$ = 555\AA, and $I-$band, $\lambda$ = 806\AA) or 15.77 mag arcsec$^{-2}$ ($L-$band, $\lambda$ = 3.6$\mu$m). The $I-$band surface brightness profile from 
\citet[][model M]{courteau2011the-luminosity} was adopted and scaled to match the NIRC2
$K-$band wide field image at a radius of 10$''$, or a distance at which
the bulge dominates the surface brightness
(Figure \ref{fig:NIRC2bulgeprofile}). The one-dimensional
Sersic profile was converted to a two-dimensional profile using the model M bulge
ellipticity. We used the bulge PA of 6.6$^{\circ}$ from
\citet{dorman2013a-new-approach}, as they fit the bulge profile with
the same data set; the differences with their best fit parameters
are negligible. Both 
the two-dimensional bulge profile and the NIRC2 image were smoothed to match the OSIRIS
resolution and the two smoothed images were divided to obtain the ratio of bulge
luminosity to total luminosity in the nucleus
(Figure \ref{fig:sbratio}).

\begin{figure}
\begin{center}
\includegraphics[width=\columnwidth]{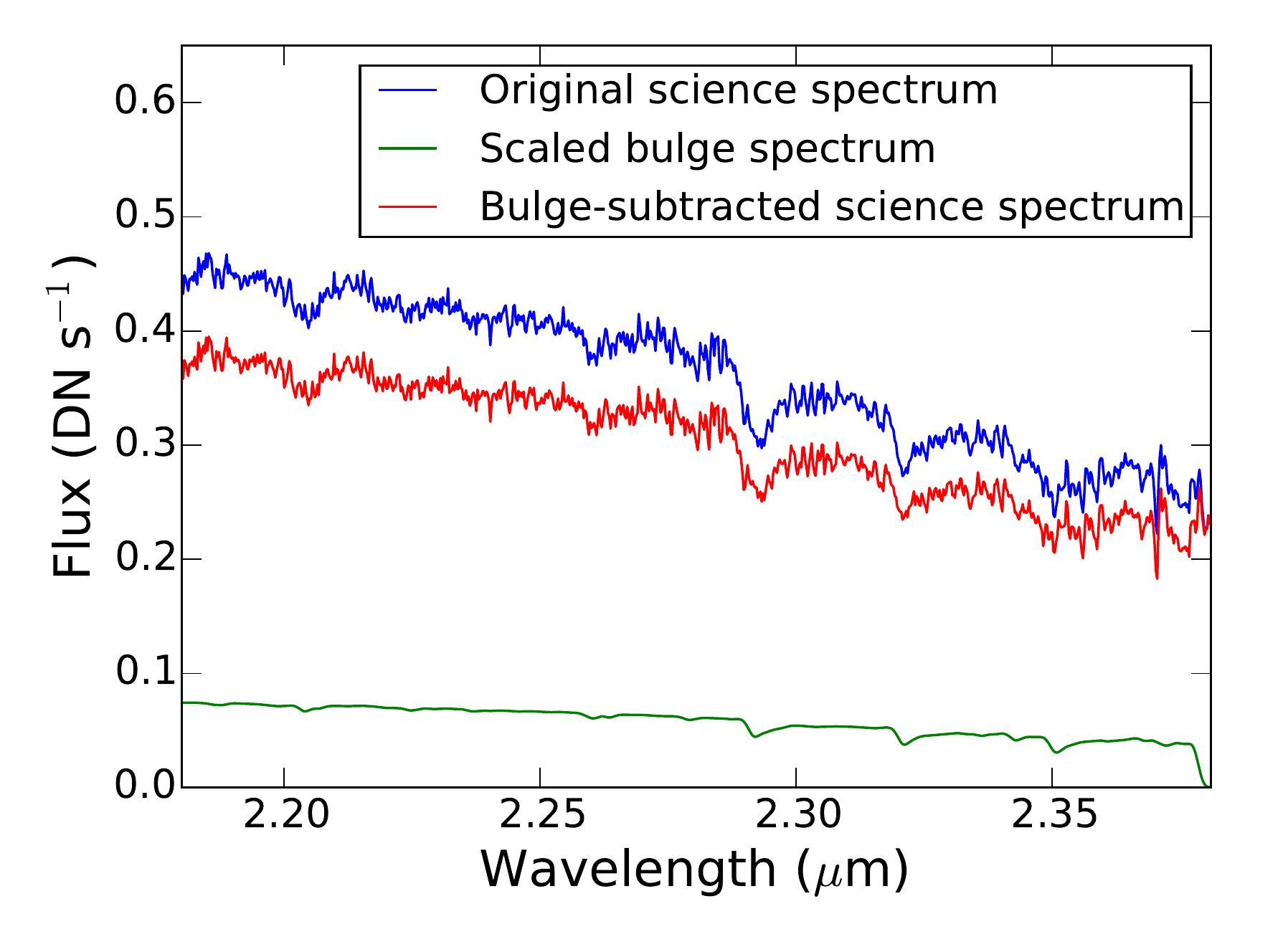}
\caption{Example science spaxel, before and after bulge
  subtraction. The spaxel is taken from near the center of the cube
  and is shown before and after bulge subtraction. The bulge spectrum,
  convolved with the bulge LOSVD and scaled per the ratio given in
  Figure \ref{fig:sbratio} for the example spaxel, is also shown.}

\label{fig:exbs}
\end{center}
\end{figure}

Bulge-dominated spaxels were selected based on the bulge surface brightness ratio map.
As no spaxels had a majority contribution from
the bulge (Figure \ref{fig:sbratio}), those spaxels in which
the bulge surface brightness ratio is at least 0.42 were selected to be representative of the bulge. This 
ensures that enough spaxels, roughly 160, were obtained to derive a high-quality bulge
template spectrum.

Next, the intrinsic spectrum of the integrated bulge
light was estimated using the pPXF package
\citep{cappellari2004parametric}, which fits a linear combination of
stellar templates convolved with a line of sight velocity distribution (LOSVD). pPXF was also configured to  fit and add a 4th-degree Legendre polynomial to the convolved spectrum, to account for continuum shape differences between the stellar templates and the science spectrum (see \S\ref{ssec:ppxf} for more 
details). The bulge-dominated spaxels were fit with pPXF. The output best-fit linearly-combined stellar templates for each spaxel, along with the additive Legendre polynomial, were used to create a template stellar spectrum for each bulge-dominated spaxel. Each resulting template spectrum reflected the flux level of the corresponding spaxel in the data cube. To normalize these spectra, the template spectrum for each spaxel was divided by its median. The median of all normalized bulge template 
spectra at each wavelength was then taken to obtain a median stellar template for the bulge.

\begin{figure*}[!t]
\begin{center}
\includegraphics[width=\textwidth]{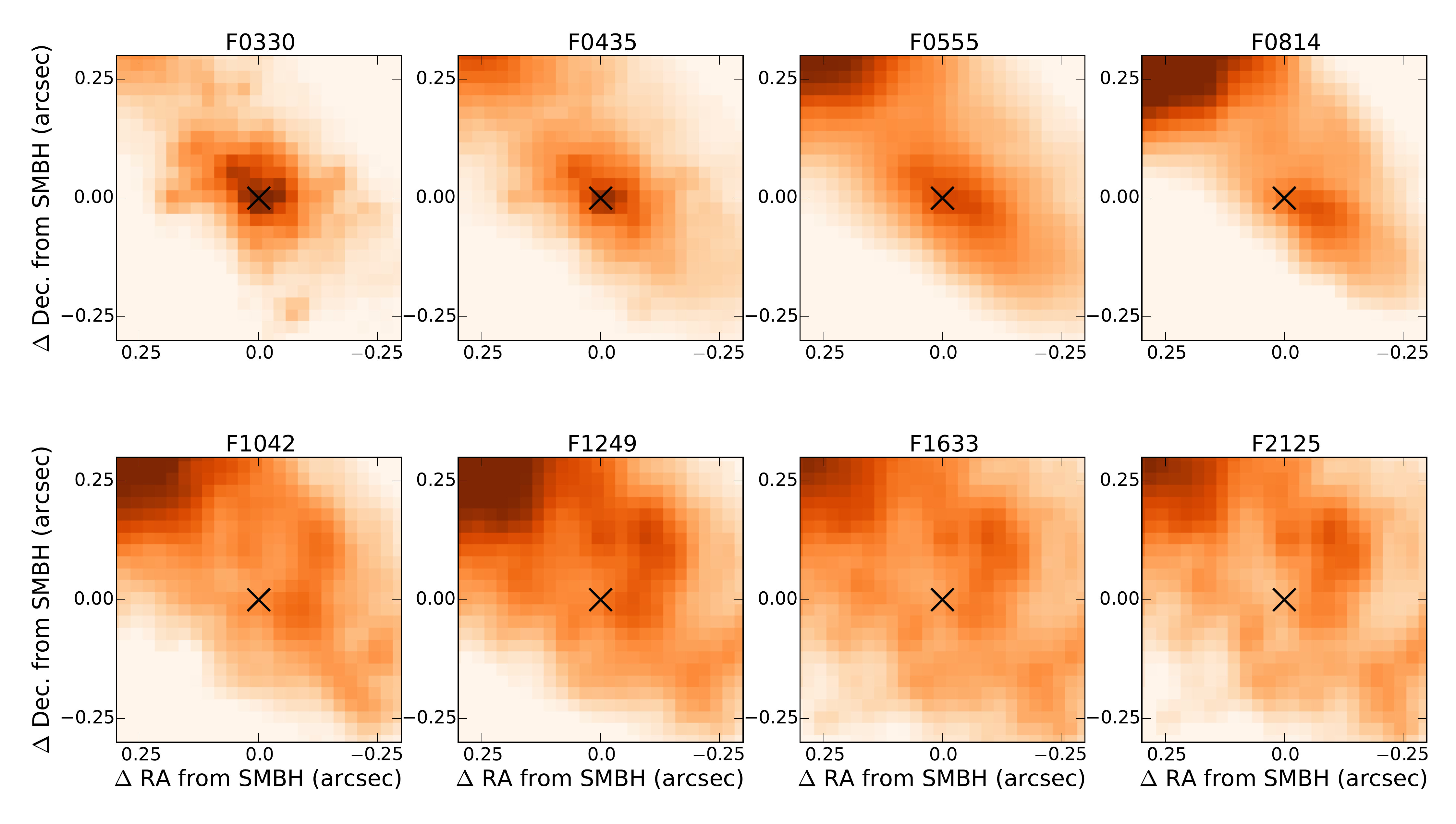}
\caption{Multi-wavelength images show the changing stellar population
  from 330 nm (F0330) to 2.1 $\mu$m (K'). The position of P3, assumed to be
  coincident with the SMBH, is marked with the black cross and is
  accurate to 0.033$''$ in the K'-band frame. The color scaling has been
  adjusted for each frame to emphasize the structure around the
  SMBH. North is up and East is left and the pixel scale
  is 0.025$''$ pixel$^{-1}$.}
\label{fig:bhcutall}
\end{center}
\end{figure*}

The median bulge spectrum was dominated by two stellar template
spectra: a late-K giant and a late-K dwarf. The late-K giant is likely
more representative of the actual bulge population, which is estimated
to be roughly the age of the galaxy
\citep{saglia2010the-old-and-heavy}. However, the late-K dwarf star
has stronger Na lines than that of the giant. The center of M31 is estimated to have
enhanced metallicity compared to Milky Way \citep{saglia2010the-old-and-heavy} and thus the
inclusion of the dwarf stellar template in the fit may be compensating
for this different abundance pattern. Alternatively, varying Na line strengths may 
represent variation in the initial mass function of the old stellar population
\citep[and references therein]{mcconnell2015radial}.

Finally, the bulge spectrum was convolved with a Gaussian LOSVD
with a fixed velocity of $-$340
  km s$^{-1}$, at the systemic velocity of the galaxy (R. Bender, private communication), and a
dispersion of 110 km s$^{-1}$ at all points in the field, or the dispersion in the bulge-dominated spaxels before bulge subtraction. No bulge
rotation was included since previous work by
KB99 found that the bulge is rotating slowly
at 2.65(r/1$''$) km s$^{-1}$ and that including this rotation does not
appreciably affect the bulge subtraction within the compact nuclear
region. The final bulge spectrum was multiplied by the median flux in
each spaxel and the appropriate surface brightness ratio to create a
bulge cube. This cube was then subtracted from the observed spectral
cube. An example spectrum, before and after bulge subtraction, along
with the scaled bulge template spectrum is shown in
Figure \ref{fig:exbs}. 

There are systematic sources of error in the bulge subtraction that we do not fully explore but which 
should be kept in mind. The edges of the OSIRIS FOV are subject to much more bulge subtraction than 
the center (Figure \ref{fig:sbratio}). Subtracting so much flux ends up amplifying the noisiness of the 
bulge-subtracted spectrum. This noise is not purely Poissonian, but
also includes noise from imperfectly-corrected sky and telluric
features, which introduce larger systematic errors into the spectra at
the edge 
of the FOV. In addition, the bulge is typically modeled as a Sersic profile, which is sharply peaked at the 
origin. However, \citet{bacon2001the-m-31-double} also fit the bulge using a Multi-Gaussian Expansion 
(MGE), which models the bulge surface brightness as the sum of three Gaussians. Their MGE bulge 
profile is much shallower in the inner few arcsec, with a difference of a few tenths of a magnitude 
at the origin 
from the equivalent Sersic profile. Alternative bulge subtraction methods may subtract more or less flux 
than does our method, which may impact the kinematics derived from the
bulge-subtracted spectra. 
In Appendix \ref{sec:bulge_subtraction}, we explore in more detail several
alternative methods of bulge subtraction and show that their impact
does not change the overall conclusions if this paper.

\subsection{Position of the supermassive black hole}
\label{ssec:bhpos}

\begin{deluxetable}{llccl}
\tablecaption{SMBH alignment}
\tablehead{
\colhead{Telescope} & \colhead{Instrument} & \colhead{Filter} & \colhead{N$_{\text{sources}}^a$} & 
\colhead{Error$^b$ ($''$)}
}
\startdata
HST & ACS & F330W & ... & ...\\
HST & ACS & F435W &  16 & 0.0093 \\
HST & WFPC2 & F555W & 26 & 0.013 \\
HST & WFPC2 & F814W &  16 & 0.018 \\
HST & WFPC2 & F1042M &  19 & 0.016 \\
Keck & NIRC2 & $J-$band &  19 & 0.015 \\
Keck & NIRC2 & $H-$band &  18 & 0.0065 \\
Keck & NIRC2 & $K'-$band &  18 & 0.0063 \\
\enddata
\tablenotetext{a}{Number of sources used to align the given frame with the next bluer frame.}
\tablenotetext{b}{Alignment error in the SMBH position from each coordinate
  transformation.}
\label{tab:align}
\end{deluxetable}

We identified the position of the SMBH in our
OSIRIS and NIRC2 data using the position of P3, the very compact
\citep[$r\sim$~0\farcs06,][]{lauer1998m32-/--1} young nuclear cluster
that contains the SMBH
\citep{bender2005hst-stis}. Unlike the old stellar
population, which is bright in the NIR, P3 is NIR-dark and thus
effectively invisible in our OSIRIS and NIRC2 data. However, P3 is
bright in the UV; we adopted as the SMBH position the location of source
11, or the brightest UV source in the \citet{lauer2012the-cluster}
analysis of HST/ACS F330W and F435W
images. The
NIR image cannot be directly aligned to the F330W image because there
are no compact sources that are bright in both the blue and NIR
filters that can be used to determine the transformation. Instead, we
aligned images at successively longer wavelengths using an evolving set
of compact, bright sources to fit the translation, scale, and rotation
(Table \ref{tab:align}), aligning adjacent-bandpass frames using only
the sources common to both frames (see Appendix \ref{sec:appa}). At least 16
bright compact sources are used for aligning each pair of
frames. After alignment, each frame was transformed into the reference
frame, F330W. The error on each frame's alignment was taken as the
residual error, converted to arcseconds. The total error in alignment from
F330W to the NIRC2 $K-$band image is 0\farcs033, which is smaller than the
0\farcs05 spaxels in the OSIRIS data. The OSIRIS data cube was aligned to the NIRC2 
$K-$band image by cross-correlating the OSIRIS cube, collapsed in the wavelength
direction, with the final registered NIRC2 $K-$band image.

Figure \ref{fig:bhcutall} shows the region within 0\farcs3 of P3 in
each of the eight wavelength images after alignment. The pixel scaling was
adjusted in each frame to match that of the F330W image. The SMBH coordinates were taken from 
the background-subtracted, subsampled F435W image 
\citep[][private communication]{lauer2012the-cluster} and converted to the aligned
$K'-$band image coordinates. The SMBH position is marked with the black cross in each panel in 
Figure \ref{fig:bhcutall}.

\subsection{Spatial binning}
\label{ssec:bin}
Fitting the precession rate of the disk requires a very high
signal-to-noise ratio (SNR). The SNR of the 
data varies over the FOV and was calculated empirically. For each
spaxel, we fit a straight line in a region 
of the spectrum dominated by the stellar continuum between 2.270 $\mu$m and 2.280 $\mu$m.
After dividing by the fitted line, the SNR was estimated as the mean of the normalized flux divided by the 
standard deviation. The SNR map is shown in Figure \ref{fig:dataquality}, center panel.

\begin{figure}
\begin{center}
\includegraphics[width=\columnwidth]{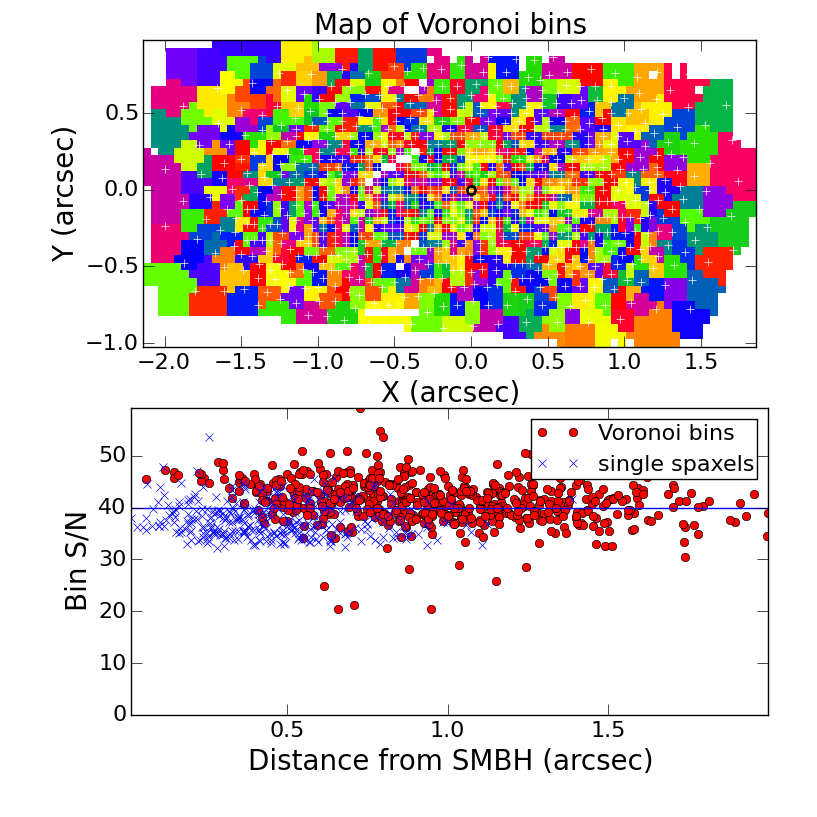}
\caption{\emph{Top:} Spatial bins 
determined via Voronoi tessellation
  \citep{cappellari2003adaptive}. Spaxels with an initially high SNR
  remain unbinned, while spaxels with a lower initial SNR are binned
  until their final combined SNR is roughly 40. The colors are
  randomized to emphasize bin boundaries. \emph{Bottom:} The final SNR
  for each bin as a function of distance from the SMBH.}
\label{fig:tess}
\end{center}
\end{figure}

The data was spatially binned using a Voronoi tessellation algorithm  
\citep{cappellari2003adaptive}, that optimally sums spaxels until a
target SNR is reached while enforcing a roundness criterion on the
resulting bins to ensure the resulting bins are as compact as possible. Spaxels already at or above the 
target SNR remain
unbinned. We used the SNR map thus generated to spatially bin the
spaxels to a target SNR of 40. The resulting bins and their SNR are
shown in Figure \ref{fig:tess}. Spaxels near P1, P2, and the SMBH
remained unbinned, while the fainter outer regions of the eccentric disk and
the innermost bulge were optimally binned. Spectra in the binned
spaxels were summed and their errors combined in quadrature. Spaxels
with a very low initial SNR ($<$2) on the very edge were masked and
discarded in the following analysis.

\begin{figure*}[!ht]
\begin{center}
\includegraphics[width=\textwidth]{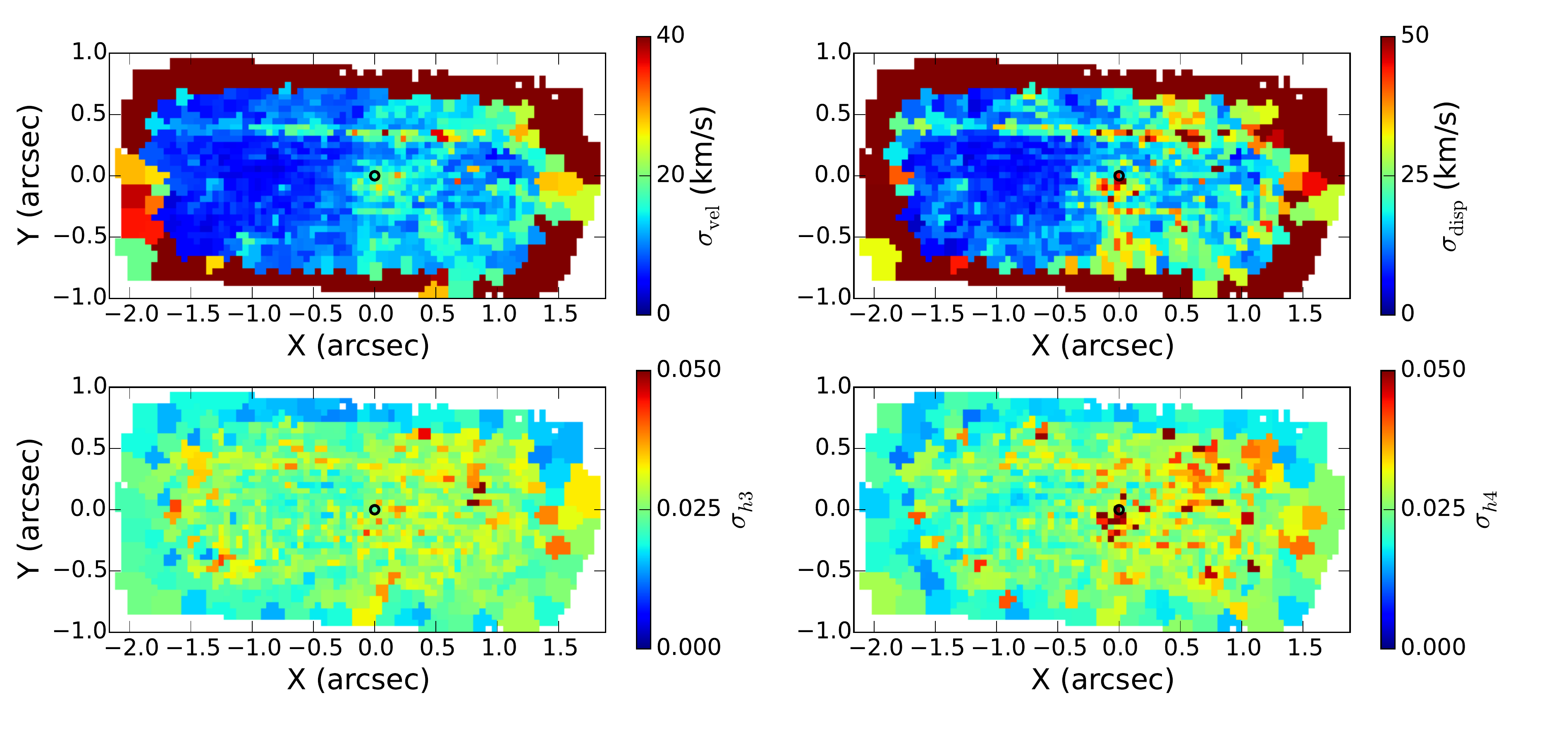}
\caption{Monte Carlo error maps for the velocity ({\em top left}),
  dispersion ({\em top right}), $h3$ ({\em bottom left}), and $h4$ ({\em
    bottom right}). In all panels, the SMBH
  position is marked by the black circle.}
\label{fig:MCerr}
\end{center}
\end{figure*}

\subsection{Kinematic fitting}
\label{ssec:ppxf}

\begin{figure}
\begin{center}
\includegraphics[width=\columnwidth]{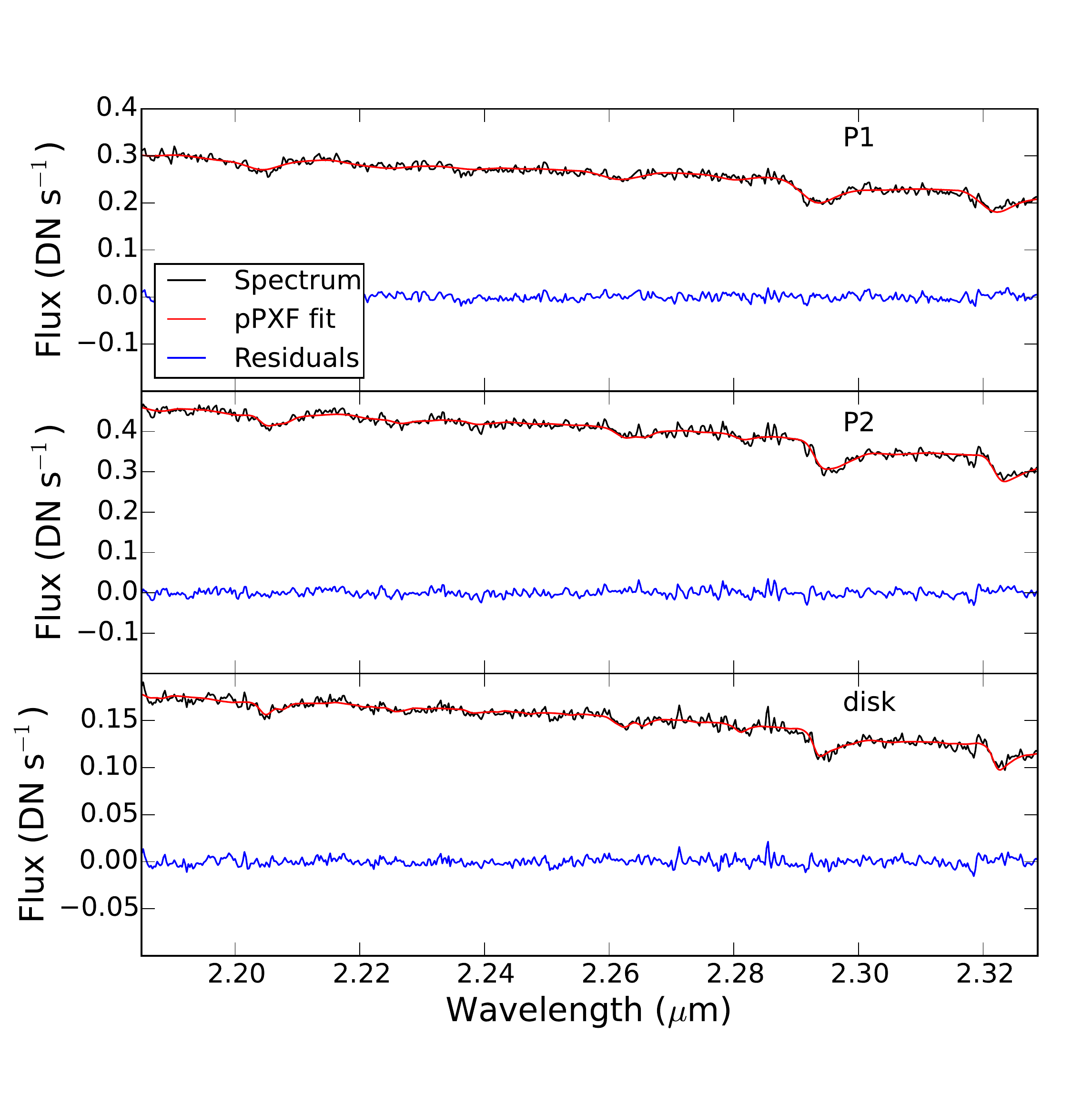}
\caption{pPXF fits and residuals for three sample spectra. In all, the input science spectrum is shown in black, the best-fit pPXF fit is shown in red, and the residuals between the two are shown in blue. \emph{Top:} Sample spaxel from near the P1 flux peak. \emph{Middle:} Sample spaxel from the P2 region. \emph{Bottom:} Sample spaxel from outside either of these regions, near the zero-velocity line.}
\label{fig:exppxf}
\end{center}
\end{figure}

Kinematics were extracted using the penalized pixel fitting method
(pPXF) described in \citet{cappellari2004parametric}. This method fits
the observed spectrum with a linear combination of spectral templates,
convolved by a line of
sight velocity distribution (LOSVD), which we parameterized as a fourth-order
Gauss-Hermite expansion of a Gaussian profile \citep[Appendix  A
of][]{van-der-marel1993a-new-method}. A fourth-degree Legendre
polynomial was fit and added to the convolved spectrum to correct for
mismatch between the continuum of the template and that of the science
spectrum. The moments
of the best-fit LOSVD (including $v$, $\sigma$, and the higher order
moments $h_3$ and $h_4$) and the weights of the spectral templates were
returned. The Gemini GNIRS spectral template library
\citep{winge2009the-gemini}, a library of evolved and main sequence
late-type stars observed in the $K'-$band at $R\sim$5900, was used as inputs to pPXF. Only
those templates observed in both the blue (2.15--2.33~$\mu$m) and the
red (2.24--2.43~$\mu$m) modes were used, for a total of 23 template stars. The
final template library covers G4--G5 II, F7--M0 III, K0 IV, and G3--K8
V stars. Independent fits were performed for each spaxel over the range
2.185--2.381~$\mu$m to obtain measurements of $v$, $\sigma$, $h_3$ and
$h_4$. Example fits for spaxels near P1, P2, and in the disk away from these regions are shown in Figure \ref{fig:exppxf}.

The templates preferred by the pPXF fits on the bulge-subtracted stellar population are
nearly identical to those found in the fits to the bulge-dominated spaxels. Similar to 
the bulge population, the old eccentric disk is best fit by a late-K giant and a late-K dwarf 
template spectrum. This implies, as found by previous work \citep{saglia2010the-old-and-heavy}, that the old 
eccentric disk has a similar stellar population to that of the bulge. However, we defer a full 
analysis of the eccentric disk stellar population for future work.

Measurement errors on the LOSVD moments were assessed via a Monte
Carlo (MC) analysis using 100 simulations. In each simulation, the
input spectrum was randomly perturbed within the 1$\sigma$ flux errors
and the pPXF fits were recalculated. The uncertainty of each LOSVD
moment (i.e., mean $v$, $\sigma$, $h3$, $h4$) was taken as the standard error of
the mean of the fitted moments from the MC trials. Error maps are
shown in Figure \ref{fig:MCerr}. The median errors are 11.3 km
s$^{-1}$ on $v$ and 16.7 km s$^{-1}$ on $\sigma$.

\begin{figure*}
\begin{center}
\includegraphics[width=\textwidth]{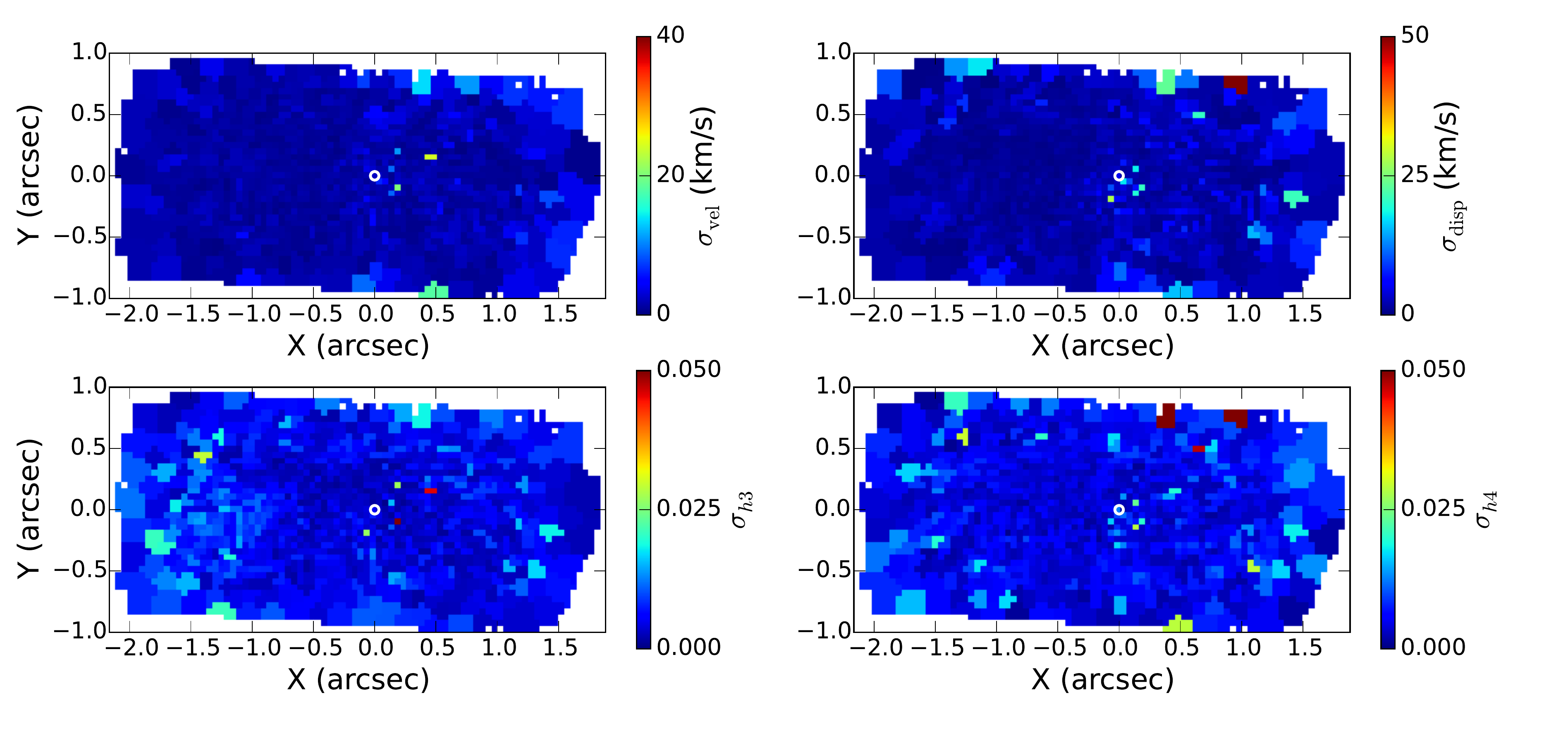}
\caption{Jackknife error maps, to test for template
  mismatch to the data for the velocity ({\em top left}),
  dispersion ({\em top right}), $h3$ ({\em bottom left}), and $h4$ ({\em
    bottom right}). In all panels, the SMBH position is
  marked by the white circle. }
\label{fig:JKerr}
\end{center}
\end{figure*}

In addition to measurement errors, mismatch between the spectral
templates and the intrinsic stellar population can provide a source of
systematic error. The magnitude of this error was assessed via a
jackknife analysis. One spectral template at a time was dropped from
the library used in the fitting routine. pPXF fits were run for each
of these 23 template library subsets, and the systematic errors from
template mismatches were calculated as the standard error of the mean
of the resulting outputs. Error maps are shown in
Figure \ref{fig:JKerr}, and the median errors are 1.3 km s$^{-1}$ on
$v$ and 1.6 km s$^{-1}$ on $\sigma$. These errors are nearly an order
of magnitude lower than the measurement errors, and thus negligible
compared to the observational error. We disregard this source of
error in the remainder of the analysis.

\section{Results}
\label{sec:res}

\subsection{Kinematics of the eccentric disk}
\label{sec:res_kinematics}

\begin{figure*}
\begin{center}
\includegraphics[width=\textwidth]{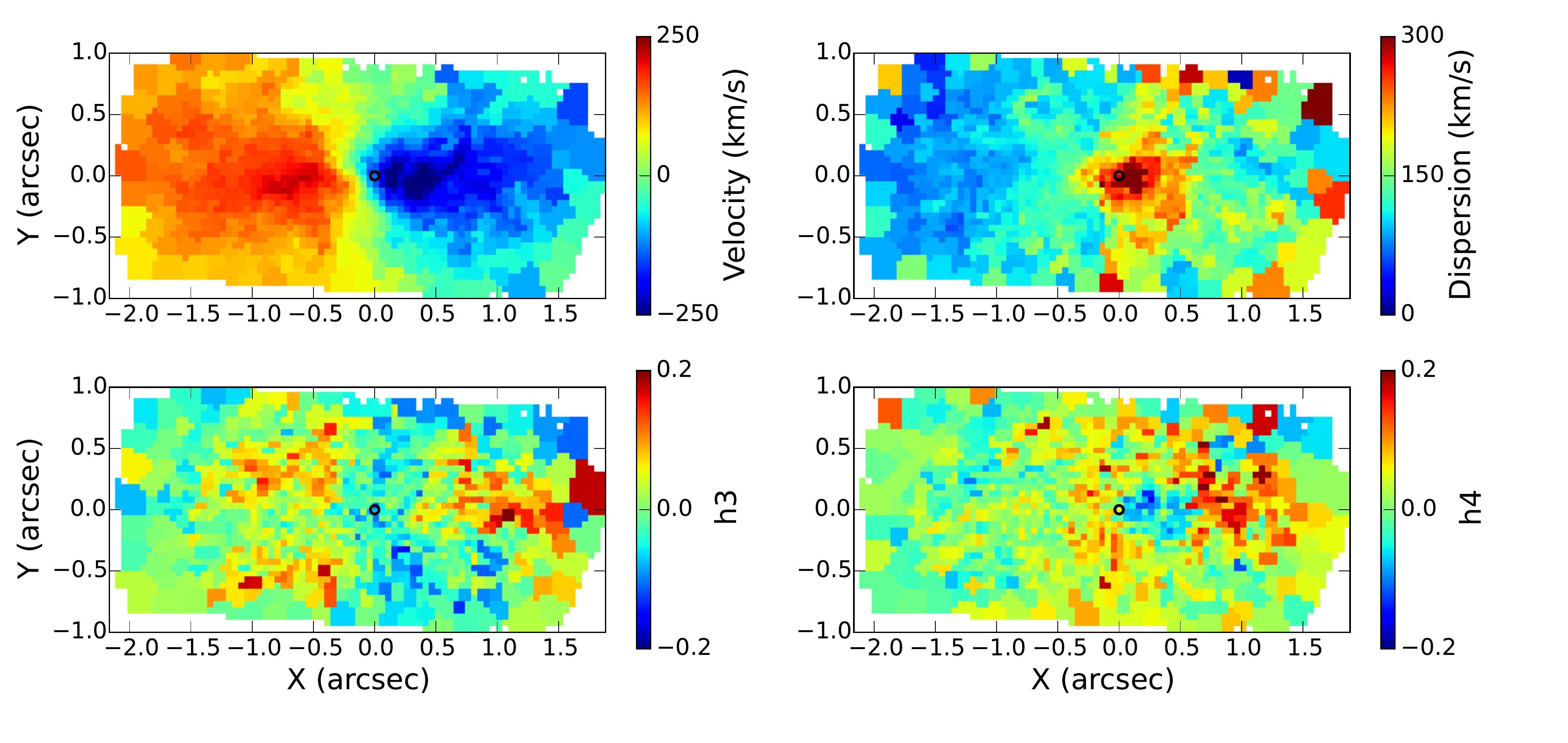}
\caption{Kinematic maps, as calculated with pPXF (see text). In all
  panels, the SMBH position is marked by the black
  circle. \emph{Top left:} Velocity map, shifted by the systemic
  velocity. \emph{Top right:} Velocity dispersion. \emph{Bottom left:}
  $h3$, or the first asymmetric higher-order moment of the
  LOSVD. \emph{Bottom right:} $h4$, or the first symmetric
  higher-order moment of the LOSVD.}
\label{fig:kinmaps}
\end{center}
\end{figure*}

\begin{figure}
\begin{center}
\includegraphics[width=\columnwidth]{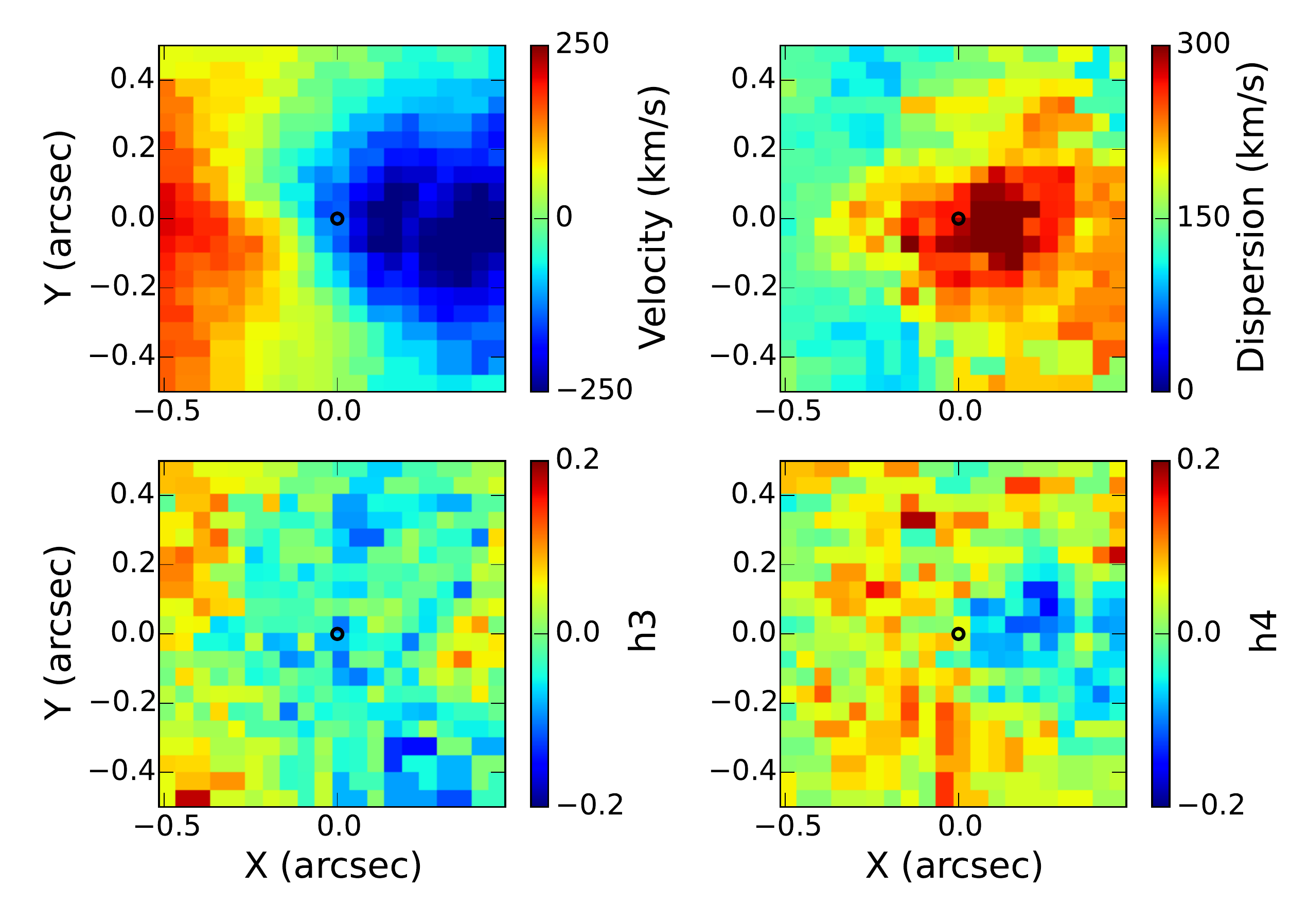}
\caption{Kinematic maps from Figure \ref{fig:kinmaps}, zoomed in to
  the central arcsec around the SMBH (black circle). \emph{Top left:}
  Velocity map, shifted by the systemic velocity. \emph{Top right:}
  Velocity dispersion. \emph{Bottom left:} $h3$ \emph{Bottom right:}
  $h4$.}
\label{fig:kinmapszoom}
\end{center}
\end{figure}

\begin{figure}
\begin{center}
\includegraphics[width=\columnwidth]{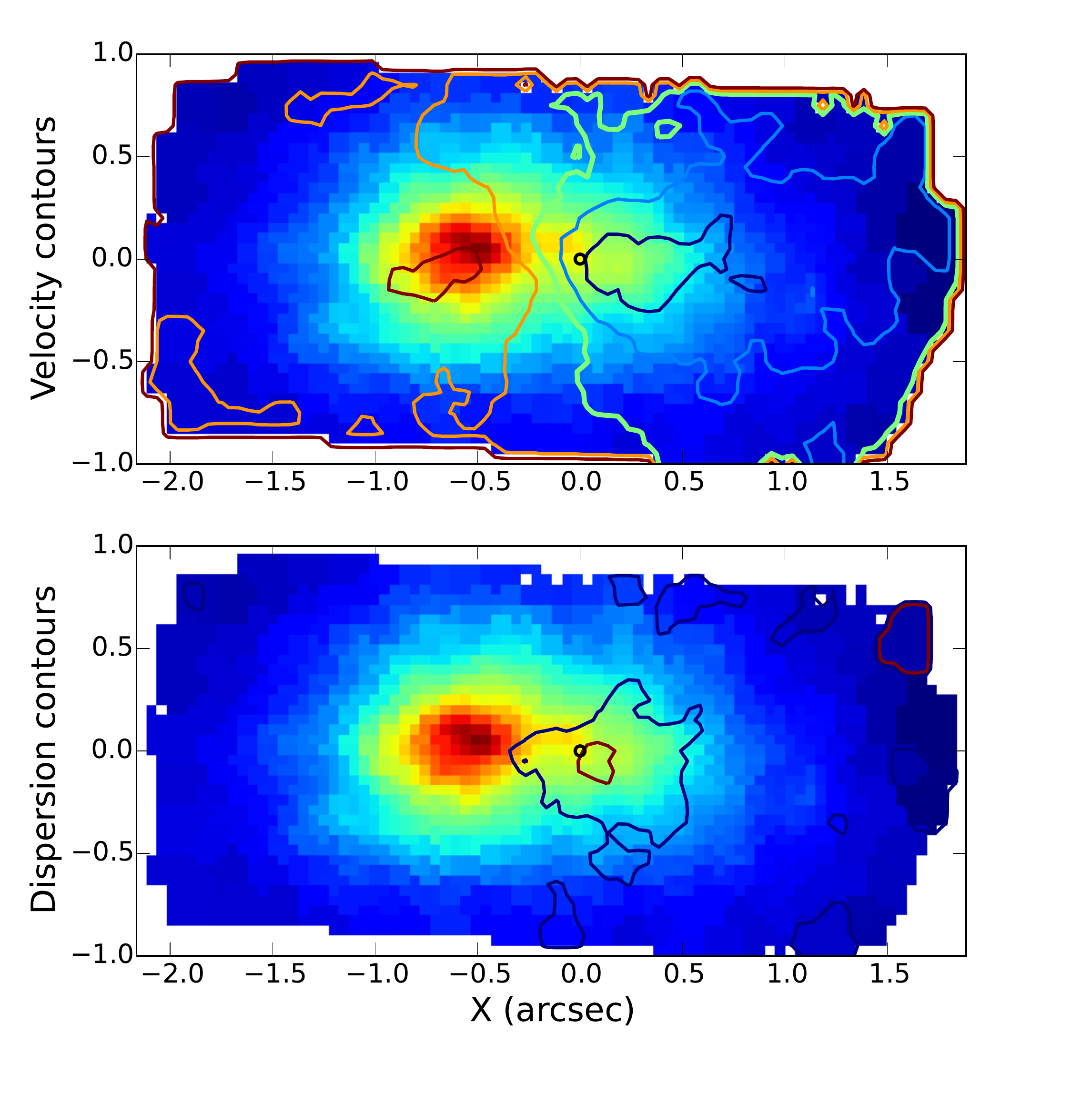}
\caption{Kinematic contours overlaid on the OSIRIS cube, collapsed in the wavelength direction. In both, the contours have been slightly smoothed and the SMBH is marked with a black circle. \emph{Top:} Velocity contours, in steps of 100 km s$^{-1}$, with the lowest contour at $-$200 km s$^{-1}$ shown in dark blue. \emph{Bottom:} Dispersion contours, in steps of 100 km s$^{-1}$, with the lowest contour at 200 km s$^{-1}$ shown in dark blue.}
\label{fig:contours}
\end{center}
\end{figure}

The results of the pPXF kinematic fitting are shown in Figure
\ref{fig:kinmaps} and Figure \ref{fig:kinmapszoom}. 
The velocity map shows a clear rotation signature but with a slight asymmetry in the morphology of the 
velocity peaks. In addition, the peak of the velocity is offset from
the P1 peak flux (Figure \ref{fig:contours}). There is also an asymmetry
in the magnitude of the velocities on either side of the disk, with 
maxima of 223$\pm$7 km s$^{-1}$ on the northeast side and
$-$312$\pm$16 km s$^{-1}$ on the southwest side. In addition, the zero of the 
velocity gradient does not coincide with the SMBH position, but is offset
by 0\farcs15 towards P1, which is consistent with previous
long-slit measurements \citep{bender2005hst-stis}.

The dispersion is peaked close to but not at the position of the SMBH as
is shown in the zoomed kinematic maps in 
Figure \ref{fig:kinmapszoom} and in the contour plot in Figure \ref{fig:contours}. The dispersion reaches a maximum of
381$\pm$55 km s$^{-1}$ at an 
offset in ($x$, $y$) of (+0\farcs09, $-$0\farcs1) from the SMBH, or a distance
of 0\farcs13 on the P2 side. The wider dispersion peak ($\sigma >$ 200 km s$^{-1}$) is concentrated on the P2 side and extends towards P1.

Overall, the maps for the higher order moments $h3$ and $h4$ are quite
noisy and trends are difficult to 
discern. There is a possible enhancement of $h4$ on the P1 side of the
SMBH, but the results are not robust. We find that the significance of
the $h3$ and $h4$ maps are particularly sensitive to the bulge subtraction.

\subsection{Comparison to previous multi-wavelength imaging}
\begin{figure*}
\begin{center}
\includegraphics[width=\textwidth]{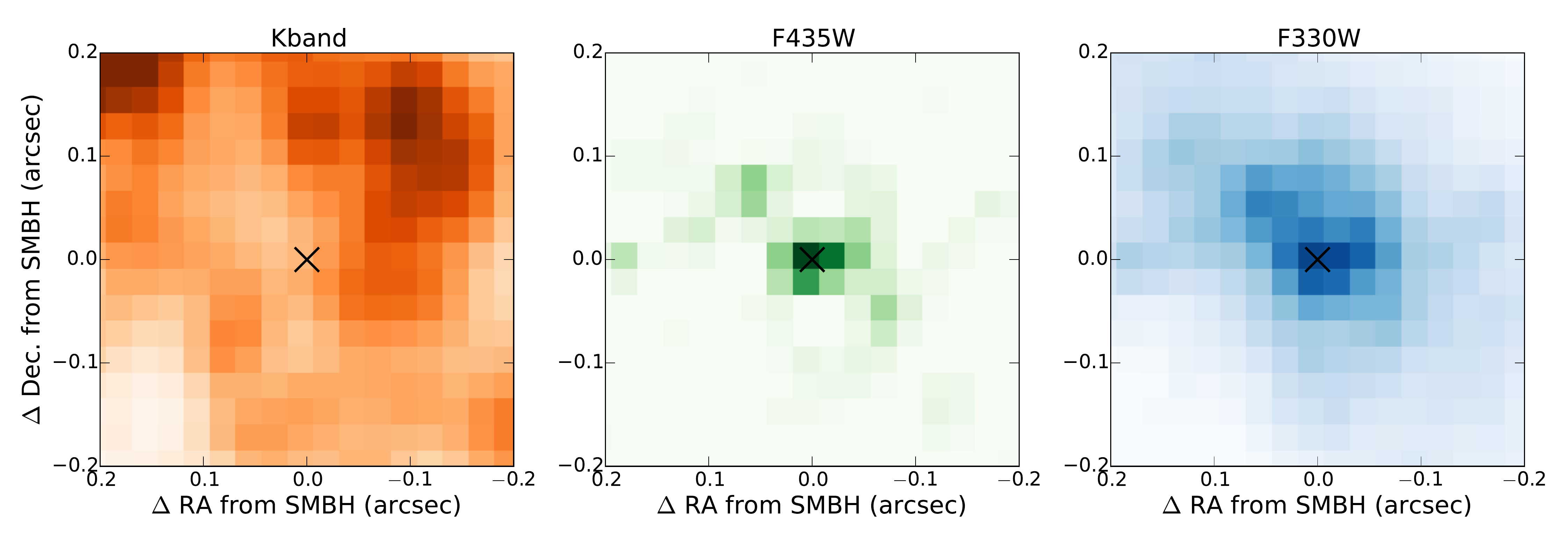}
\caption{Position of the SMBH, taken as source 11 in HST/ACS F435W in
  \citet{lauer2012the-cluster} and calculated in the NIR by aligning
  with the UV data (see text). The SMBH is shown as the black cross in
  each panel. Note that P3, coincident with the SMBH and bright in
  the bluer bandpasses, is not seen above the detection limit in the
  $K-$band image. \emph{Left:} NIRC2 $K-$band image, \emph{center:}
  background-subtracted F435W image (Lauer, private communication),
  \emph{right:} F330W image. All frames have been aligned and scaled
  to the F330W image (0.025$''$ pixel$^{-1}$).}
\label{fig:align}
\end{center}
\end{figure*}

Closer examination of Figure \ref{fig:bhcutall} shows that there are clear differences in the structure of the 
stellar population at each wavelength; notably, P3 is bright and compact only in the F330W and
F435W frames and is dark in the NIR. Figure \ref{fig:align} shows the inner 0\farcs2 in the F330W, 
background-subtracted F435W \citep[][private communication]{lauer2012the-cluster}  and $K'$ images, 
with 
the SMBH position marked. In contrast with the bluer images, there is little to no flux at the position of P3 
and the SMBH in the NIR, indicating a separation of the young and old
stellar populations in the nucleus.

\begin{figure}
\begin{center}
\includegraphics[width=\columnwidth]{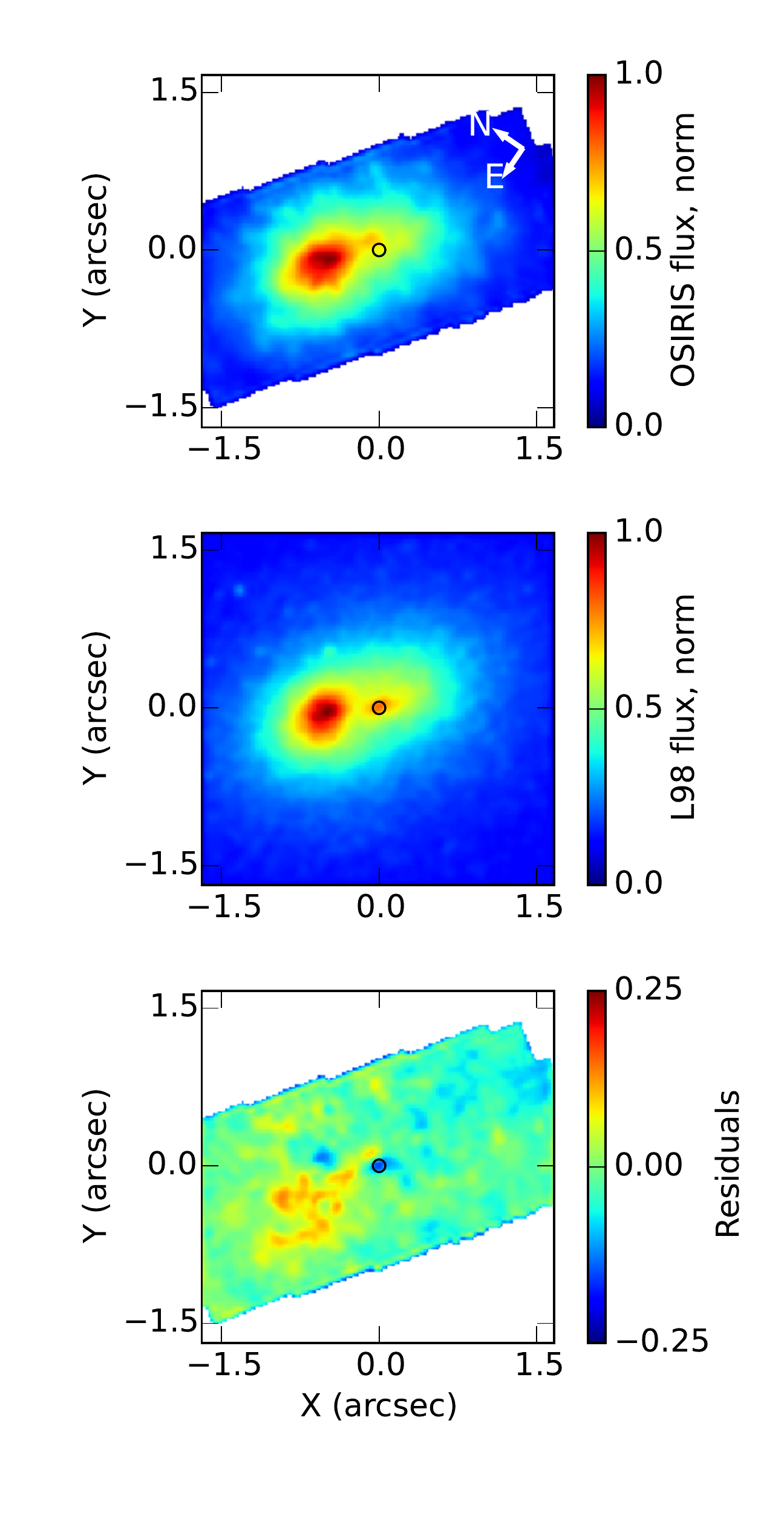}
\caption{Comparison of OSIRIS flux with that of the F555W flux from \citet{lauer1998m32-/--1}. \emph{Top:} OSIRIS data cube collapsed in the wavelength direction, rotated to a PA of 55.7$^{\circ}$ and normalized so the peak flux is equal to 1. \emph{Middle:} F555W flux map from \citet{lauer1998m32-/--1}, normalized so the peak flux is equal to 1, and convolved to match the OSIRIS resolution. \emph{Bottom:} Residuals, taken as the OSIRIS flux minus the F555W flux. In all, the SMBH position is marked with the black circle.}
\label{fig:l98}
\end{center}
\end{figure}

A clearer comparison of the difference between the flux maps at two different wavelengths can be seen in Figure \ref{fig:l98}. \citet{lauer1998m32-/--1} observed the nuclear eccentric disk with HST/WFPC2 in three filters: F300W, F555W, and F814W. We show their F555W flux map, convolved to the OSIRIS spatial resolution, in comparison with the OSIRIS flux map. There are clear structural differences visible between the two flux maps, particularly around the P1 peak. In addition, the morphology differences seen around the SMBH in Figure \ref{fig:bhcutall} can been seen here: the secondary P2 peak is more northerly in the OSIRIS data than in the F555W image.

\begin{figure}
\begin{center}
\includegraphics[width=\columnwidth]{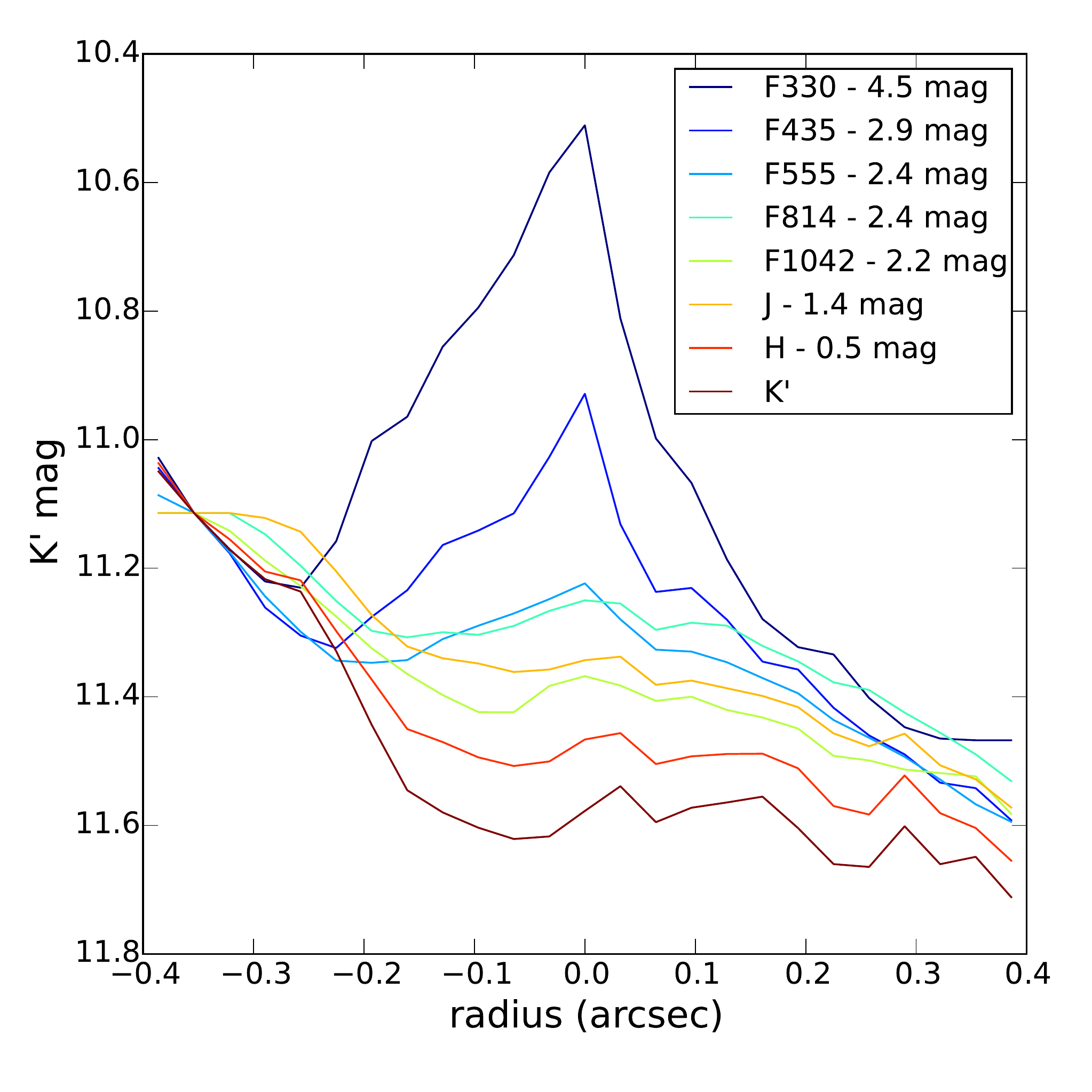}
\caption{One-dimensional cut across each of the eight wavelength
  images, at a PA of 39$^{\circ}$ (e.g., the STIS slit
  PA, \S\ref{ssec:comp}). Offsets have been added to the cuts
  in each wavelength, except for the K'-band curve, the brightest
  wavelength. The offsets were chosen so the flux cuts aligned at
  roughly the position of P1. The additive magnitude offsets are
  indicated in the legend.}
\label{fig:fluxcut}
\end{center}
\end{figure}

This structural difference can also be seen in one-dimensional cuts
across each image (Figure \ref{fig:fluxcut}). 
A 1D cut has been taken across each frame in Figure \ref{fig:bhcutall} at the PA of the 
HST/STIS long slit observations (see \S\ref{ssec:comp}) and passing through the SMBH. Each 1D cut 
has been flux calibrated according to the published instrumental zero-points. The nucleus is 
much brighter in the NIR than at UV or optical wavelengths, so offsets in magnitudes (shown in the figure 
legend) have been added to each cut to equalize them on the P1 side. The UV peak at the origin in the 
F330W and F435W frames is apparent. The optical bandpasses show an extended, shallower flux peak 
roughly coincident with the SMBH position and extending slightly to the southwest. However, the $H$ and $K'$ 
flux cuts do not show this shallower peak and instead are flat from the SMBH position to the cutoff at the 
southwest edge.

\subsection{Comparison to previous spectroscopy}
\label{ssec:comp}
The nucleus of M31 has been well studied with both ground and space-based spectroscopic observations. Several 
recent studies are summarized in Table \ref{tab:res}. In this section, we compare the OSIRIS 
kinematics with two high resolution data sets: HST/STIS long-slit observations reported 
in \citet[hereafter B05]{bender2005hst-stis} and ground-based IFS observations 
reported in \citet[hereafter B01]{bacon2001the-m-31-double}.

\begin{deluxetable}{lccccr}
\tablecaption{Data Quality Comparison
\label{tab:res}}
\tablehead{
\colhead{Instrument} & \colhead{Filter} & \colhead{FWHM} &
\colhead{IFU?} &
\colhead{Reference}
}
\startdata
HST  FOC & $\sim B$ & $\sim$0\farcs05 & No & \citet{statler1999stellar} \\
CFHT  SIS & $\sim I$ & 0$\farcs$63 & No & KB99 \\
CFHT  OASIS & $\sim I$ & 0$\farcs$41--0$\farcs$5 & Yes & B01 \\
HST  STIS & $\sim I$ & 0$\farcs$12 & No & B05 \\
Keck  OSIRIS & $K$ & 0$\farcs$12 & Yes & this work \\
\enddata
\end{deluxetable}

\begin{figure}
\begin{center}
\includegraphics[width=\columnwidth]{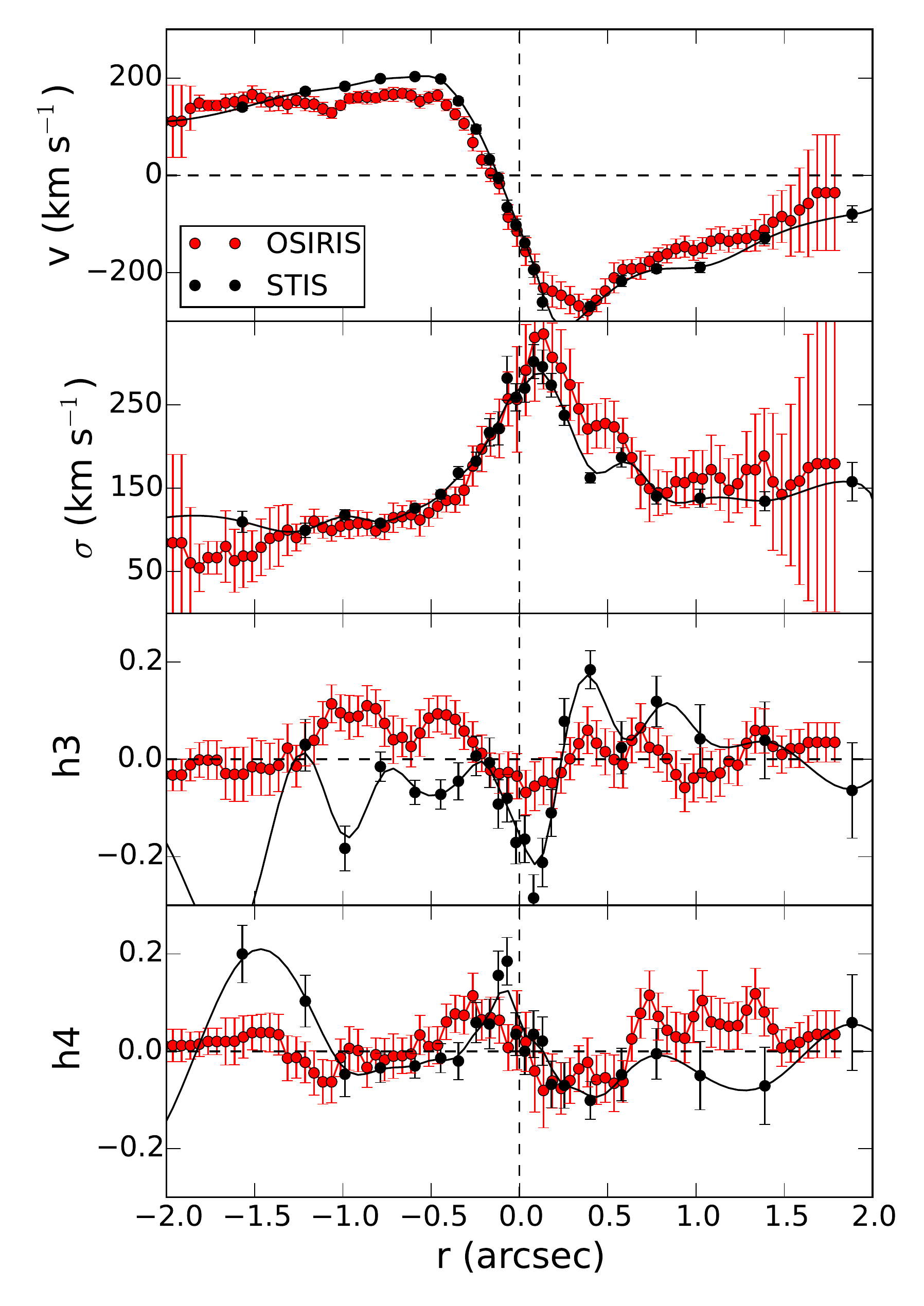}
\caption{Comparison of OSIRIS and STIS kinematics. The STIS data were
  taken at a PA of 39$^{\circ}$ and are shown as black points. The
  STIS data have been smoothed along the slit by the OSIRIS PSF; the
  smoothed STIS data are shown as the solid black line. A 0\farcs1
  width cut, to represent the STIS slit, has been taken across the
  OSIRIS kinematics at the STIS PA; those data, along with the Monte
  Carlo errors, are shown in red. All data have been bulge
  subtracted. The SMBH position is at 0\farcs0, and radius increases
  towards the southwest.}
\label{fig:STIScomp}
\end{center}
\end{figure}

\subsubsection{Comparison with STIS}
\label{ssec:stis}
HST/STIS long-slit spectroscopy was obtained for the nucleus of M31 in
July 1999; the full analysis was published in
B05. Spectra were taken both in blue
(2900--5700 \AA) and red (8272--8845 \AA, the Ca triplet) modes; we
confine our discussion solely to the red mode. The slit was oriented
at a PA of 39$^{\circ}$, along the major axis of the optical disk. The
slit width was 0\farcs1 and the estimated resolution is FWHM =
0\farcs12. 

We compare the kinematics reported in B05 with
our OSIRIS kinematics in Figure \ref{fig:STIScomp}. In addition to the
original STIS kinematics, we show the STIS kinematics smoothed along
the slit by a Gaussian kernel derived using the OSIRIS PSF. The
comparison to the OSIRIS kinematics was derived by taking a cut across
the kinematic maps in Figure \ref{fig:kinmaps} at the STIS PA. Each
point along the STIS slit is represented by the mean of three OSIRIS
spaxels orthogonal to the slit to match the STIS slit width, and the errors are taken as the Monte Carlo errors of the
same spaxels, added in quadrature.  

Though overall the kinematics reported by STIS and by OSIRIS are
similar, there are several inconsistencies, particularly in the
velocity and dispersion. Overall, while the velocity gradient
from the OSIRIS data is well matched to that from STIS, the OSIRIS
data show lower peak velocities on both the red- and blue-shifted
sides of the nuclear disk than the STIS kinematics. In addition, the
wings of the velocity profile from OSIRIS are lower than those from
STIS, which may point to residuals arising from differences in the bulge subtraction
methods. The peak dispersion values are the same, within the errors,
but the OSIRIS dispersion peak is broader, with a possibility of a
secondary dispersion peak at 0\farcs5, on the P2 side. 
This extended enhanced dispersion in the NIR-bright old stellar population 
could be probing the apoapse of the eccentric disk.

\subsubsection{Comparison with OASIS}
\label{ssec:oasis}

\begin{figure}
\begin{center}
\includegraphics[width=\columnwidth]{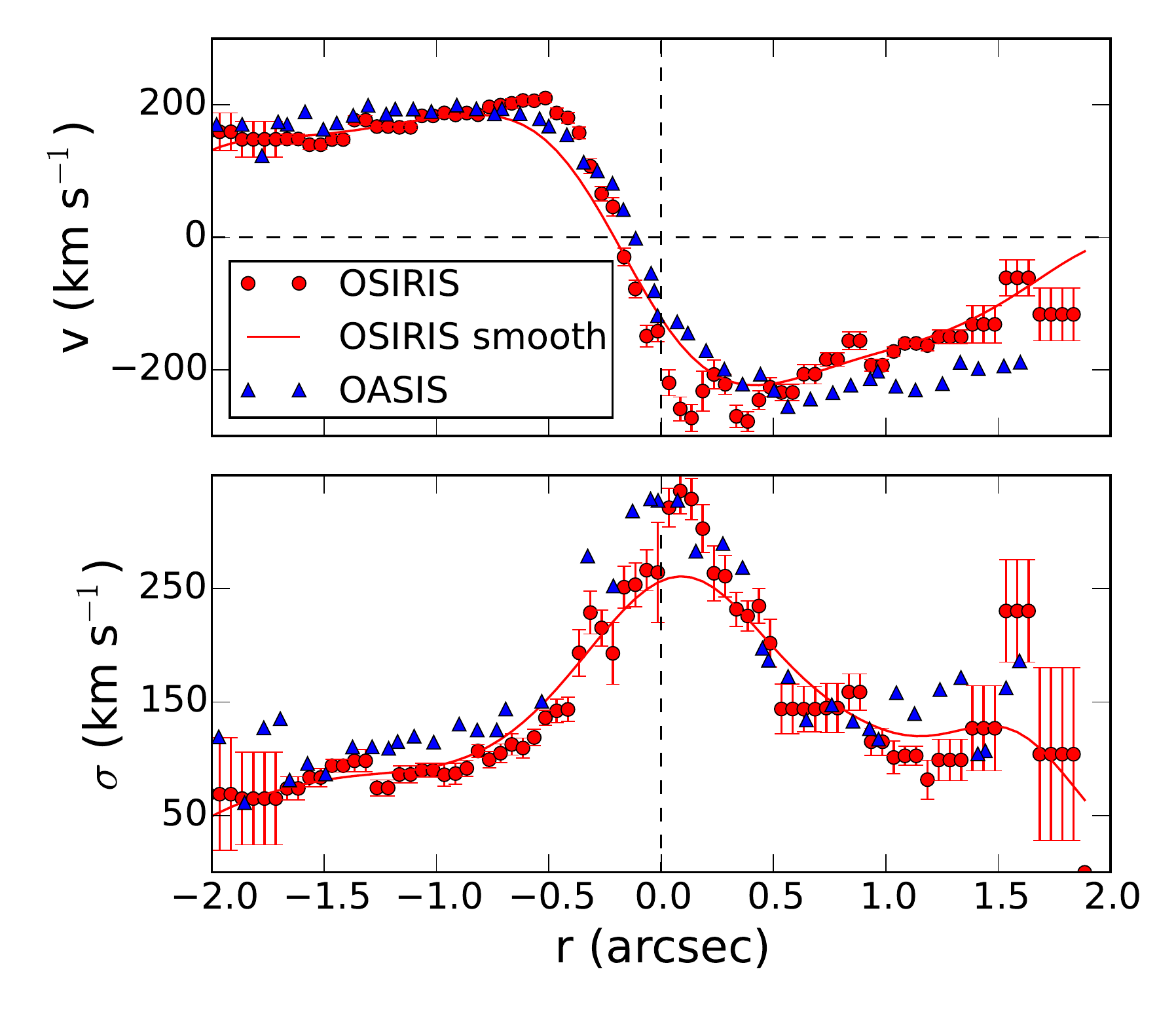}
\caption{Comparison of OSIRIS and OASIS kinematics. The OASIS data,
  shown in blue and taken from their Fig. 10, are taken at a PA of
  56.4$^{\circ}$, which they identify as their kinematic major
  axis. The cut across the OSIRIS data, shown in red, is taken at the
  same PA, through our SMBH position and parallel to the long axis of
  the OSIRIS FOV. The red curve shows the OSIRIS kinematics smoothed
  to the OASIS resolution. All data have been bulge subtracted. Our
  SMBH position is at 0\farcs0, and radius increases towards the
  southwest. The OASIS data have been shifted by $-$0\farcs1 to align the
  velocity gradients.}
\label{fig:OASIScomp}
\end{center}
\end{figure}

AO corrected integral field spectroscopy was obtained by the
OASIS instrument on CFHT and first presented in
B01. Two different mosaics of different
spatial resolution were presented, with resolution ranging from
0\farcs41 to 0\farcs5. Similar to the STIS data, the chosen wavelength
range was 8351--9147 \AA~to optimize observations of the Ca
triplet. OASIS provides a 4$''$ $\times$ 3$''$ FOV.

We show the bulge-subtracted OASIS kinematics in
Figure \ref{fig:OASIScomp}. These kinematics were extracted along the
OASIS kinematic major axis (PA = 56.4$^{\circ}$) and were initially
presented in B01 (their Fig. 10, solid line). We also show a cut across the OSIRIS
kinematics at the same PA. The red points are the original OSIRIS
kinematics and the red curve is the OSIRIS kinematics smoothed to the
OASIS resolution. The B01 velocity and dispersion profiles have been
shifted by $-$0\farcs1, to better align the velocity gradient with that
of the OSIRIS data and to account for differing SMBH positions.
Overall, the OASIS velocity profiles agrees well with the smoothed
OSIRIS data, and is similar to the correspondence seen between the
OASIS and STIS kinematics, as reported in B01 (their
Fig. 13). However, the OASIS and OSIRIS kinematics are not quite
aligned along the radial direction. This offset is potentially due to
a difference in SMBH position between the OASIS and OSIRIS data;
shifting the comparison cut in the OSIRIS data by up to 0\farcs1 from
our SMBH position bring the data into closer alignment along the
radial direction, though the correspondence is still not
perfect. Also, smoothing the OSIRIS kinematics to match the OASIS
resolution brings the two velocity profiles into agreement, but the
reported OASIS dispersion peak is higher than that seen in the
smoothed OSIRIS data. The peak dispersion in the smoothed
two-dimensional OSIRIS dispersion map is less than 250 km s$^{-1}$, or
significantly lower than that seen by OASIS. It is unclear how the
OASIS data are able to probe the dispersion at such high resolution
given that beam smearing should lower the peak resolved
dispersion. Differences in bulge subtraction may be the culprit, or
else the OASIS resolution may be better than reported.

\section{Eccentric Disk Models}
\label{sec:mod}

We compare the OSIRIS data to the flux and kinematic models from
\citet[][hereafter PT03]{peiris2003eccentric-disk}. These Keplerian
models were originally fit to HST photometry and ground-based
long-slit data (KB99) to determine the orientation of the eccentric
disk with respect to the larger-scale galactic disk. In this work, we
add rigid-body rotation to the models and determine the best-fit
precession rate and orientation using the OSIRIS data. Details of the
model fitting are described in \S\ref{ssec:mfit} and results
on the disk orientation and precession are presented in
\S\ref{ssec:orient} and \S\ref{ssec:precess}, respectively.

\subsection{Model fitting}
\label{ssec:mfit}

The original disk models consisted of $\sim 10^7$ particles in disk-plane
coordinates, which were rotated to two different orientations as described in PT03: the
{\em aligned} model, matched to the orientation of the large-scale galactic disk, and the
{\em non-aligned} model, with orientation parameters left free
(Table \ref{tab:modfit}, columns 2 and 3).
The aligned models were a poor fit to both our data and that of PT03,
thus we drop the aligned models from the discussion and
focus on the nonaligned models. 
Throughout the model fitting, we adopt the best-fit parameters
of the PT03 non-aligned model, 
including a black hole mass of M$_\bullet = 1.02\times10^8$
\msun, and keep the parameters fixed unless otherwise specified.
In order to fit the model to the OSIRIS data, the particles from
the nonaligned model were first rotated to an orientation of our
choosing as specified by three angles: 
the inclination of the disk with respect to the sky ($\theta_i$), 
the angle of the ascending node in the sky plane ($\theta_l$), 
and the angle from the ascending node to the periapse vector
($\theta_a$). 
A solid body rotation was also introduced by adding a fixed 
rotation speed, $\Omega_P$ (in units of km s$^{-1}$ pc$^{-1}$), to the disk-plane 
model velocities. A righthanded coordinate system was assumed, so 
positive precessions are counterclockwise. For a given set of model
parameters, the particles were randomly perturbed in their sky-plane
positions using a Gaussian kernel that matched the OSIRIS resolution
in order to smooth out features at higher resolutions.
The particles were then spatially binned using the
tessellated pattern used for the data (\S\ref{ssec:bin}) and the line-of-sight
velocities were binned in increments of 5 km s$^{-1}$. The resulting
model LOSVD was fit using the Gauss-Hermite expansion in the same
manner as the observations.

In order to find the best-fit
orientation and precession with our new data, a grid search was performed
over $\Omega_P$, $\theta_i$, $\theta_l$, and $\theta_a$. 
The grid of angles was centered on the original best-fit nonaligned
values and spread over a range of $\pm$15$^{\circ}$ with a step of 5$^{\circ}$ for each
angle. Precession values in the coarse grid ranged from 
$-$30 to +30 km s$^{-1}$ pc$^{-1}$ in steps of
5 km s$^{-1}$ pc$^{-1}$. 

The goodness of fit was tested by calculating the $\chi^2$, or the sum 
of the squared residuals between the OSIRIS flux or kinematics 
and that of the model, weighted by the squared flux or MC errors. 
Only the flux, velocity, and dispersion moments were used in the
fitting process, as the $h3$ and $h4$ 
measurements have lower S/N. For the flux, the OSIRIS data and model 
were first normalized by dividing by the sum of the flux in each image. Only the inner 1\farcs3 
was used for the calculation. The resulting $\chi_m^2$ for each
moment, $m$, 
was divided by the number of spaxels (1746) within the inner region
minus the number of free parameters (4), to obtain a reduced
$\tilde{\chi}_m^2$. 
The best fitting model across all moments was selected based on a weighted sum 
$\tilde{\chi}^2 = \sum_m \tilde{\chi}_m^2 w_m$  over the flux, velocity, and dispersion
moments. The weights used were the inverse of the minimum $\tilde{\chi}_m^2$ value for
each moment, which is equivalent to error re-scaling, in order to
prevent any one moment from dominating the fit. 

The errors on the best-fit parameters were estimated by performing two Monte Carlo
analyses. First, errors were calculated using 100 samples of the data. In each, 
the flux and kinematic maps were randomly perturbed using a normal
distribution matched to the MC errors for the data and the
$\tilde{\chi}^2$ values were recalculated for every model in the
grid. Notably, all Monte Carlo samples yielded the same best-fit
model parameters as the original, non-perturbed data; however, the best-fit
$\tilde{\chi}^2$ for each MC sample varied. To define a 68\%
(1$\sigma$) confidence interval from the data error analysis, we adopted the standard
deviation of the best-fit $\tilde{\chi}^2$ values, or $\Delta\tilde{\chi}_{data}^2$ = 0.01. 
MC errors were also calculated for the models using 100 simulations.
In each realization, the model $x$ and $y$ sky plane position of each particle was
perturbed randomly within a Gaussian kernel that matched the OSIRIS
resolution. For each MC simulation,
the model particles in the best-fit orientation were randomly perturbed within the same
Gaussian kernel. To define a 68\% (1$\sigma$) confidence interval from
the model stochasticity, we adopted the standard
deviation of the $\tilde{\chi}^2$ values from the best-fit models, or $\Delta\tilde{\chi}_{model}^2$ = 0.08.
The errors from the data and the model MC simulations were added in quadrature, for a total
error of $\Delta\tilde{\chi}^2$ = 0.08. Errors on each of the fitted model parameters were 
derived using this error.

\begin{figure*}
\begin{center}
\includegraphics[width=\textwidth]{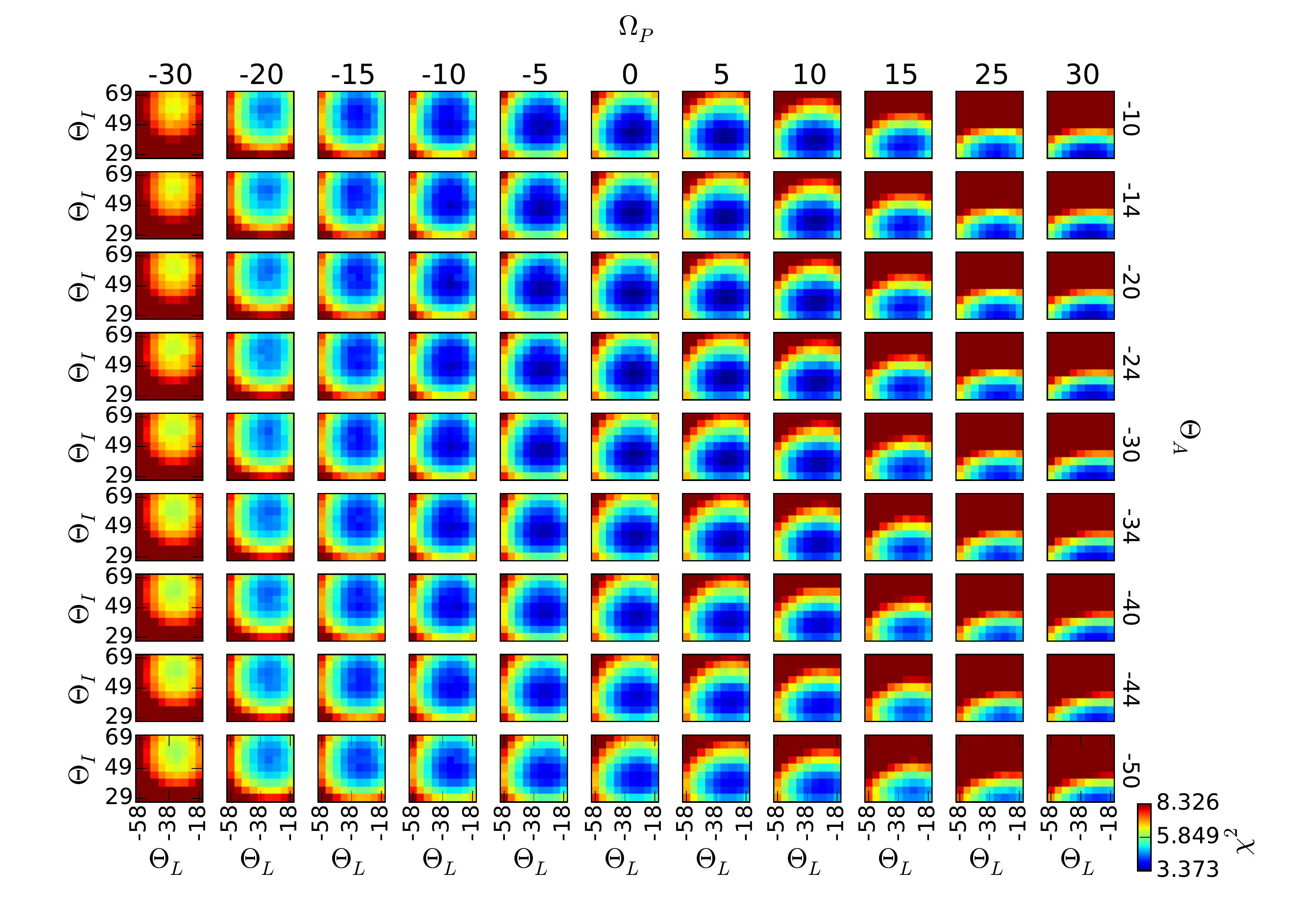}
\caption{Map of the weighted sum of the reduced $\tilde{\chi}^2$ values for the full 
range of all four modeled parameters 
($\theta_l$, $\theta_i$, $\theta_a$, and $\Omega_P$). Each small box shows the range
of $\theta_i$ (y-axis) and $\theta_l$ (x-axis), while $\theta_a$ increases along the large-scale
y-axis and $\Omega_P$ increases along the large-scale x-axis. The color stretch in
each panel has been adjusted to emphasize the $\tilde{\chi}^2$ minimum. }
\label{fig:chi2map}
\end{center}
\end{figure*}

\begin{figure*}
\begin{center}
\includegraphics[width=\textwidth]{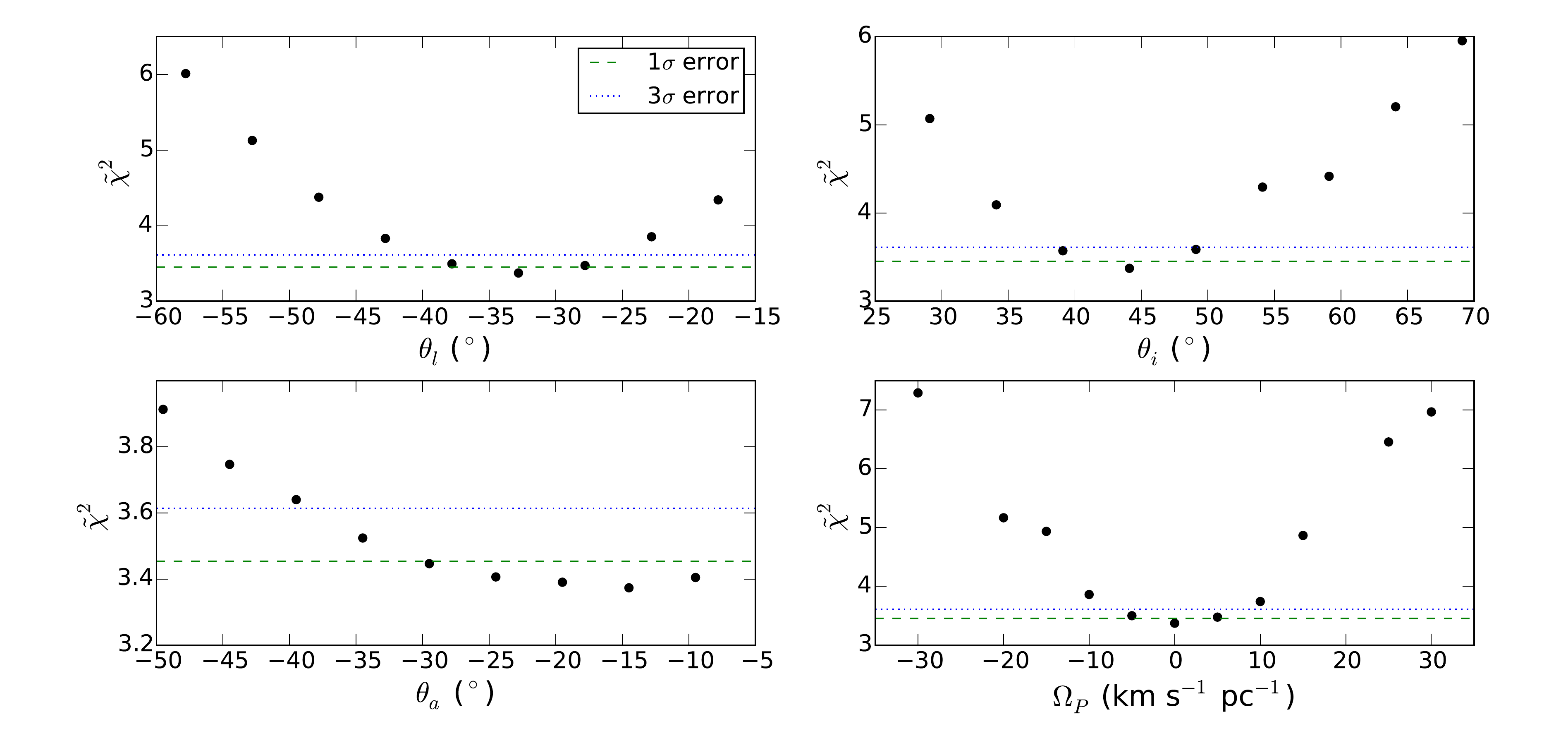}
\caption{The reduced, weighted, and summed $\tilde{\chi}^2$ is shown as a function of the four fitted 
parameters: $\theta_l$, $\theta_i$, $\theta_a$, and $\Omega_P$. For each, the three non-plotted 
parameters are held constant at their best-fit value and only the given parameter is varied. 
The 1$\sigma$ error, taken as the standard deviation of the minimum reduced, weighted, and 
summed $\tilde{\chi}^2$ from the MC simulations, is shown as the dashed horizontal line. 
The 3$\sigma$ errors from the same is shown as the dotted horizontal line.}
\label{fig:1Dchi2}
\end{center}
\end{figure*}

Figure \ref{fig:chi2map} maps the sum of the weighted, reduced $\tilde{\chi}^2$
values for the full range of model parameters that were explored.
Figure \ref{fig:1Dchi2} shows $\tilde{\chi}^2$ values as a function of each parameter
in 1D with the other parameters held constant at their best fit values.
The best-fit $\tilde{\chi}_m^2$ for each moment and the combined
$\tilde{\chi}^2$ are shown in Table \ref{tab:chi2}.

\begin{figure*}
\noindent
\centering
\begin{minipage}[t]{0.49\textwidth}
\includegraphics[width=\columnwidth]{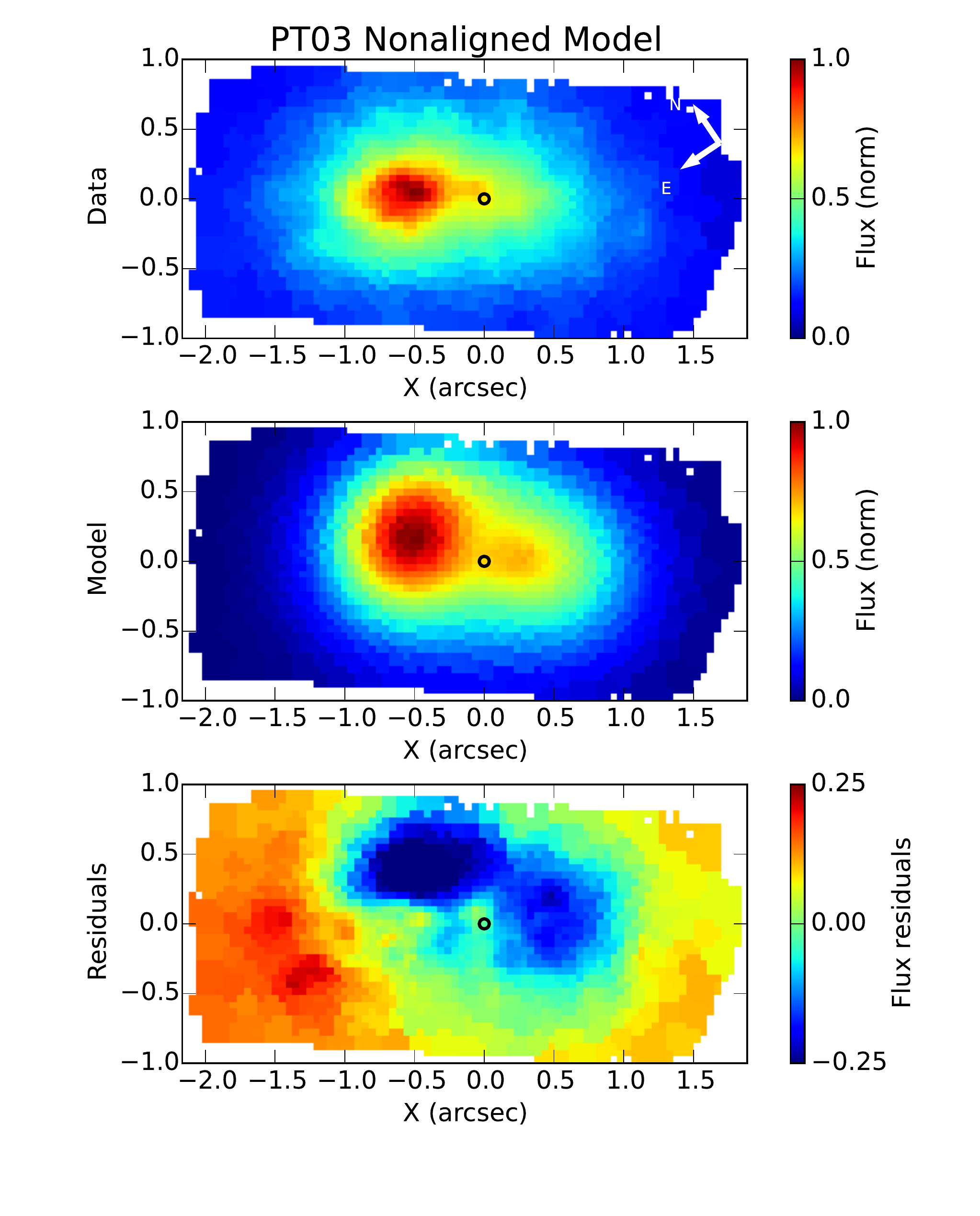}
\end{minipage}
\begin{minipage}[t]{0.49\textwidth}
\includegraphics[width=\columnwidth]{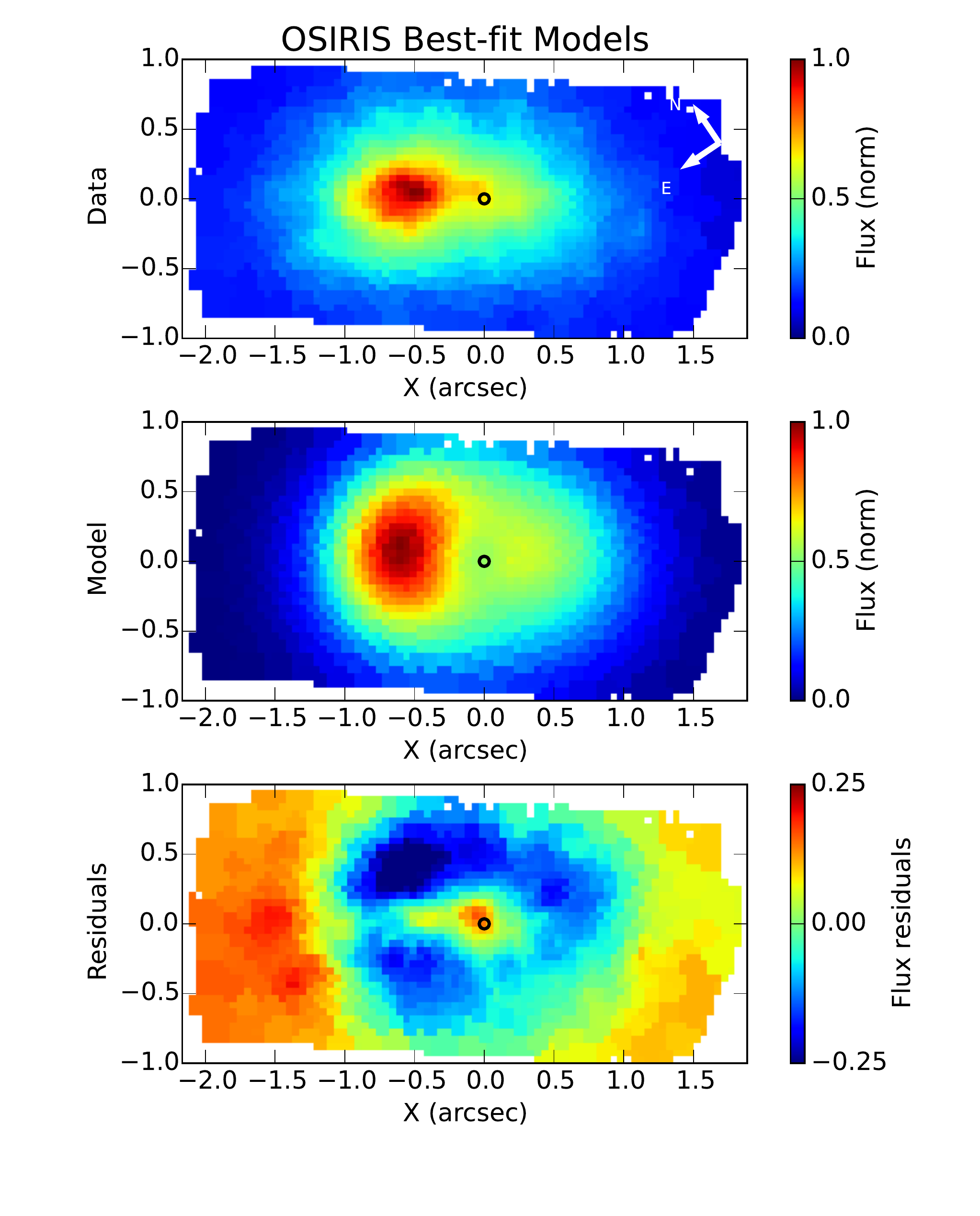}
\end{minipage}
\caption{Comparison of data with modeled flux. In both, the \emph{top} panels shows the normalized flux in the collapsed OSIRIS data cube, the \emph{center} panels shows the modeled number of stars per
  spatial bin (normalized, taken to be equivalent to the flux), and the \emph{bottom} panels show the residuals, data minus model. \emph{Left:} Comparison of data with nonaligned models from \citet{peiris2003eccentric-disk}.
\emph{Right:} Comparison of data with best-fit models. Orientation
angles and precession of the models are given in Table \ref{tab:modfit}.}
\label{fig:modflux}
\end{figure*}

\begin{figure*}
\noindent
\centering
\begin{minipage}[t]{0.49\textwidth}
\includegraphics[width=\columnwidth]{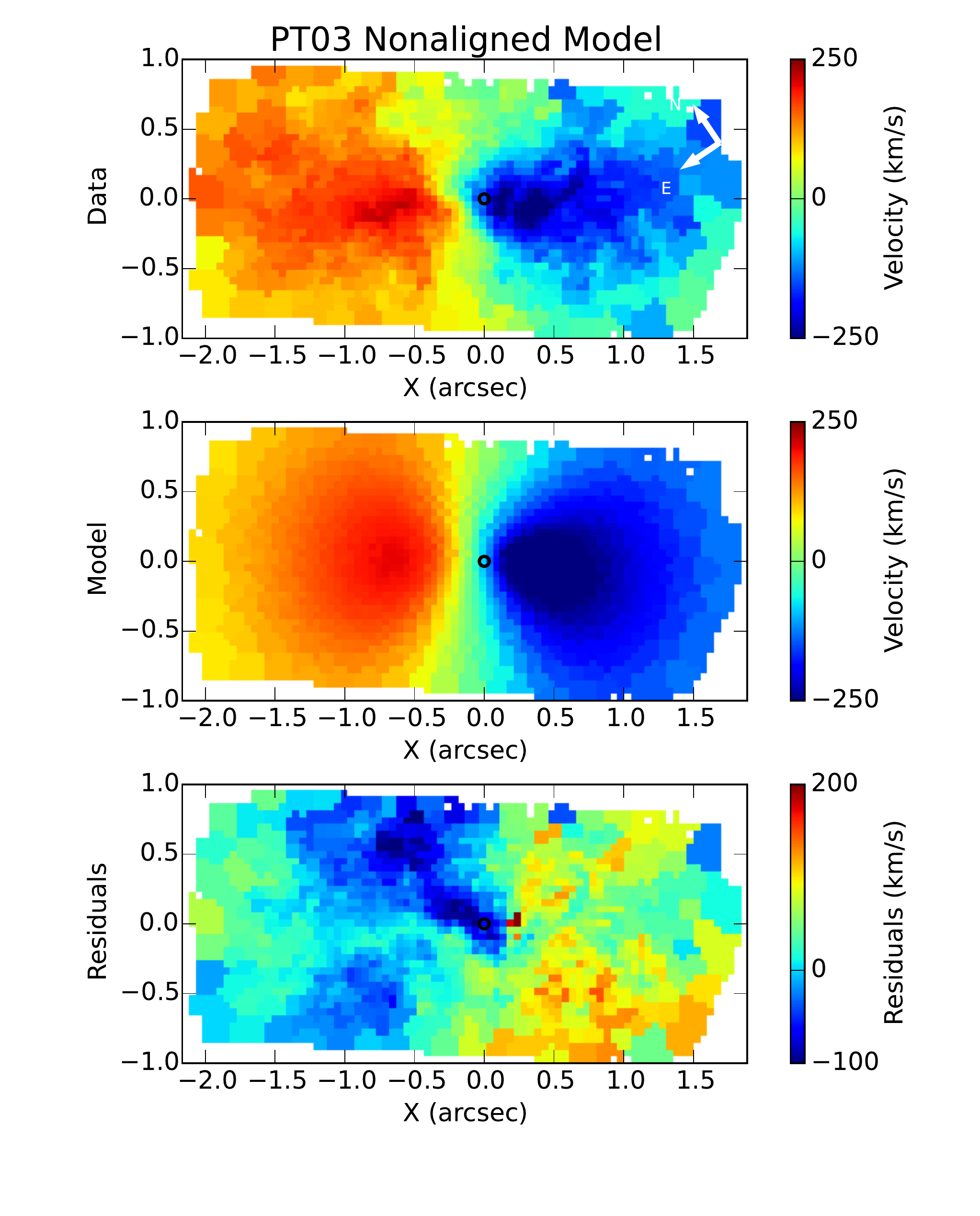}
\end{minipage}
\begin{minipage}[t]{0.49\textwidth}
\includegraphics[width=\columnwidth]{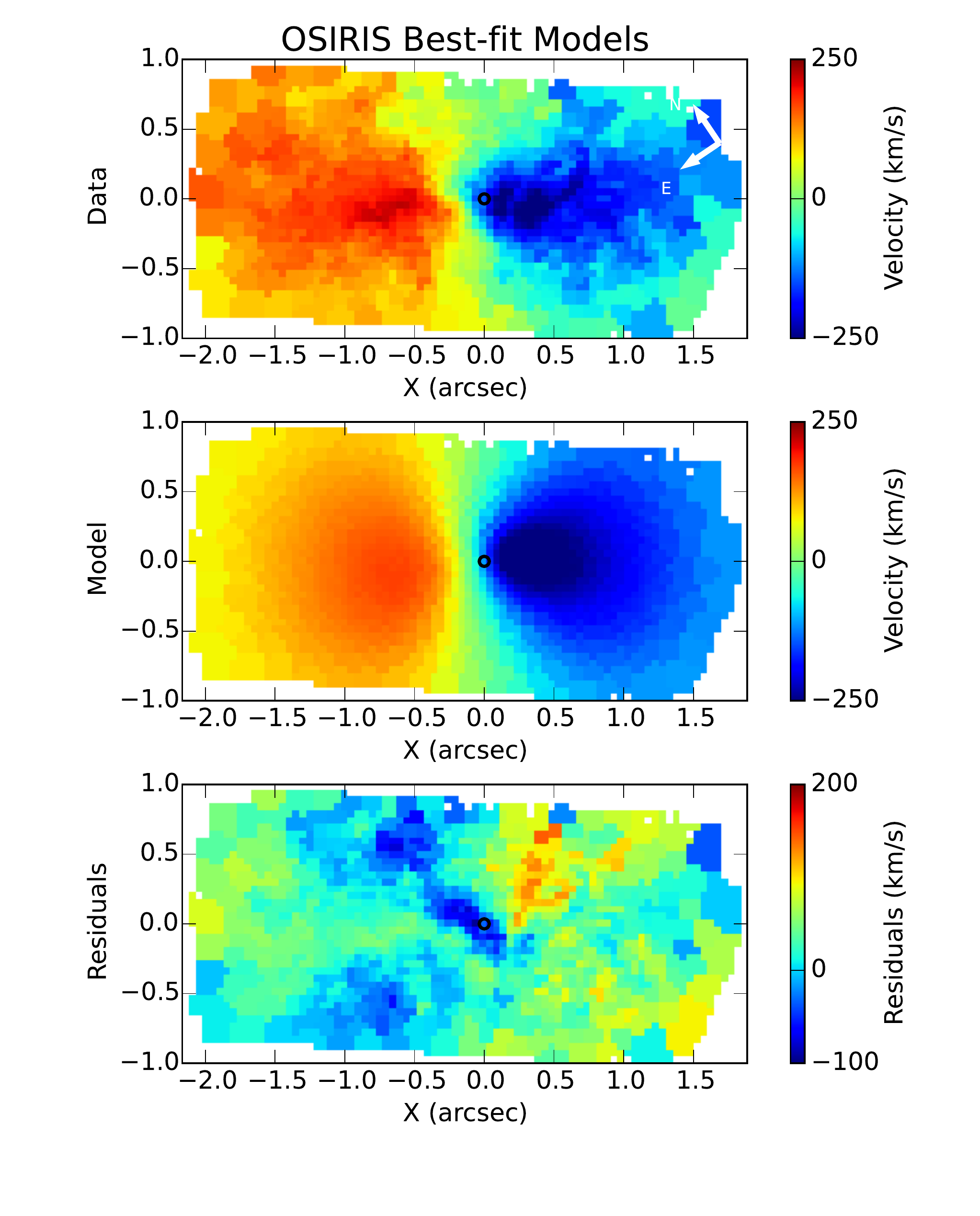}
\end{minipage}
\caption{Comparison of OSIRIS velocity with modeled velocity. Panels are similar to those in Figure \ref{fig:modflux}. \emph{Left:} Comparison of velocity data and PT03 nonaligned models. \emph{Right:} Comparison of velocity data and best-fit models.}
\label{fig:modvel}
\end{figure*}

\begin{figure*}
\noindent
\centering
\begin{minipage}[t]{0.49\textwidth}\label{fig:nonalignedsigma}
\includegraphics[width=\columnwidth]{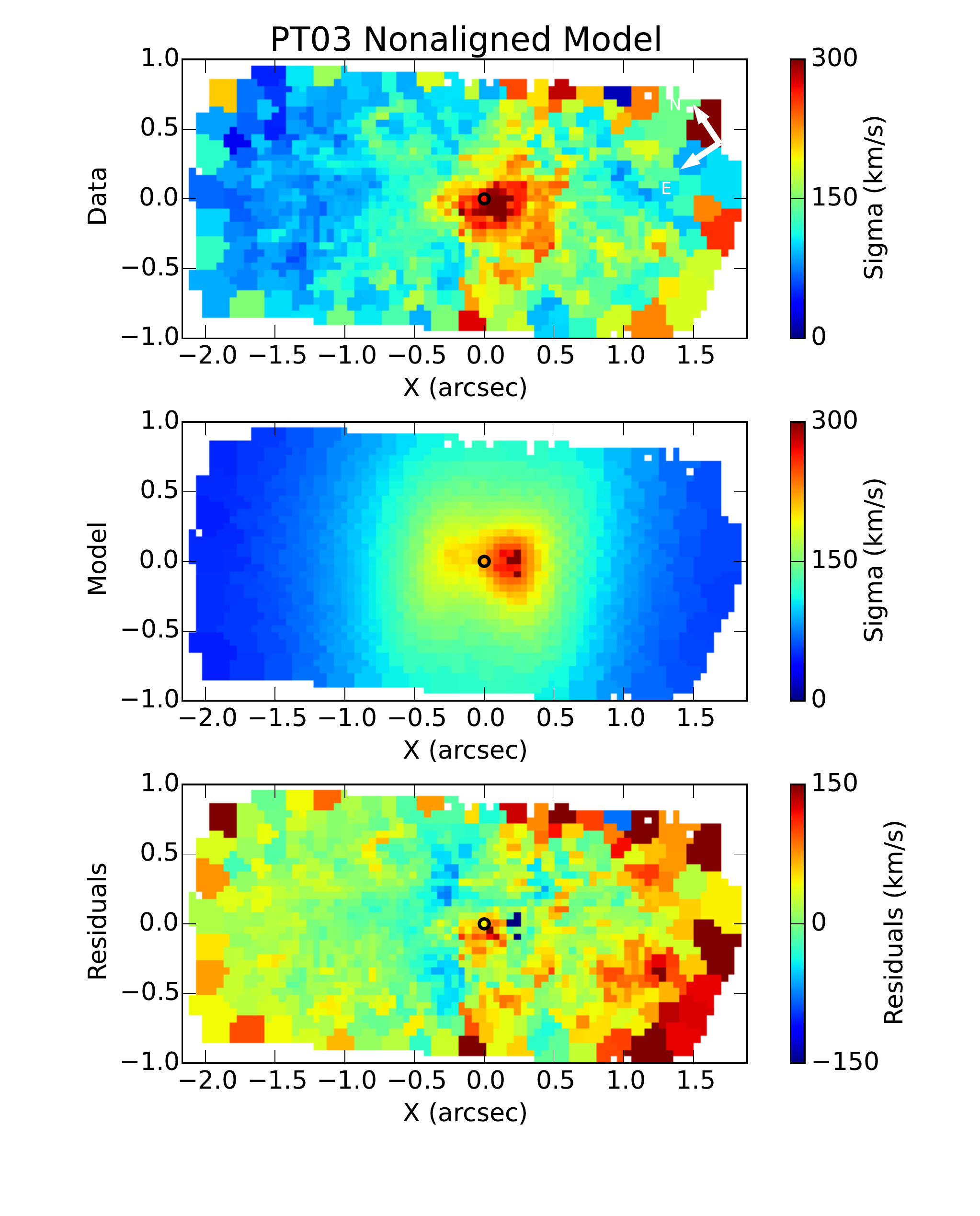}
\end{minipage}
\begin{minipage}[t]{0.49\textwidth}\label{fig:nonaligned052}
\includegraphics[width=\columnwidth]{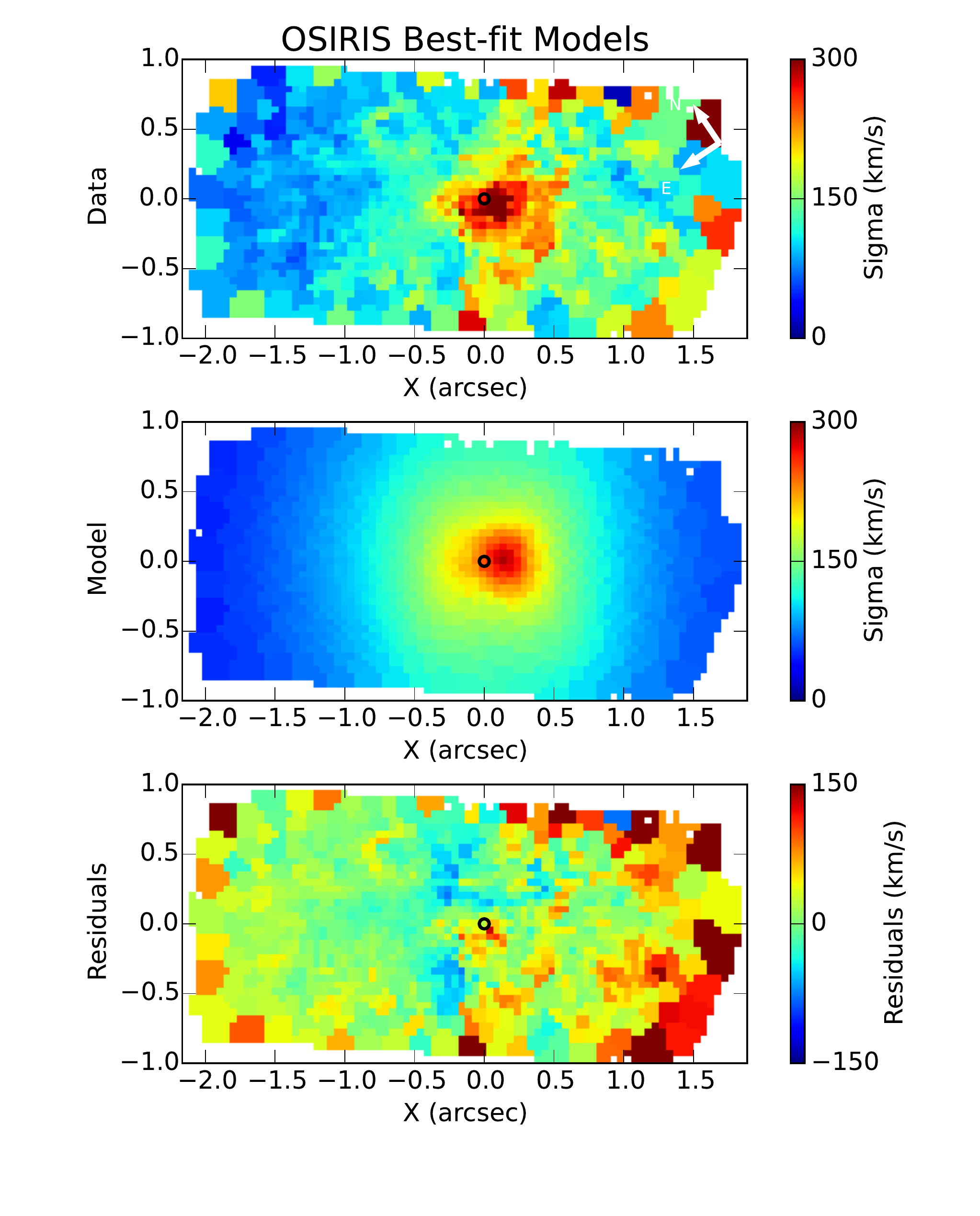}
\end{minipage}
\caption{Comparison of OSIRIS dispersion with modeled dispersion. Panels are similar to those in Figure \ref{fig:modflux}. \emph{Left:} Comparison of dispersion data and PT03 nonaligned models. \emph{Right:} Comparison of dispersion data and best-fit models.}
\label{fig:modsig}
\end{figure*}



\subsection{Orientation}
\label{ssec:orient}

Our flux, kinematics, and residuals from the original nonaligned models of
PT03 are shown in the left panels of
Figure \ref{fig:modflux}, Figure \ref{fig:modvel}, and Figure \ref{fig:modsig}.
We note some discrepancies, particularly in the flux and 
velocity comparisons. (Following PT03, we note that comparisons between the data and 
models are only valid within the inner 1\farcs3.) The flux residuals 
may be due to the difference in morphology of the stellar population 
at different wavelengths (\S\ref{ssec:bhpos}), as the flux models 
were originally fit to imaging data taken by HST in filter F555W (originally 
reported in \citealt{lauer1998m32-/--1}) and the kinematics were fit to 
Ca triplet long slit data from KB99. 



The best-fit models and residuals are shown in the right panels of 
Figure \ref{fig:modflux}, Figure \ref{fig:modvel}, and Figure \ref{fig:modsig}.
The best-fit orientation is [$\theta_l$, $\theta_i$, $\theta_a$] = 
[$-$33$^{\circ}\pm4^{\circ}$, 44$^{\circ}\pm2^{\circ}$, 
$-$15$^{\circ}\pm15^{\circ}$]. The 1$\sigma$ errors are derived using the 
$\Delta\tilde{\chi}^2$ above.
Overall, the fitting preferred smaller values of $\theta_i$ and larger 
values of $\theta_l$ and $\theta_a$ than that found in PT03 (see Table
\ref{tab:modfit} 
for a comparison).

\begin{deluxetable}{crrr}
\tablecaption{Model Fitting Results
\label{tab:modfit}}
\tablehead{
\colhead{Parameter} & \colhead{Aligned$^a$} & \colhead{Nonaligned$^a$} & \colhead{Best fit$^b$}
}
\startdata
$\theta_l$ & $-$52.3$^{\circ}$ & $-$42.8$^{\circ}$ & $-$33$^{\circ}\pm$4$^{\circ}$ \\
$\theta_i$ & 77.5$^{\circ}$ & 54.1$^{\circ}$ & 44$^{\circ}\pm$2$^{\circ}$ \\
$\theta_a$ & $-$11.0$^{\circ}$ & $-$34.5$^{\circ}$ & $-$15$^{\circ}\pm$15$^{\circ}$ \\
$\Omega_P^c$ & ... & ... & 0.0$\pm$3.9 \\
\enddata
\tablenotetext{a}{PT03}
\tablenotetext{b}{This work}
\tablenotetext{c}{Not fit in PT03, units are km s$^{-1}$ pc$^{-1}$}
\end{deluxetable}

\begin{deluxetable}{lrr|rr}
\tablecaption{Minimum reduced $\tilde{\chi}_m^2$ by model
\label{tab:chi2}}
\tablehead{
\colhead{} & \colhead{} & \colhead{} & \colhead{Best fit} & \colhead{Total} \\
\colhead{Moment} & \colhead{PT03} & \colhead{TW $\Omega_P$} & \colhead{per moment} & \colhead{best fit} 
}
\startdata
Flux & 154.8 & 99.6 & 95.6 & 99.7 \\
Velocity & 21.9 & 190.8 & 11.3 & 14.1 \\
Dispersion & 4.0 & 3.6 & 3.1 & 3.3 \\
$h3^a$ & 13.2 & 11.6 & 7.7 & 9.5 \\
$h4^a$ & 8.2 & 7.4 & 6.5 & 6.9 \\
\hline
Weighted total & 4.7 & 18.9 & ... & 3.4 \\
\enddata
\tablenotetext{a}{Not included in weighted fit}
\end{deluxetable}

\subsection{Precession}
\label{ssec:precess}

The reduced $\tilde{\chi}^2$ values for the flux and velocity
residuals preferred a positive precession, while the dispersion
residuals preferred a negative precession, but overall the 
preferred precession is $\Omega_P = 0.0 \pm 3.9$ km s$^{-1}$ pc$^{-1}$. We show the data, model, and residuals 
for the best fitting orientation and precession in the right panels of
Figure \ref{fig:modflux}, Figure \ref{fig:modvel}, and Figure \ref{fig:modsig}. 

We also fit the precession using the 1D method formulated in
\citet[hereafter TW]{tremaine1984a-kinematic} and modified by
\citet{sambhus2000the-pattern}. Briefly, the original method requires
1D profiles of both the surface brightness and the LOS velocity along
the line of nodes, or the intersection of the sky and disk planes. The
modified method allows for the profiles to be taken along the P1$-$P2
line. In either, the formulation is 
\[ \Omega_P~\text{sin}~\theta_i \int_{-\infty}^{\infty} r~\Sigma(r)~dr = \int_{-\infty}^{\infty} v_{\mathrm{LOS}}(r)~\Sigma(r)~dr \]
where $r$ is the projected distance from the black hole, $\Sigma(r)$
is the surface brightness, and $v_{\mathrm{LOS}}(r)$ is the line-of-sight velocity. 

We take the surface brightness and LOS velocity profiles along the P1$-$P2 line, or parallel to the long axis of our FOV, and calculate the precession along the profile intersecting with the SMBH position. The method allows determination of the precession along strips parallel to this line. The precession along two strips on either side of the profile intersecting the SMBH position is also calculated, for a total of 5 determinations. The standard deviation is taken as the error. Using the best-fit value for $\theta_i$, we derive $\Omega_P$ = $-$18$\pm$5 km s$^{-1}$ pc$^{-1}$. 

However, the original method by TW assumes that the disk is thin; the
models from PT03 show that the best fit disk is quite thick ($h/r
\sim$ 0.4). We check the results of the TW method by calculating the
precession for the models of the same orientation as the best fit
model into which a known precession value has been injected. We find
that the values obtained using the TW method for these models are
systematically too high. A line is fit to the output TW precessions to
calibrate the method for the best fit orientation. We extrapolate the
fit to obtain the calibrated TW precession for our data: 62$\pm$5 km
s$^{-1}$ pc$^{-1}$. A model is generated using this precession along
with the best-fit orientation angles and the goodness of fit to the
data is calculated (Table \ref{tab:chi2}); this precession gives a
significantly worse fit to the data than the best-fit precession does,
particularly to the velocity map. A comparison of the data and the
models derived using the TW precession is shown in Appendix \ref{sec:appc}.

\section{Discussion}
\label{sec:disc}
\citet{chang2007the-origin} derived the maximum value for the precession of the eccentric disk, $
\Omega_P$, in order for gas released via stellar winds to end up on crossing orbits. Gas on these 
crossing orbits is able to collide, shock, cool, and fall into orbit around the SMBH on timescales faster 
than the viscous timescale. There, it collects into an accretion disk around P3 until it acquires 
enough mass to reach the Toomre instability limit and collapse to form stars. Based on a range of values 
for the disk thickness ($h/r$ = 0.1--0.3), they calculated that the maximum precession that would allow 
these crossing orbits was $\Omega_P \lesssim$ 3--10 km s$^{-1}$ pc$^{-1}$. 

\begin{figure*}
\begin{center}
\includegraphics[width=\textwidth]{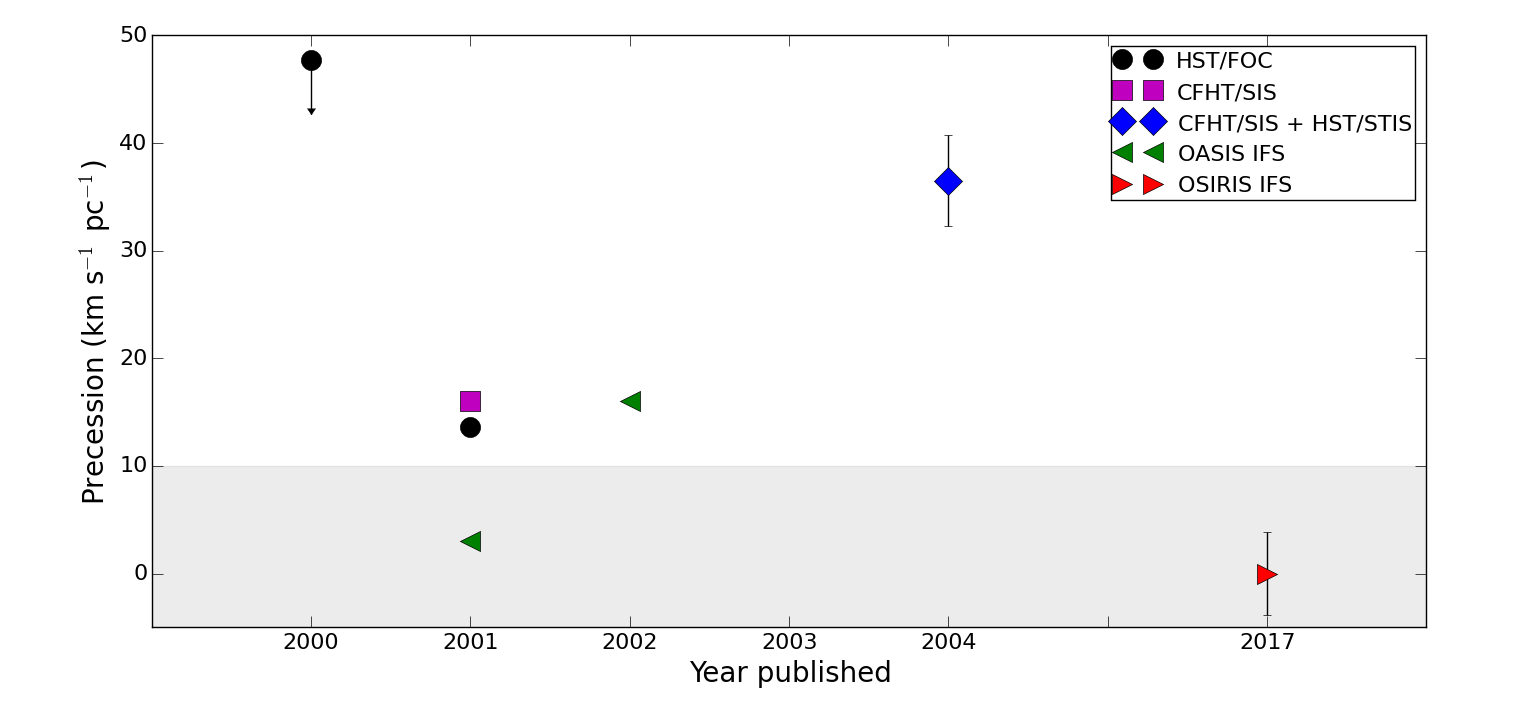}
\caption{Previous estimates of the precession value of the eccentric disk, as a function of year published. Markers represent the source of the observations used in the analysis; all are long-slit observations except for the OASIS IFS green triangles and the red triangle, representing the current work. Error bars and upper limits are shown where provided in their respective papers. The precession value from \citet{sambhus2000the-pattern} was given as a function of disk inclination and has been adjusted to match our best-fit disk inclination. The shaded gray box represents the theoretical limits from \citet{chang2007the-origin}.}
\label{fig:litprecess}
\end{center}
\end{figure*}

The model fitting in \S\ref{ssec:mfit} yielded a best-fit value for the precession equivalent to zero: 
$\Omega_P = 0.0 \pm 3.9$ km s$^{-1}$ pc$^{-1}$. This is smaller than other values found for the 
precession of the disk (see Figure \ref{fig:litprecess}), but benefits
from a combination of 2D FOV coverage and high spatial resolution 
unmatched by other observational studies. 
We note that there are many different modeling
approaches in the literature and that our adopted models of PT03
have some limitations, which may impact the derived precession
rate. Notably, the models do not include
self-gravity \citep[][\S 5]{peiris2003eccentric-disk}, nor do
we include precession in a self-consistent fashion. However, the 3D SPH
simulations of \citet{hopkins2010the-nuclear}, which include
self-gravity and precession, also predict a slow (1-5 km s$^{-1}$
pc$^{-1}$), rigid-body precession, in agreement with our results.
Out of prior observational constraints on the precession, only 
B01 obtains a similarly small value of $\Omega_P$ = 3 km s$^{-1}$ pc
$^{-1}$ with their OASIS IFS data and
three-dimensional $N$-body
modeling. \citet{sambhus2002dynamical} also fit their models to the
IFS observations from B01 and obtain a higher value of $\Omega_P$ = 16
km s$^{-1}$ pc$^{-1}$; but, they use Schwarzschild-type modeling
and a thin disk assumption.

Other modeling studies \citep{jacobs2001long-lived,
 salow2001eccentric, salow2004self-gravitating} have fit their models
to long-slit spectroscopic data and obtain values of $\Omega_P$
ranging from 14 km s$^{-1}$ pc$^{-1}$ to over 30 km s$^{-1}$
pc$^{-1}$. 
While several of the models used in these works include 
self-gravity (unlike PT03), they also assume a cold, thin disk, which 
may influence the derived precession rate.
\citet{sambhus2000the-pattern} use a modified
\citet{tremaine1984a-kinematic} method with long-slit spectroscopy to
estimate $\Omega_P \lesssim$ (sin 77$^{\circ}$/sin $i$) 30 km s$^{-1}$
pc$^{-1}$. However, we found that this method produces precession
values that are systematically offset from the known input model
values and hence this method appears to be unreliable for this system.
Given the discrepancies between precession rates
derived from different models and with new observational constraints
from our data set, it is clear that further developments in
dynamical models are needed. 

Our best-fit precession is in line with the \citet{chang2007the-origin} theory. The lack of precession in the 
eccentric disk allows gas released via stellar winds to quickly move into the vicinity of P3 and collect 
there, until enough gas has collected to collapse and form stars every 500 Myr.

\begin{deluxetable*}{crrrrrrr}
\tablecaption{Literature Parameters
\label{tab:modfit}}
\tablehead{
\colhead{Parameter} & \colhead{[1]} & \colhead{[2]} & \colhead{[3]} & \colhead{[4]} & \colhead{[5]} & \colhead{[6]} & \colhead{[7]}
}
\startdata
$\theta_l$ & - & $-$48$^{\circ~b}$ & - & - & $-$27.34$^{\circ~b}$ & $-$48$^{\circ~a,b}$ & $-$35$^{\circ}$ \\
$\theta_i$ & 77$^{\circ~a}$ & 55$^{\circ}$$\pm$5$^{\circ}$ & - & 77$^{\circ~a}$ & 51.54$^{\circ}$ & 52.5$^{\circ}$ & 57$^{\circ}$ \\
$\theta_a$ & - & - & $-$21$^{\circ~a,b}$ & - & - & - & $-$34$^{\circ}$ \\
$\Omega_P^c$ & $\lesssim$47.7$^{d}$ & 3 & 16 & 13.6 & 16 & 36.5 & - \\
\enddata
\tablenotetext{}{References: [1] \citet{sambhus2000the-pattern}, [2] B01, [3] \citet{jacobs2001long-lived}, [4] \citet{salow2001eccentric}, [5] \citet{sambhus2002dynamical}, [6] \citet{salow2004self-gravitating}, [7] \citet{brown2013three-dimensional}}
\tablenotetext{a}{Assumed, from the literature}
\tablenotetext{b}{Converted from format given in literature}
\tablenotetext{c}{Units are km s$^{-1}$ pc$^{-1}$}
\tablenotetext{d}{Given as a function of sin $\theta_{\mathrm{i}}$; adjusted to match our best-fit value}
\end{deluxetable*}

Our best-fit disk orientation is tilted with respect to both the PT03
nonaligned orientation and to the larger-scale galactic disk. It is
more misaligned with the larger-scale galactic disk than was the PT03
nonaligned model. Examination of Figure \ref{fig:l98}
and Figure \ref{fig:modflux} (right panel) shows that this is partly due to
the differential morphology of the secondary brightness peak P2 in the
NIR (see Appendix \ref{sec:appb} for a comparison of the nonaligned
models with the data they were fit to). In previous observations in
the optical, the PA of a line connecting P1 and P2 is 43$^{\circ}$
\citep{lauer1993planetary}. However, in our NIR data, the PA of the
P1$-$P2 line is aligned with the long axis of the OSIRIS FOV, or a PA
of $\sim$56$^{\circ}$. The angle $\theta_l$ has decreased compared to that found by PT03 to
compensate for this shift in morphology with wavelength. Our best-fit
value for $\theta_l$ is more in line with that found by
\citet[$\theta_l$ = $-$35$^{\circ}$]{brown2013three-dimensional} in
their modeling of the disk using observations from HST/WFPC2
\citep{lauer1998m32-/--1}, HST/STIS (B05), and OASIS (B01) or by
\citet[$\theta_l$ = $-$27.34$^{\circ}$]{sambhus2002dynamical} using
the OASIS (B01) observations (see Table \ref{tab:modfit}). In
addition, the P1 peak is narrower and more elongated along the P1$-$P2
line in the NIR than in optical observations. The best-fit value for
$\theta_i$, which controls the inclination of the disk, is equivalent
to a more face-on disk orientation than that of PT03 or other analyses in the literature. The angle $\theta_a$ roughly
controls the inclination of the disk along the minor axis, and thus
effectively adjusts the brightness at P1 and P2 to compensate for the
change in the other angles.  The source of the morphological
differences between the optical and infrared is not understood;
however, the 2D OSIRIS spectroscopy can be used to investigate the
nature of the stellar populations in the eccentric disk. 

The NIR data also allow insight into the distribution of the old
stellar population in the nuclear region. Figure \ref{fig:bhcutall} shows that the UV-bright P3 gives way to a hole in the stellar distribution at redder wavelengths. This lack of cusp at the SMBH is similar to that seen in the Milky Way \citep{do2013three-dimensional}. However, the inner core (or, potentially, hole) seen in the Milky Way's nuclear star cluster has a radius of at least 0.5 pc. At the distance of M31, this is equivalent to a core radius of 0\farcs13 on the sky, which is larger than that observed here ($<$ 0\farcs1). Future modeling efforts will aim to better model the size of the central hole in the stellar distribution in the NIR.

Previous high spatial resolution kinematics for this system were obtained using long-slit spectroscopy with HST. We find a similar kinematic structure to that seen in the STIS long-slit measurements from B05, though we note some discrepancies with our data, which may be attributable to different bulge subtraction methods. The previous highest spatial resolution kinematics with full 2D FOV coverage were limited by early generation AO correction and the resolution obtained, a factor of 4 poorer than that of the OSIRIS observations presented here, was a small improvement over seeing-limited observations. Comparison of the OASIS data from B01 with our OSIRIS kinematics, smoothed to their resolution, shows some discrepancies in alignment of the profiles, which may be attributable to differing SMBH positions. In addition, the dispersion peak reported by OASIS is much higher than that in our smoothed data; differing bulge subtraction methods may be the culprit. The results presented here show the value of the combination of the power of full 2D kinematic mapping with high spatial resolution and will be valuable for future modeling.

\section{Conclusion}
\label{sec:con}

We present high-resolution, 2D kinematic maps of the nucleus of M31, enabling us to measure the 
orientation and precession rate of the
eccentric disk of old stars. In comparison with previous HST optical
images, the infrared data presented here shows no signs of a central 
star cluster and rather shows a hole in the infrared light at the
black hole position. The 2D kinematics are largely in agreement with 
previous long-slit kinematic measurements. However, the new NIR flux
maps favor a different orientation for the eccentric nuclear disk than the previous best-fit model 
fitted to optical data of
\citet{peiris2003eccentric-disk}. The best-fit orientation for the morphology seen in the NIR is
 [$\theta_l$, $\theta_i$, $\theta_a$] = [$-$33$^{\circ}\pm$4$^{\circ}$, 
 44$^{\circ}\pm$2$^{\circ}$, $-$15$^{\circ}\pm$15$^{\circ}$], or 
 offset from the best-fit model from \citet{peiris2003eccentric-disk} by 
 [10$^{\circ}$, $-$10$^{\circ}$, 20$^{\circ}$]. 

We also present a measurement of the precession rate for the eccentric
disk of $\Omega_P = 0.0 \pm 3.9$ km s$^{-1}$ pc$^{-1}$. This slow
precession rate favors the scenario put forth by \citet{chang2007the-origin}
suggesting that stellar winds from the AGB and red giant stars in the
old eccentric disk provide the fuel for the starburst that produced
the young nuclear cluster.

\acknowledgements
We would like to thank Shelley Wright for help with OSIRIS
observations, and the OSIRIS Pipeline Working Group for their work on
the OSIRIS data reduction pipeline and discussions on its
improvement. We also thank Tod Lauer for sharing HST images.
We thank Scott Tremaine for his helpful comments on the manuscript.
HVP was
partially supported by the European Research Council (ERC)
under the European Community's Seventh Framework Programme
(FP7/2007-2013)/ERC grant agreement number 306478-CosmicDawn.
RMR acknowledges support from
NSF AST-121095, 13755, and 1518271.
AMG was supported by NSF AST-1412615.
NSO/Kitt Peak FTS data used here were produced by NSF/NOAO.
This research has made use of NASA's Astrophysics Data System.
Based on observations made with the NASA/ESA Hubble Space Telescope,
and obtained from the Hubble Legacy Archive, which is a collaboration
between the Space Telescope Science Institute (STScI/NASA), the Space
Telescope European Coordinating Facility (ST-ECF/ESA) and the Canadian
Astronomy Data Centre (CADC/NRC/CSA).
This research has made use of the Keck Observatory Archive (KOA),
which is operated by the W. M. Keck Observatory and the NASA Exoplanet
Science Institute (NExScI), under contract with the National
Aeronautics and Space Administration. 
The data presented herein were obtained at the W.M. Keck Observatory,
which is operated as a scientific partnership among the California
Institute of Technology, the University of California and the National
Aeronautics and Space Administration. The Observatory was made
possible by the generous financial support of the W.M. Keck
Foundation. The authors wish to recognize and acknowledge the very
significant cultural role and reverence that the summit of Mauna Kea
has always had within the indigenous Hawaiian community.  We are most
fortunate to have the opportunity to conduct observations from this
mountain.

\facility{Keck:II (OSIRIS), Keck:II (NIRC2)}

\appendix
\section{Bulge Subtraction}
\label{sec:bulge_subtraction}

\begin{figure*}
\begin{center}
\includegraphics[width=0.5\textwidth]{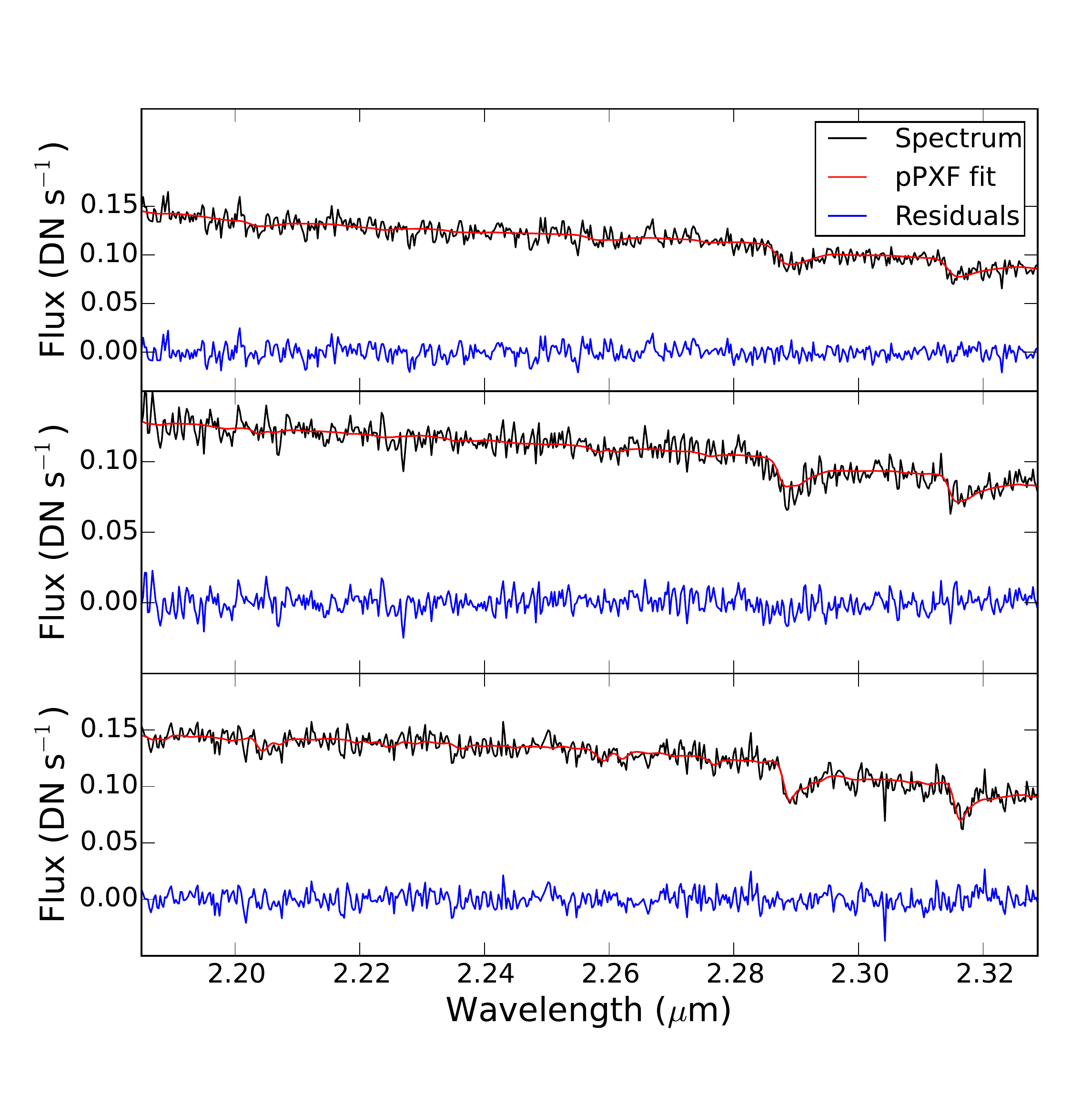}
\caption{Spectra from three representative spaxels designated as 
bulge-dominated, with the pPXF fits overlaid. Each spaxel is extracted
from a different region of the FOV and all are shown before bulge subtraction.
The pPXF fits, used to derive the bulge stellar population as part of the bulge
subtraction process, are shown in red for each spaxel. The residuals to the
fits are shown in blue.}
\label{fig:bulgespaxfit}
\end{center}
\end{figure*}

Bulge subtraction is a critical aspect of the analysis and in this
Appendix, we explore the impact of different bulge subtraction
approaches on the final kinematic fitting. 
First, we examine the pPXF fits for the spaxels determined to be
bulge-dominated; three representative spaxels are shown in 
Figure \ref{fig:bulgespaxfit} along with their pPXF fits and the fit residuals.
pPXF fits these non-bulge subtracted spaxels well.

\begin{figure*}
\begin{center}
\includegraphics[width=\textwidth]{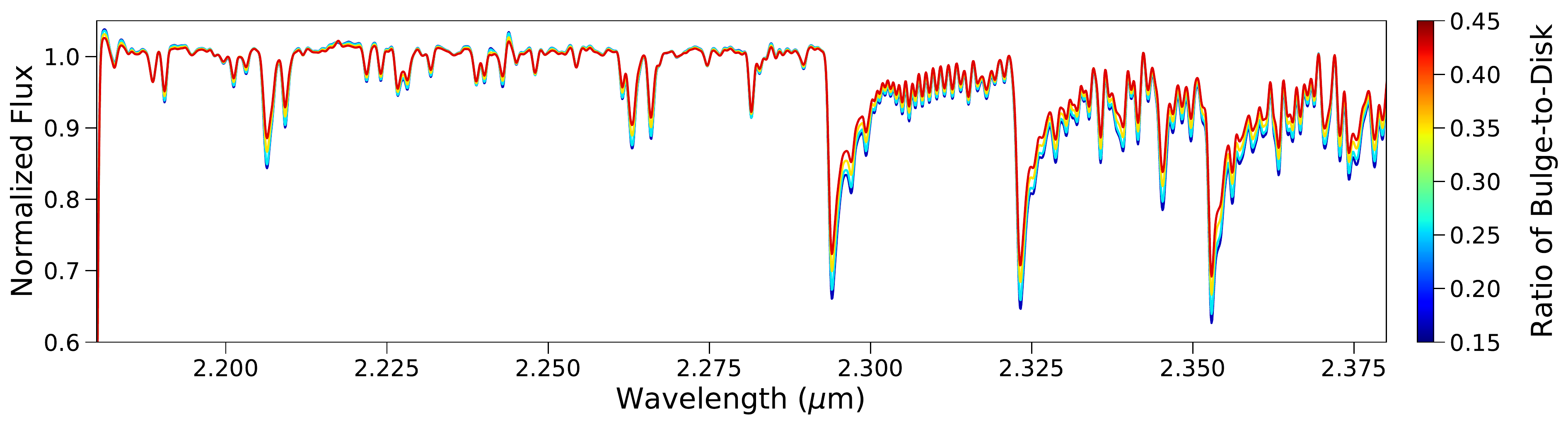}
\caption{The composite spectral template, prior to convolution
with the LOSVD, of spaxels with different bulge-to-disk ratios.
The spectral template varies minimally between spaxels
with different bulge contributions.
\label{fig:bulge_template_vary}}
\end{center}
\end{figure*}

Second, we test our 
assumption that the outer portions of the data cube give a
representative spectrum of the bulge. The family of best-fit 
templates was fairly uniform over the entire FOV. Figure 
\ref{fig:bulge_template_vary} shows the best-fit intrinsic spectrum,
before applying an LOSVD, for different bulge surface brightness
ratios: the variation is minimal. Further, the broadband
colors of the eccentric disk and the larger bulge have been shown
to be identical \citep{saglia2010the-old-and-heavy} suggesting that they contain
the same population, with the exception of the more concentrated young
nuclear cluster inside 0\farcs1. 

\begin{figure*}
\begin{center}
\includegraphics[width=\textwidth]{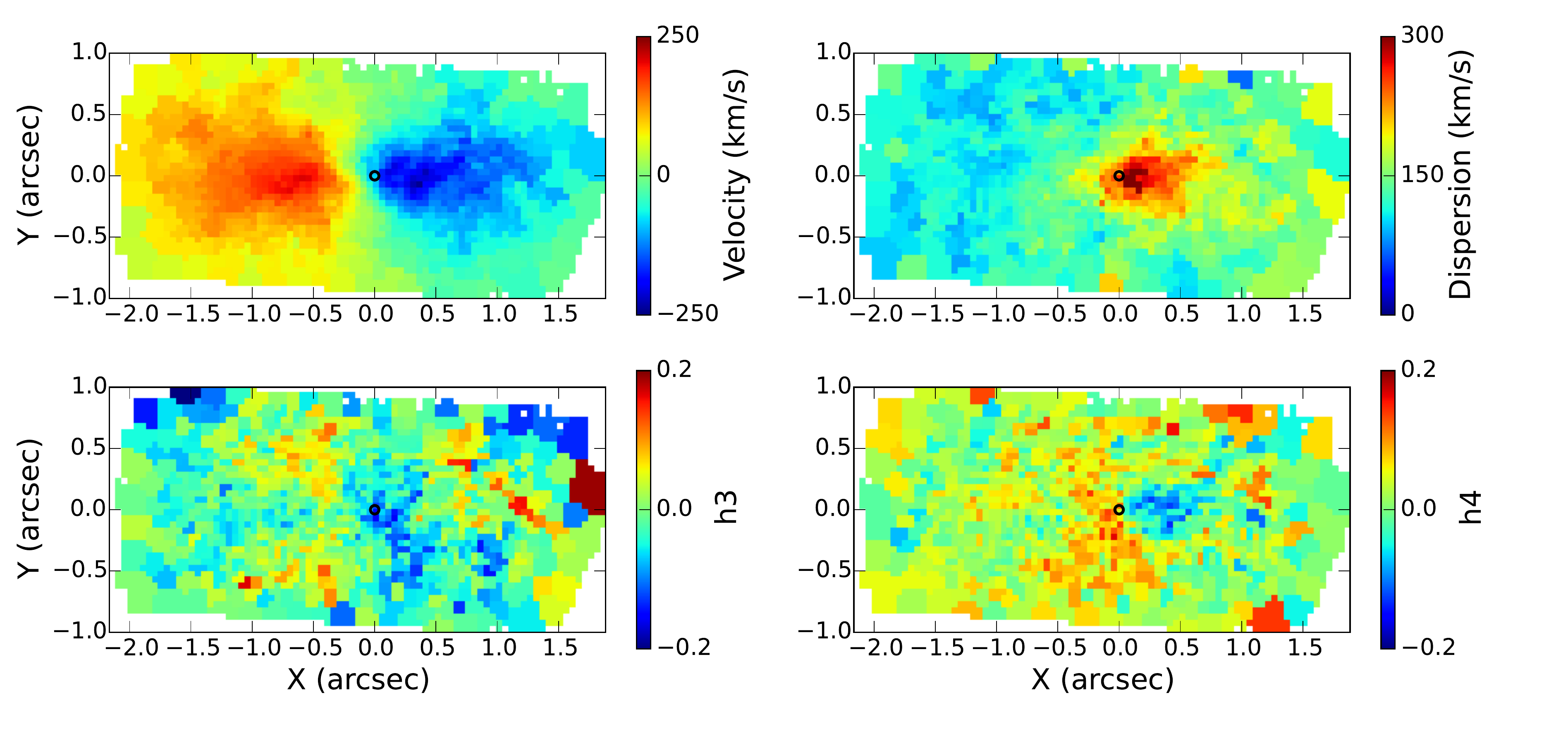}
\caption{Kinematic maps, before bulge subtraction. The tessellation
  pattern is the same as that used for the bulge-subtracted data, for
  ease of comparison with Figure \ref{fig:kinmaps}. While some
  correlations are present between the moments, the low SNR of $h3$
  and $h4$ and the presence of systematic errors make it difficult to 
  draw definitive conclusions from the higher-order moments. }
\label{fig:kinmaps_nobs}
\end{center}
\end{figure*}

\begin{figure*}
\begin{center}
\includegraphics[width=0.8\textwidth]{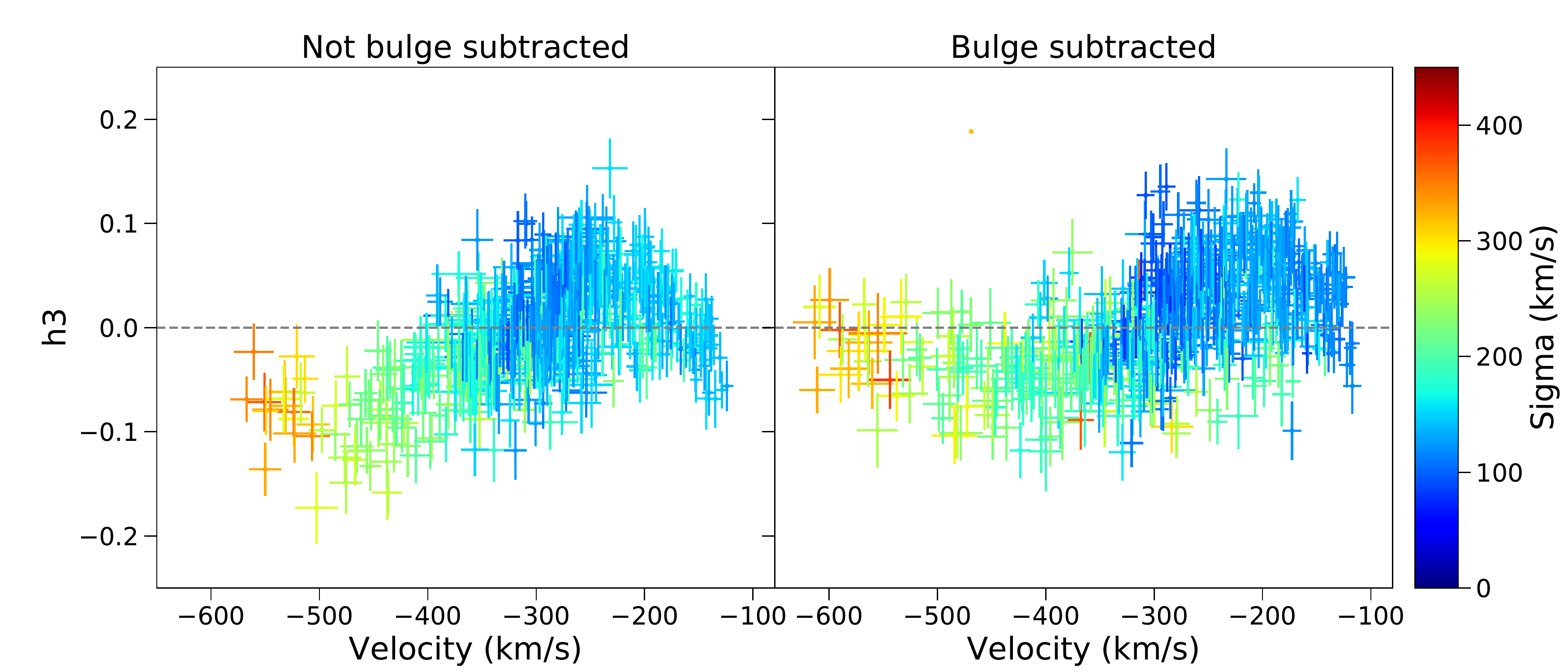}
\caption{
Correlations between $h3$ and velocity are strong before bulge
subtraction ({\em left}) and decrease significantly after
bulge subtraction ({\em right}). A similar correlation and
improvement is also seen between $h3$ and the velocity dispersion,
as shown in color.
\label{fig:vel_moment_correlations}
}
\end{center}
\end{figure*}

Next, we explore the quality of our bulge subtraction in more detail
by examining the correlation between the different velocity moments. 
Prior to bulge subtraction, there are strong correlations between
$h3$ and the velocity as seen in the kinematic maps for the non-bulge 
subtracted data, Figure \ref{fig:kinmaps_nobs}. 
This correlation is verified bin by bin in Figure
\ref{fig:vel_moment_correlations}, left panel.
Presumably, this effect comes from the fast disk-rotation shifting
absorption lines across the slowly rotating (or static, as we assume),
unsubtracted bulge spectrum, which contributes a skewness to the
resulting LOSVD. Interestingly, a similar physical explanation produced an
anticorrelation instead of a correlation between $v$ and $h3$ in at
least one other integral-field data set for a different galaxy
\citep{Menezes:2018}. 
The correlations between the lower-order
moments ($v$ and $\sigma$) and the higher order moments ($h3$ and
$h4$) are also of great interest as they can be interesting markers of other
dynamical features such as bars \citep[e.g.][]{Iannuzzi:2015}. However, we believe
it is premature to use the higher order moments $h3$ and $h4$ to draw
scientific conclusions given their large errorbars and systematic
uncertainties (see \S\ref{sec:res_kinematics} for details). 
Furthermore, after bulge subtraction, this correlation is
largely removed as shown in Figure \ref{fig:vel_moment_correlations},
right panel.

Lastly, we test an extreme case of no bulge subtraction and 
explore the impact on the resulting disk orientation and precession
speed. For the non-bulge-subtracted kinematic maps, the resulting best-fit
inclination ($\theta_i$) differs by 15$^\circ$, going from 44$^\circ$
in the bulge-subtracted case to 29$^\circ$ in the non-bulge subtracted
case, which is a very significant change relative to the uncertainties
(Table \ref{tab:modfit}). The $\theta_a$ parameter changes
from $-14.5^\circ$ to $-19.5^\circ$, within the uncertainties. 
There is no change in $\theta_l$ and the precession
rate changes from 0 km s$^{-1}$ pc$^{-1}$ to 5 km s$^{-1}$ pc$^{-1}$,
within the uncertainties. The $\tilde{\chi}^2$ is slighly higher in the non-bulge
subtracted case, going from 3.37 to 3.48.  With the exception of the
inclination, the final disk parameters are insensitive to errors in bulge subtraction.

The main conclusion of our tests is that bulge-subtraction is
essential; however, even in the most extreme case, the impact on the
resulting disk properties is negligable, except for the inclination estimate.

\section{SMBH alignment sources}
\label{sec:appa}

\begin{figure*}[htbp]
\begin{center}
\includegraphics[width=\textwidth]{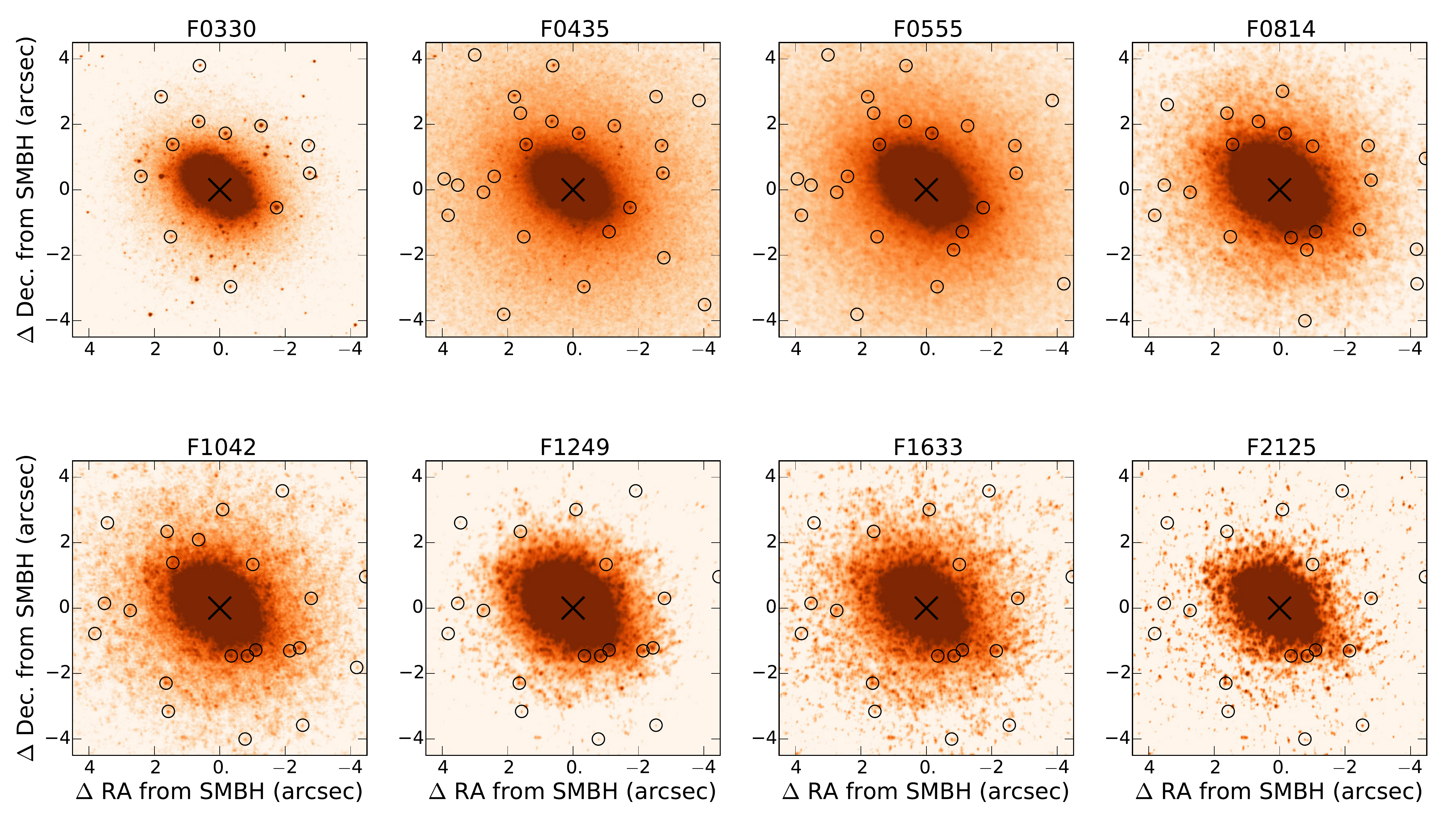}
\caption{Images are aligned as in \S\ref{ssec:bhpos}, and are
  shown scaled to match the F330W frame (pixel scale of 0.025$''$
  pixel$^{-1}$). The position of the SMBH is marked with the black
  cross and is accurate to 0.033$''$ in the K'-band frame. The color
  scaling has been chosen to emphasize the faint outer region and the
  compact sources used for alignment. The alignment sources are
  circled in black.}
\label{fig:bhcutallwide}
\end{center}
\end{figure*}

P3, the compact young cluster assumed to be coincident with M31's
SMBH, is bright in the UV but dark in the NIR. Locating the SMBH in
the OSIRIS data thus requires registering the UV and the NIR
images. However, as there are essentially no compact sources bright in
both bandpasses, we instead register pairs of frames adjacent in
wavelength. We step through a total of 8 frames, starting in the UV
and ending in the NIR $K'-$band. The alignment sources used are
circled in Figure \ref{fig:bhcutallwide}. The position of P3 is marked
with a black cross, but as the color scaling is set to emphasize the
fainter alignment sources, the inner nuclear disk structure is not
visible in this figure. 

\section{TW method model residuals}
\label{sec:appc}

\begin{figure*}
\noindent
\centering
\begin{minipage}[t]{0.49\textwidth}
\includegraphics[width=\columnwidth]{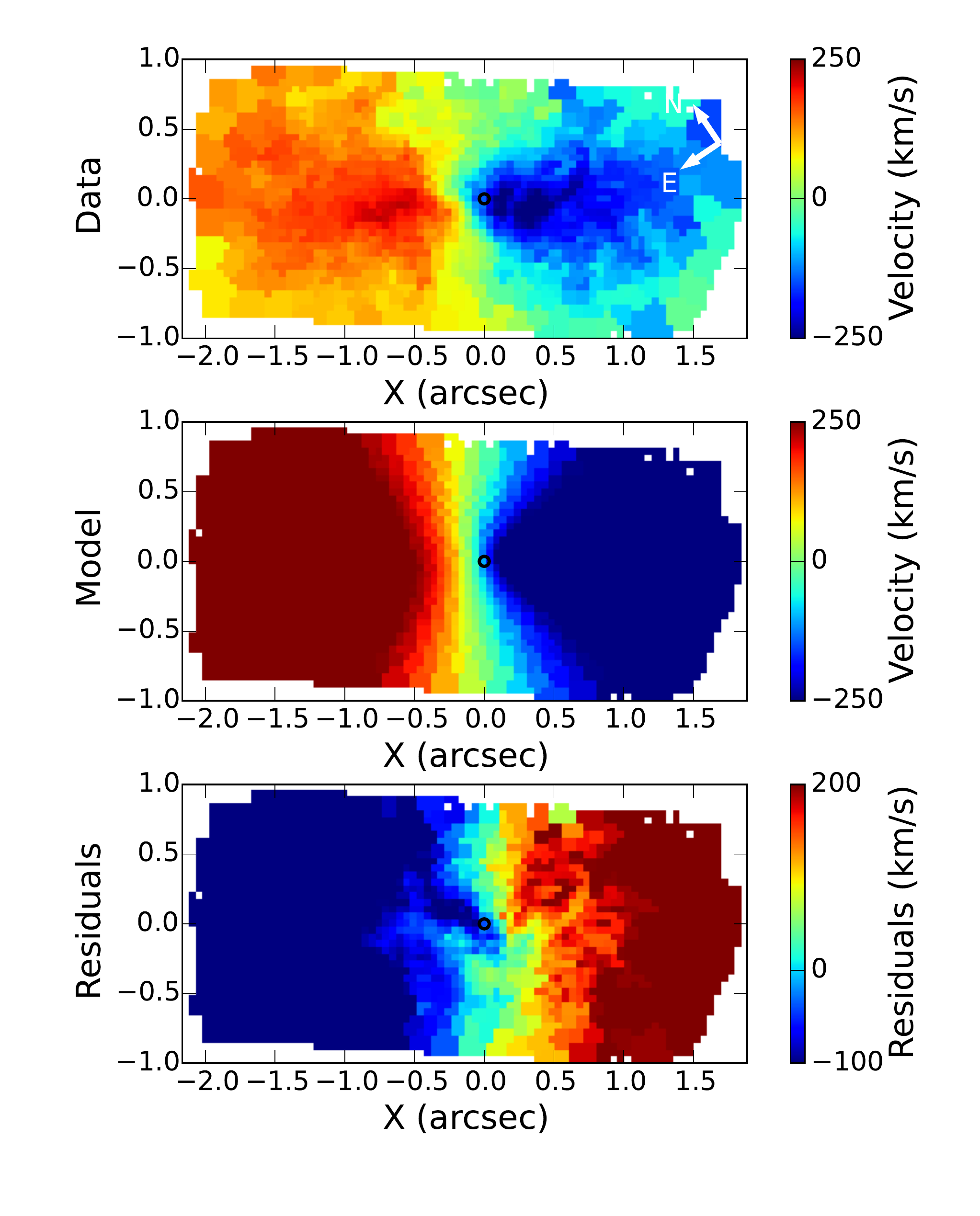}
\end{minipage}
\begin{minipage}[t]{0.49\textwidth}
\includegraphics[width=\columnwidth]{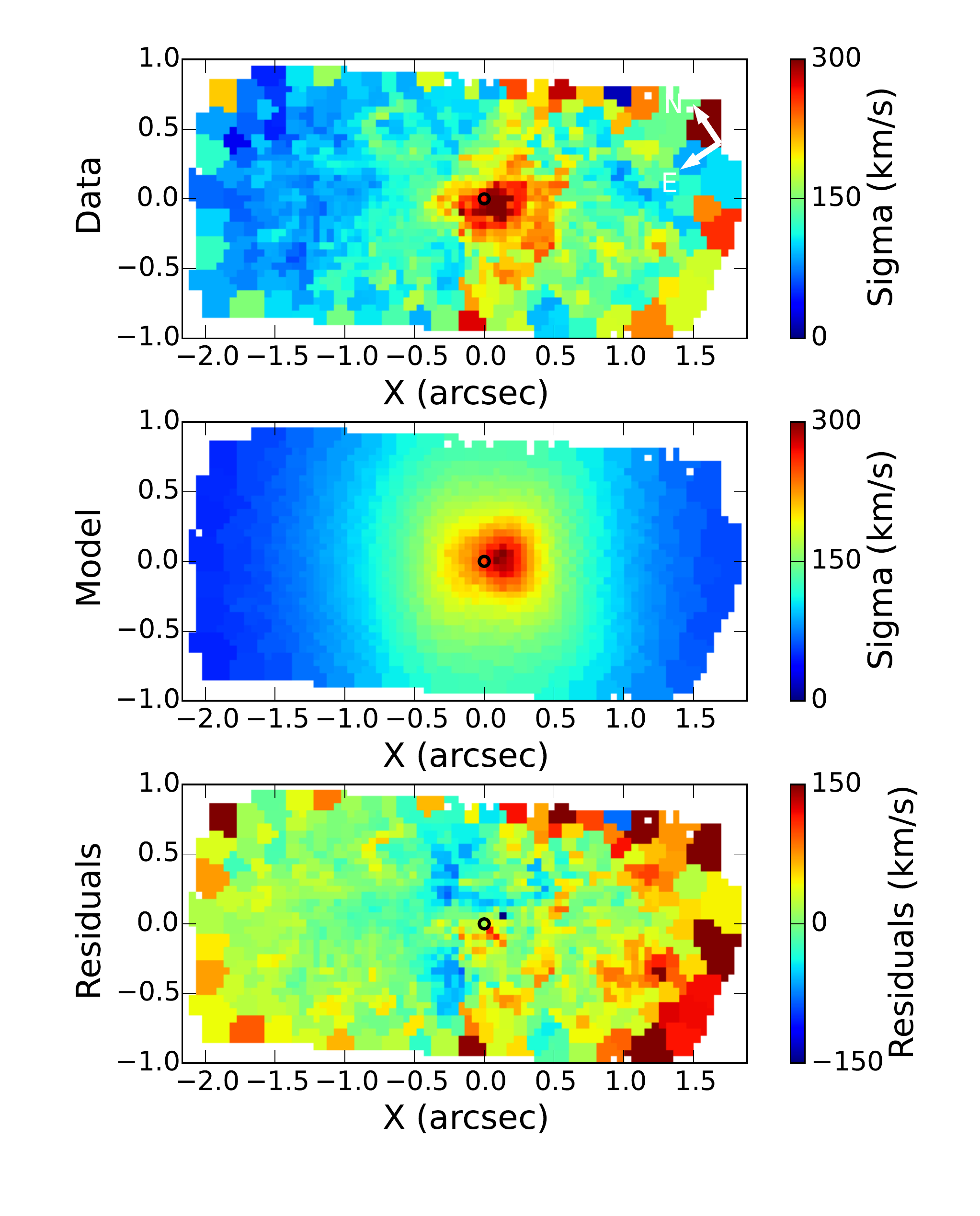}
\end{minipage}
\caption{Comparison of OSIRIS data with the models suggested by the TW method. The orientation is the same as for our best-fit models, but the precession was derived using the TW method and calibrated using models with known precessions. While the dispersion residuals are small, the velocity residuals are much higher than in with our best-fit precession value of 0 km s$^{-1}$ pc$^{-1}$. The flux residuals are unchanged by the high precession and are identical to those in the right panel of Figure \ref{fig:modflux}.}
\label{fig:twmod}
\end{figure*}

The precession of the eccentric disk is found using the method from
\citet{tremaine1984a-kinematic}, which only requires a one-dimensional
slice in both surface brightness and line-of-sight velocity. We show
in \S\ref{ssec:precess} that this produces a value for the precession,
$\Omega_P$ = 63.0$\pm$5.3 km s$^{-1}$ pc$^{-1}$, that is higher than
that derived from the two-dimensional model fitting. The
goodness-of-fit is not improved for this value of the
precession. Figure \ref{fig:twmod} shows the OSIRIS flux and kinematic maps compared to the models derived using the best-fit orientation and the precession from the TW method. While the flux and dispersion residuals are small, the velocity residuals are much higher than for the best-fit precession value of $\Omega_P = 0.0 \pm 3.9$ km s$^{-1}$ pc$^{-1}$.

\section{Nonaligned model residuals with HST/F555W imaging}
\label{sec:appb}

\begin{figure}
\begin{center}
\includegraphics[width=\textwidth]{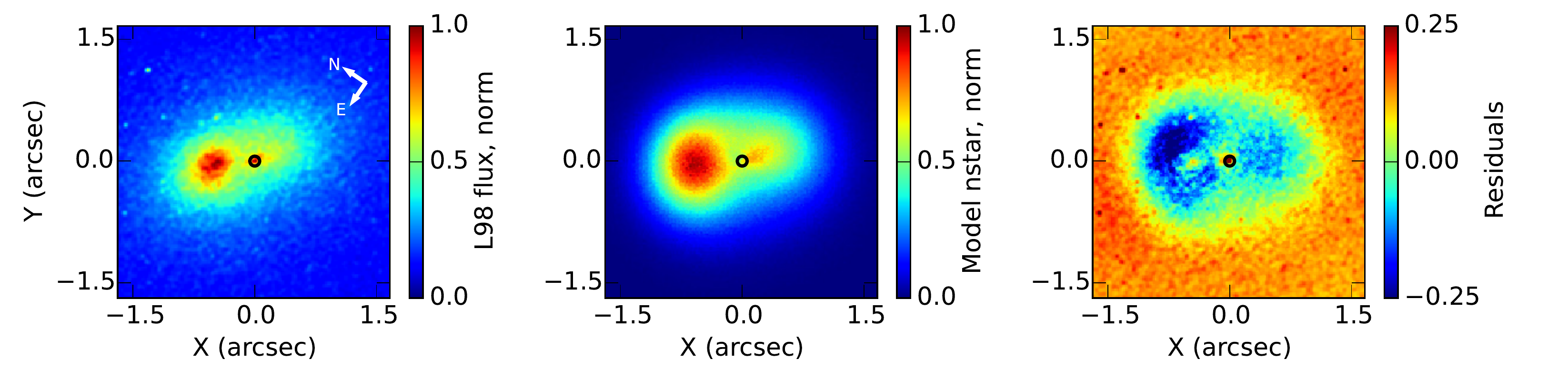}
\caption{Comparison of F555W flux map from \citet{lauer1998m32-/--1}
  with the nonaligned models from PT03. These photometric data were
  used for the fitting of the original models. In all, the SMBH
  position is marked with the black circle, and the PA =
  55.7$^{\circ}$, the original PA from \citet{lauer1998m32-/--1} (note
  that this PA differs from that of the OSIRIS figures). \emph{Left:}
  F555W flux map, scaled so the peak flux is equal to
  1. \emph{Middle:} Modeled number of stars per spatial bin, binned to
  the subsampled F555W pixel scale of 0\farcs0228 pix$^{-1}$. The
  models are oriented using the nonaligned orientation from PT03
  (angles given in Table \ref{tab:modfit}) and scaled so the peak is equal to 1. \emph{Right:} Residuals, data minus models.}
\label{fig:l98mod}
\end{center}
\end{figure}

\citet{lauer1998m32-/--1} observed the nuclear eccentric disk in M31
with HST/WFPC2 in the F300W, F555W, and F814W filters. We show the
F555W image in Figure \ref{fig:l98mod} (left panel). PT03 used this
photometry, in combination with spectroscopy from KB99, to fit the
orientation of their models. Their best-fit models to the data are not
aligned with the larger-scale galactic disk of M31 and are designated
as the nonaligned models (Table \ref{tab:modfit}). Figure
\ref{fig:l98mod} shows the nonaligned models and the residuals between
the F555W photometry and these models. Comparison of this figure with
Figure \ref{fig:modflux} shows that while the nonaligned models are a
good fit to the F555W photometry, they are not a good fit to the
OSIRIS NIR photometry.

\bibliography{main.bib}

\end{document}